\documentclass[12pt]{article}
\usepackage{epsfig}
\def\ep              {\varepsilon }
\def\MS              {$\overline{\rm MS}$ }
\def\DR              {$\overline{\rm DR}$ }
\def\gl              {{\mathrm g}}
\def\Lag             {{\cal L}}
\def\beqaa           {\begin{eqnarray}}
\def\eeqaa           {\end{eqnarray}}
\def\demu            {\partial_{\mu}}

\def\ti              {\tilde}
\def\sg              {\ti \gl}
\def\sq              {\ti q}
\def\beq             {\begin{equation}}
\def\eeq             {\end{equation}}

\def\msg             {m_{\sg}}

\newcommand{\msq}[1]   {m_{\ti q_{#1} }}

\newcommand{\GeV}{\unskip\,\mathrm{GeV}}

\def\btd {{\phantom .}\bar{\!\!\!\!\!{\phantom a}\tilde d}}
\def\btu {\!\!{\phantom l}\bar{\!\!\!\!\!{\phantom m}\tilde u}}
\def\btq {\!\!{\phantom a}\bar{\!\!\!\!{\phantom a}\tilde q}}

\newcommand{\eps}{\varepsilon}

\begin{document}

\begin{center}
{\Large \bf  Two-loop \boldmath ${\cal O}(\alpha_s^2)$ MSSM corrections
to the pole masses of heavy quarks}
\\ \vspace*{5mm} A.~Bednyakov$^a$, A.~Onishchenko$^{b,c,d}$,
V.~Velizhanin$^e$, O.~Veretin$^b$
\end{center}

\begin{center}
a) Moscow State University, Moscow, Russia \\
\vspace*{0.5cm}
b) Institut f\"ur Theoretische Teilchenphysik \\
Universit\"at Karlsruhe, D-76128 Karlsruhe, Germany \\
\vspace*{0.5cm}
c) Institute for High Energy Physics, Protvino, Russia \\
\vspace*{0.5cm}
d) Institute for Theoretical and Experimental Physics, \\
Moscow, Russia \\
\vspace*{0.5cm}
e) Bogoliubov Laboratory of Theoretical Physics, \\
Joint Institute for Nuclear Research, Dubna, Russia
\end{center}

\vspace*{0.5cm}

\begin{abstract}
We present the results for two-loop MSSM corrections to the relation
between pole and running masses of heavy quarks up to the order
${\cal O}(\alpha_s^2)$. The running masses are defined in the \DR
scheme, usually  used in multiloop calculations in supersymmetric
theories. The analysis of the value of these corrections in
different regions of the parameter space is provided.
\end{abstract}

\section{Introduction}

Recently  a lot of theoretical and experimental efforts have been spent
to find some new physics beyond that described by the Standard Model
(SM). One of the most popular ways to extend SM is to make our
world supersymmetric at some higher scale than those currently
tested by experimental facilities over the world. A way of doing it
minimally and at the same time to have renormalizable local
quantum filed theory is to consider the $N=1$ SUSY field theory. It
is obtained from SM by replacing SM fields with groups of
fields, forming representations of the $N=1$ SUSY algebra, and
introducing an additional Higgs doublet required by supersymmetry.
This way we obtain the Minimal Supersymmetric Standard Model
(MSSM) \cite{Haber:1984rc,Nilles:1983ge,Barbieri:xf}.

Huge theoretical research in this direction was made in the last
few years as well as parameter space of this model was heavily
constrained by experimentalists. Today people consider radiative
SUSY corrections to SM precision observables and try to observe
SUSY particles directly in an experiment. The precision with
which masses of heavy quarks are known at present, requires that
higher order corrections, in addition to the leading one-loop SUSY
corrections, be taken into account.

It is the aim of the present article to calculate the ``leading''
${\cal O}(\alpha_s^2)$ MSSM corrections to the pole masses of
heavy quarks and to see how they affect the SUSY particle spectra 
obtained from the renormalization group analysis with universal
boundary conditions at the GUT scale. In fact we will be interested in
the relation between the pole mass of a heavy quark and the running
heavy quark \DR mass. The latter is defined as a quantity computed
in dimensional reduction and renormalized minimally. \footnote{The \DR
scheme \cite{Siegel}, which is similar to the \MS scheme, is usually
used for calculation in supersymmetric models, because contrary to
the \MS scheme it manifestly preserves supersymmetry, at least for not
very high loop orders \cite{Avdeev:1982xy}. The details about the
renormalization scheme used will be given below.} By "leading"
corrections we mean those, which are mediated by strong
interactions and formally they should be dominant. However, as it
is known from one-loop calculations, in the case of $b$-quark
there exists a large contribution from a stop-chargino loop enhanced
by the $t$-quark Yukawa coupling, $\tan\beta$ or by the supersymmetric
Higgs mass parameter $\mu$. So, in this sense, our results for
the $b$-quark are not complete. But, they can be used to see the
typical size of these corrections and already make some
conclusions on whether they should be accounted for in an accurate
analysis or not.

To compute needed diagrams we use the large mass expansion
procedure and for the moment we restrict ourselves only to terms
up to the ${\cal O}(m_b^2/M_{\rm SUSY}^2)$ order in the case of
$b$-quark and up to the ${\cal O}(m_t^2/M_{\rm SUSY}^2)$ order in the
case of $t$-quark. It is a good approximation for the $b$-quark.
However, as our numerical analysis shows, such approximation does
not work well in the case of $t$-quark and higher terms in large
mass expansion should be taken into account. We suppose to
calculate missing corrections from stop-chargino loops to
the $b$-quark pole mass and to provide more terms in the
expansion in a relation between the top pole and \DR masses elsewhere.

The paper is organized as follows. In section 2, we define
quantities we want to compute and make comments on the choice of
our renormalization scheme. In section 3, we describe our model ---
the supersymmetric QCD (a subset of MSSM, relevant to the
calculation of $\alpha_s^n$ corrections) and present in detail the
renormalization procedure. Section 4 contains our results and
numerical analysis of the effect produced by our correction on
supersymmetric particle spectra. The latter were computed with the
help of the SoftSUSY program \cite{SoftSUSY} under universal boundary
conditions at GUT scale. Finally, section 5 contains our
conclusion.

\section{Pole mass and choice of renormalization scheme}

First, we would like to note that there are several quark mass
definitions. This is mainly due to the fact that quarks were not
ever observed as free particles. Therefore, the definition of their
masses rely heavily on theoretical constructions. Different
definitions exist referring to different renormalization schemes
used in quantum field theories. We will mention only a few most
popular ones --- \MS mass $\overline m$, pole mass $M$ and \DR mass
$m$. In the nonsupersymmetric QCD the relations between these
three masses are known (to two-loops between \DR and pole masses
\cite{Avdeev:1997sz} and to two
\cite{Gray:1990yh,Fleischer:1998dw} and to three
\cite{Chetyrkin:1999ys,Melnikov:2000qh} loops between \MS and pole
masses) and in the supersymmetric QCD case there exists only one-loop
relation between \DR and pole masses of heavy quarks
\cite{Hempfling:1993kv,Hall:1993gn,SUSY94&Wright,Donini95,Pierce:1996zz}.
In this work, we will establish the two-loop relation between \DR
and pole masses in supersymmetric QCD.

In \DR scheme, particle masses and couplings depend explicitly on the
renormalization scale $\bar{\mu}$ which is taken in applications equal
to the characteristic scale of a studied process. The renormalization
scale dependence of these quantities could be conveniently
described by renormalization group equations. They are known very
well for MSSM and supersymmetric QCD. It should be noted that \DR
mass is a short distance quantity. That is, it is sensitive only
to short distance effects. To describe processes with the
characteristic scale of the order of quark mass itself, a different
on-shell mass definition is used and here pole mass of a particle
comes into play.

The particle pole mass
is defined by the singularity of
the corresponding two-point function. As explicit perturbative
calculations showed, the pole mass of a quark is an infrared finite
and gauge invariant quantity. For this reason it is considered as
a physically meaningful quantity \cite{else}.  We will restrict
ourselves to perturbation theory only and will not analyze the
exact nature of the two-point function singularity which may
involve some nonperturbative dynamics. In the present paper, we
employ the definition of the pole quark mass, where it equals 
the value of $p$, at which the inverse quark propagator turns to zero.

The pole mass $M$ of a quark is defined as a formal solution for
$\hat p$ (in the Minkowski metric) at which the reciprocal of the
connected full propagator equals 
zero
\begin{equation}
\hat p - m - \Sigma(\hat p, m) = 0,
\label{pole}
\end{equation}
where $\Sigma(\hat p, m) = m \Sigma_1 (p^2,m)
+ (\hat p - m)\Sigma_2 (p^2,m)$ is the
one-particle-irreducible two-point function, $m$ may stand for the
bare or renormalized mass, $m_{\rm bare}$ or $m_{\rm ren}$,
depending on the prescription used in evaluating
$\Sigma$. The solution to eq.~(\ref{pole}) is sought order by order in
 perturbation theory. To two loops we have
\begin{equation}
M = m +\Sigma^{(1)}(m,m) +\Sigma^{(2)}(m,m)
+ \Sigma^{(1)}(m,m)~ \Sigma^{(1)'}(m,m) + {\cal O}(\Sigma^{(3)}),
\label{2loop}
\end{equation}
where $\Sigma^{(L)}$ is the $L$-loop contribution to $\Sigma$, and
the prime denotes the derivative with respect to the first
argument. In what follows, we will be interested in the relation
between the pole quark mass $M$ and the running \DR mass $m$ computed in
MSSM up to the ${\cal O}(\alpha_s^2)$ order. Technically, to solve
this problem, we need to evaluate $\approx 200$ two-loop
propagator type diagrams involving many different mass scales (see
Fig.~\ref{twoloopdiag}). In general, it is quite a complex problem to be solved
exactly. However, in our case all mass scales  can be divided into
two groups denoted by $m_{\rm soft}$ and $m_{\rm hard}$, such that
scales from $m_{\rm soft}$ are much less than each of the scales from the
$m_{\rm hard}$ group

\begin{center}
\begin{tabular}{lcc}
& $m_{\rm soft}$ & $m_{\rm hard}$ \\
case of $b$-quark: & $m_b$ & $m_t$, $m_{\tilde q}$, $m_{\sg}$ \\
case of $t$-quark: & $m_t$ &  $m_{\tilde q}$, $m_{\sg}$
\end{tabular}
\end{center}
\noindent
Here $m_{\sg}$ is the gluino mass, $m_{\tilde q}$ stands
for different squark masses. We also explicitly denoted
which mass scales belong to soft or hard groups in the cases
of $b$- and $t$-quarks.

  Given this mass hierarchy, one can employ the large mass expansion
procedure to reduce evaluation of multi scale two-loop integrals
to the calculation of single scale on-shell two-loop integrals,
two-loop tadpole integrals with two scales and products of
one-loop on-shell integrals and one-loop tadpole diagrams. To
compute two-loop single scale on-shell integrals, we made use of the
ONSHELL2 package \cite{onshell2}. For evaluation of two-loop
tadpole diagrams with two scales, the recurrence relations of
\cite{Davydychev} were used. Calculation of one-loop
self-energies, including their derivatives with respect to
momentum, is an easy task and we will not make further comments on
it.

Now let us make some comments on the choice of a renormalization
scheme. Here we use the same renormalization prescription, as in
\cite{Avdeev:1997sz}. To be more specific, we
use regularization by dimensional reduction, a modification of the
conventional dimensional regularization, originally proposed in
\cite{Siegel}. In this procedure, the vector and spinor algebras in
the numerator of Feynman diagrams are four-dimensional, which is
the requirement imposed by supersymmetry. However, to make sense
from divergent momentum integrals we need to regularize them by
noninteger space-time dimension $d=4-2\varepsilon$. 
For
quantum corrections not to break gauge invariance, we need a
cancellation of squared momenta in both the numerator and the denominator
of Feynman integrals. Thus, the momenta should form 
$d$-dimensional subspace in four dimensions. As the momenta become
$d$-dimensional, four-vectors naturally split into true
$d$-vectors and so-called $\varepsilon$-scalars, obtained in the
process of dimensional reduction of the original four-dimensional
Lagrangian. The $\varepsilon$-scalars are nothing, but matter
fields. Their appearance is the only difference from the
conventional dimensional regularization.

We would like to note that, in general, renormalization of
$\varepsilon$-scalars and their interactions is not identical to
that of vectors, so the original four-covariance may be spoiled
by quantum corrections. Moreover, quantum corrections may also
generate a mass for $\varepsilon$-scalars. There is
arbitrariness in choosing the renormalization scheme for this mass
\cite{jones}. A consistent way is to choose the finite
$\varepsilon$-scalar mass counterterm, so that the pole (and
renormalized) mass of the $\varepsilon$-scalars equals zero.  The
$\varepsilon$-scalar field renormalization is left minimal. Other
renormalizations are done minimally, that is by subtracting only
poles in $\varepsilon$.

\section{Definition of the model and renormalization procedure}

For our purposes here we do not need the complete MSSM, but only
its part --- supersymmetric extension of QCD (SUSY QCD).
In the superfield formulation SUSY QCD Lagrangian
consists of two parts: the rigid supersymmetric part and the part containing
soft supersymmetry breaking terms.
\begin{eqnarray}
{\cal L}_{\rm rigid} &=& \int d^2\theta \, \frac{1}{2}\,{\rm  Tr}
W^{\alpha}W_{\alpha} + \int d^2\bar{\theta} \, \frac{1}{2}\,{\rm  Tr}
\bar{W}_{\dot \alpha}\bar{W}^{\dot \alpha} \label{rigidlag} +
\int d^2\theta \,{\cal W} + \int d^2\bar{\theta} \, \bar{\cal W} \nonumber\\
&+& \int d^2\theta \,d^2\bar{\theta} \,
\left[\bar Q e^{2g_3T^a V^a}Q +
\bar U e^{-2g_3(T^a)^* V^a}U +
\bar D e^{-2g_3(T^a)^* V^a}D\right] , 
\end{eqnarray}
where $Q=Q_L$, $U=U_R$, $D=D_R$,
$T^a=\lambda^a/2$ ($a$=1,...,8) with $\lambda^a$ being the Gell-Mann matrices
\footnote{Note that U and D are colour antitriplets
      and the anticolour generator is $-(T^a)^*$}
 and the gauge field strength tensors are
 $$W_{\alpha}=-\frac {1}{8g_3}\bar D^2e^{-2g_3 T^a V^a}D_\alpha
    e^{2g_3 T^a V^a},
   \ \ \ \bar W_{\dot \alpha}=-\frac {1}{8g_3} D^2e^{2g_3 T^a
V^a}\bar D_{\dot \alpha} e^{-2g_3 T^a V^a}$$

The superpotential of the ordinary SUSY QCD contains only
quadratic mass terms for chiral (antichiral) superfields.
Considering SUSY QCD as a part of MSSM, where all masses of
particles are generated through the Higgs mechanism, the
superpotential takes the following form

\begin{equation}
{\cal  W}=\epsilon^{ij}\left[y_d D H_1^i Q^j  +
y_u U H_2^j Q^i +\mu H_1^j H_2^i \right],\label{rigidsuppot}
\end{equation}
where the $y_d$ and $y_u$ are the Yukawa coupling constants carrying the
generation indices which have been suppressed (as well as group
indices). In this work we will need the Higgs fields only to
generate masses of our particles and will not be interested in the
detailed structure of the MSSM Higgs sector.

To  perform the supersymmetry breaking that satisfies the requirement of
"softness", we introduce a gluino mass term,
soft masses of  scalar superpartners of quarks and
soft trilinear couplings with Higgs fields,
which may be written in terms of $N=1$ superfields, provided
one introduces external spurion superfields~\cite{spurion}
\begin{eqnarray}
{\cal L}_{\rm soft} &=&
-\frac 12\int d^2\theta \, 2 M_3\theta^2 {\rm Tr}W^{\alpha}W_{\alpha}
- \frac 12\int d^2\bar{\theta} \, 2\bar{M}_3\bar{\theta}^2 {\rm
Tr}\bar{W}_{\dot \alpha}\bar{W}^{\dot \alpha}  \nonumber \\
&-&\int d^2 \theta \, d^2\bar{\theta}
\left[M^2_{\tilde q}\bar Q e^{2g_3 T^a V^a} Q
+ M^2_{\tilde u}\bar U e^{-2g_3 (T^a)^* V^a} U
+ M^2_{\tilde d}\bar D
e^{-2g_3 (T^a)^* V^a} D \right] \theta^2 \bar{\theta}^2 \nonumber \\
&-&\int d^2\theta \,
    \epsilon^{ij}
\left[y_d A_d D H_1^i Q^j  +
y_u A_u U H_2^j Q^i \right]\theta^2 + {\rm h.c.}
\label{lagsoft}
\end{eqnarray}

The chiral superfields have the following component fields
representations
\begin{eqnarray}
Q&=&\tilde q_L-i\,\theta\sigma^\mu
\bar\theta\partial_\mu\tilde q_L-\frac14
\theta\theta\bar\theta\bar\theta
\partial^\mu\partial_\mu\tilde q_L+\sqrt{2}\theta q_L-\frac{i}{\sqrt{2}}
\theta\theta\bar\theta\bar\sigma^\mu\partial_\mu q_L+
\theta\theta F_q \,,\\
U&=& \btu_R-i\,\theta\sigma^\mu
\bar\theta\partial_\mu \btu_R-\frac14
\theta\theta\bar\theta\bar\theta
\partial^\mu\partial_\mu \btu_R+\sqrt{2}\theta u_R-
\frac{i}{\sqrt{2}}
\theta\theta\bar\theta\bar\sigma^\mu\partial_\mu u_R+
\theta\theta F_u \,,\\
D&=& \btd_R-i\,\theta\sigma^\mu
\bar\theta\partial_\mu \btd_R -\frac14
\theta\theta\bar\theta\bar\theta
\partial^\mu\partial_\mu \btd_R +\sqrt{2}\theta d_R-
\frac{i}{\sqrt{2}}
\theta\theta\bar\theta\bar\sigma^\mu\partial_\mu d_R+
\theta\theta F_d
\end{eqnarray}
and for the vector superfields in the Wess-Zumino gauge we have
\begin{eqnarray}
V^a&=&\theta\sigma^\mu\bar\theta
G^a_\mu+\theta\theta\bar\theta{\bar{\tilde \gl}}^a+\bar\theta\bar\theta\theta{\tilde \gl}^a
+\frac12\theta\theta\bar\theta\bar\theta D^a.
\end{eqnarray}
Then, rewriting everything in the component fields, the Lagrangian
of SUSY QCD takes the form ($q_R=(u_R,d_R)$)
\begin{eqnarray}
{\cal L}&=&-\frac 14 G^a_{\mu\nu}G^{a\;\mu\nu}
+i{\bar {\tilde \gl}}^a \bar \sigma^\mu D_\mu {\tilde \gl}^a
+i\bar q_R \bar \sigma^\mu D_\mu q_R
+i\bar q_L \bar \sigma^\mu D_\mu q_L\nonumber\\
&+& \left(D^\mu{\tilde q}_R\right)^\dagger D_\mu  {\tilde q}_R
+\left(D^\mu{\tilde q}_L\right)^\dagger D_\mu  {\tilde q}_L
-g_3\btq_R T^a D^a {\tilde q}_R
+g_3\btq_L T^a D^a {\tilde q}_L
\nonumber\\
&+&\sqrt{2}  g_3{q}_R T^a {\tilde \gl}^a {\tilde q}_R
-\sqrt{2}  g_3{\bar q}_L T^a \bar{\tilde \gl}^a {\tilde q}_L
+\sqrt{2}  g_3\btq_R T^a \bar{\tilde \gl}^a { \bar q_R }
-\sqrt{2}  g_3\btq_L T^a {{\tilde \gl}}^a q_L
\nonumber\\
&+&\frac 12 D^a D^a
+\frac 12 D^i D^i
+\frac 12 D' D'
+g_2\btq_L \frac{\sigma^i}{2} D^i {\tilde q}_L
+g_2\bar h_j \frac{\sigma^i}{2} D^i h_j
+g_1\bar h^0_j \frac{{Y}_{h_j}}{2} D' h^0_j
\nonumber\\
&+&g_1\btq_L \frac{{Y}_{{\tilde q}_L}}{2} D' {\tilde q}_L
-g_1\btq_R \frac{{Y}_{{\tilde q}_R}}{2} D' {\tilde q}_R
+ \bar F_{h^0_1}F_{h^0_1}
+ \bar F_{h^0_2}F_{h^0_2}
+ \bar F_{q_L}F_{q_L}
+ \bar F_{q_R}F_{q_R}
\nonumber\\
&-&\left[\mu h^0_1 F_{h^0_2}
+\mu F_{h^0_1} h^0_2
-y_d \btd_R F_{h^0_1} \tilde d_L
-y_u \btu_R F_{h^0_2} \tilde u_L\right .
\nonumber\\
&-&\left . y_d(F_{d_R} h^0_1 \tilde d_L+ \btd_R h^0_1 F_{d_L})
-y_u(F_{u_R} h^0_2 \tilde u_L+ \btu_R h^0_2 F_{u_L})\right .
\nonumber\\
&+&\left . y_d d_R h_1^0 d_L
+y_u u_R h_2^0 u_L
+y_d A_d {\tilde d}_R h_1^0 {\tilde d}_L
+y_u A_u {\tilde u}_R h_2^0 {\tilde u}_L
+ {\rm h.c.} \right]
\nonumber\\
&-&\frac{M_3}{2}{\tilde \gl}^a {\tilde \gl}^a
-\frac{M_3}{2}\bar{\tilde \gl}^a \bar{\tilde \gl}^a
-{M}^2_{\tilde q}\btq_L{\tilde q}_L
-{M}^2_{\tilde u}\btu_R{\tilde u}_R
-{M}^2_{\tilde d}\btd_R{\tilde d}_R
\nonumber,\end{eqnarray}
where the gauge covariant derivative and field strength are defined as
\begin{eqnarray}
D_\mu & = & \partial_\mu+i g_3 T^a G^a_\mu \,, \nonumber\\
G^a_{\mu\nu} & = & \partial_\mu G^a_\nu - \partial_\nu G^a_\mu
-g_3 f^{abc}G^b_\mu G^c_\nu \,. \nonumber
\end{eqnarray}
Here we also added auxiliary $SU(2)\times U(1)$ fields $D^i$ and $D'$,
which are necessary to describe correctly the mass generation mechanism.
$\sigma^i$ denote  Pauli matrices and $Y_q$ is a hypercharge of a particle $q$.
Eliminating auxiliary fields $F_q$, $\bar F_q$, $D^a$, $D^i$,
and $D'$, and introducing four-component spinor notation
similar to~\cite{Haber:1984rc,Kuroda}, one gets
\begin{eqnarray}
{\cal L}&=&-\frac 14 G^a_{\mu\nu}G^{a\;\mu\nu}
+\frac {i}{2}{\bar {\tilde \gl}}^a \gamma^\mu D_\mu {\tilde \gl}^a
+i\bar q\gamma^\mu D_\mu q
+\left(D^\mu{\tilde q}_R\right)^\dagger D_\mu  {\tilde q}_R
+ \left(D^\mu{\tilde q}_L\right)^\dagger D_\mu  {\tilde q}_L
\nonumber\\
&+&\sqrt{2}g_3{\bar q}P_L T^a {\tilde \gl}^a {\tilde q}_R
-\sqrt{2}g_3{\bar q}P_R T^a {\tilde \gl}^a {\tilde q}_L
+\sqrt{2}g_3\btq_R T^a \bar{\tilde \gl}^a P_R q
-\sqrt{2}g_3\btq_L T^a \bar{\tilde \gl}^a P_L q
\nonumber\\
&-&\frac{g_3^2}{2}(\btq_L T^a {\tilde q}_L
 -\btq_R T^a {\tilde q}_R)
 (\btq_L T^a {\tilde q}_L
 -\btq_R T^a {\tilde q}_R)
-\frac{M_3}{2}\bar{\tilde \gl}^a {\tilde \gl}^a
+{\cal L}_{M} \,,
\label{lagsqnd} \\
{\cal L}_{M}&=&
-{M}^2_{\tilde q}\btq_L{\tilde q}_L
-{M}^2_{\tilde u}\btu_R{\tilde u}_R
-{M}^2_{\tilde d}\btd_R{\tilde d}_R
- y_d h^0_1 \bar d d
- y_u h^0_2 \bar u u
\label{lagsqndm} \nonumber\\
&-&\left(\frac{g_2^2}{2} I^3_{\tilde q_L}\btq_L\tilde q_L
- \frac{g_1^2}{4} {Y}_{\tilde q_L}\btq_L\tilde q_L
+ \frac{g_1^2}{4} {Y}_{\tilde q_R}\btq_R\tilde q_R\right)
\left(\bar h^0_1h^0_1 - \bar h^0_2h^0_2 \right)
\nonumber\\
&-& y_d^2\left(h^0_1\tilde d_L\right)^\dagger h^0_1\tilde d_L
- y_d^2 \bar h^0_1 h^0_1 \btd_R \tilde d_R
- y_u^2\left(h^0_2\tilde u_L\right)^\dagger h^0_2\tilde u_L
- y_u^2 \bar h^0_2 h^0_2 \btu_R \tilde u_R
\nonumber\\
&-&
\left[
y_d h^0_1A_d \btd_R {\tilde d}_L
+y_u h^0_2 A_u \btu_R {\tilde u}_L
-y_d\mu h^0_2 \btd_R\tilde d_L
-y_u\mu h^0_1 \btu_R\tilde u_L + {\rm h.c} \right]
\label{lagsqndmnot}
\end{eqnarray}
where $P_L=(1-\gamma_5)/2$ and $P_R=(1+\gamma_5)/2$ are the chiral
and antichiral projectors.

After Higgsing and taking into account that
\begin{eqnarray}
&&m_d= y_d v_1,\ \ \ \ \ m_u= y_u v_2,\ \ \ \ \
\tan \beta =\frac{v_2}{v_1}, \ \ \  \ \
\sin^2\theta_W=\frac{g_1^2}{g_1^2+g_2^2},
\nonumber\\
&&m_Z^2=\frac 12 (g_1^2+g_2^2)(v_1^2+v_2^2)
=\frac 12 \frac{g_1^2}{\sin^2 \theta_W}(v_1^2+v_2^2),\nonumber\\
&&\cos 2\beta = \frac{v_1^2-v_2^2}{v_1^2+v_2^2},\ \ \ \ \
e_q=\frac{Y_q}{2}+I^3_q
\nonumber,
\end{eqnarray}
it is convenient to rewrite ${\cal L}_M$ from eq.~(\ref{lagsqndm})
in the following form \cite{Ellis:1983ed}
\begin{equation}
\label{lagmsq}
{\cal L}_M=-m_q \bar q q
- m^2_{{\tilde q}_L} \btq_L {\tilde q}_L
- m^2_{{\tilde u}_R} \btu_R {\tilde u}_R
- m^2_{{\tilde d}_R} \btd_R {\tilde d}_R
- a_q m_q \btq_L {\tilde q}_R
- a_q m_q \btq_R {\tilde q}_L
\end{equation}
with
\begin{eqnarray}
&&  \msq{L}^2 = M^2_{\ti q}
 + m_Z^2 \cos 2\beta\,(I^3_{{\tilde q}_L} - e_q\sin^2\theta_W) + m_q^2 \,,
\nonumber\\
&&m_{{\tilde u}_{R}}^2=M^2_{\ti u}+e_u\,m_Z^2\cos 2\beta\,\sin^2\theta_W+
m_u^2 \,,
\qquad a_u = A_u - \mu \, \cot\beta
\nonumber\\
&& m_{{\tilde d}_{R}}^2=M^2_{\ti d}+e_d\,m_Z^2\cos 2\beta\,\sin^2\theta_W+
m_d^2 \,,
\qquad a_d = A_d- \mu \,  \tan\beta
\nonumber
\end{eqnarray}
where $e_q$ and $I^3_q$ are the electric charge and the third
component of the weak isospin of a particle $q$, $m_q$ is the mass
of the partner quark and $\mu$ is a Higgs mass parameter. The $M_{\ti
q}$, $M_{\ti u}$, and $M_{\ti d}$ are soft SUSY breaking masses,
$A_u$ and $A_d$ are trilinear couplings as in eq.~(\ref{lagsoft}).
The family indices have been suppressed.

Now we must diagonalize the mass matrix ${\cal M}^2_{\tilde q}$ of
 squarks to determine the physical mass eigenstates
\begin{equation}
  {\cal M}_{\sq}^2 =
  \left( \begin{array}{cc}  \msq{L}^2 & a_q m_q \\
                            a_q m_q   & \msq{R}^2
  \end{array}\right) \;=\;
  ({\cal R}^{\sq})^\dagger \left( \begin{array}{cc}  \msq{1}^2 & 0 \\
                                          0         & \msq{2}^2
                \end{array}\right) {\cal R}^{\sq} \,.
\label{msqmat}
\end{equation}

According to eq.~(\ref{msqmat}) ${\cal M}_{\sq}^2$ is diagonalized by a
unitary matrix ${\cal R}^{\sq}$. Assuming that CP violating phases
occur only in the CKM matrix, we choose ${\cal R}^{\sq}$ to be real.
The weak eigenstates $\sq_L^{}$ and $\sq_R^{}$ are
thus related to their mass eigenstates $\sq_1^{}$ and $\sq_2^{}$ by
\begin{equation}
  {\sq_1^{} \choose \sq_2^{}} = {\cal R}^{\sq}\,{\sq_L^{} \choose \sq_R^{}},
  \hspace{8mm}
{\cal R}^{\sq}=\left(\begin{array}{rr}\cos\theta_{\sq} &\sin\theta_{\sq} \\
       -\sin\theta_{\sq} & \cos\theta_{\sq} \end{array}\right) ,
\label{squarkrotation}
\end{equation}
with $\theta_{\sq}$ being the squark mixing angle.
The mass eigenvalues are given by
\begin{equation}
  m^2_{\sq_{1\!,2}} = \frac12 \left( \msq{L}^2 + \msq{R}^2
  \mp \sqrt{(\msq{L}^2 - \msq{R}^2)^2 + 4\, a^2_q m_q^2 } \,\right).
\end{equation}
By convention, we choose $\sq_1^{}$ to be the lightest mass
eigenstate. Notice that $\msq{1}\leq\msq{L,R}\leq\msq{2}$.

  For the mixing angle $\theta_{\sq}$ we require $0\leq\theta_{\sq} <\pi$.
Then one can write
\begin{equation}
  \cos\theta_{\sq} =
  \frac{- a_q m_q}{\sqrt{(\msq{L}^2-\msq{1}^2)^2 + a_q^2 m_q^2}},
  \hspace{8mm}
  \sin\theta_{\sq} =
  \frac{\msq{L}^2-\msq{1}^2}
       {\sqrt{(\msq{L}^2-\msq{1}^2)^2 + a_q^2 m_q^2}}.
\end{equation}
Moreover, $|\cos\theta_{\sq}| > \frac{1}{\sqrt 2}\,$ if
$\msq{L}<\msq{R}$ and
$|\cos\theta_{\sq}| < \frac{1}{\sqrt 2}\,$ if $\msq{R}<\msq{L}$.

\begin{center}
\begin{figure}[t]
\hspace*{1.5cm}
\includegraphics[scale=0.72]{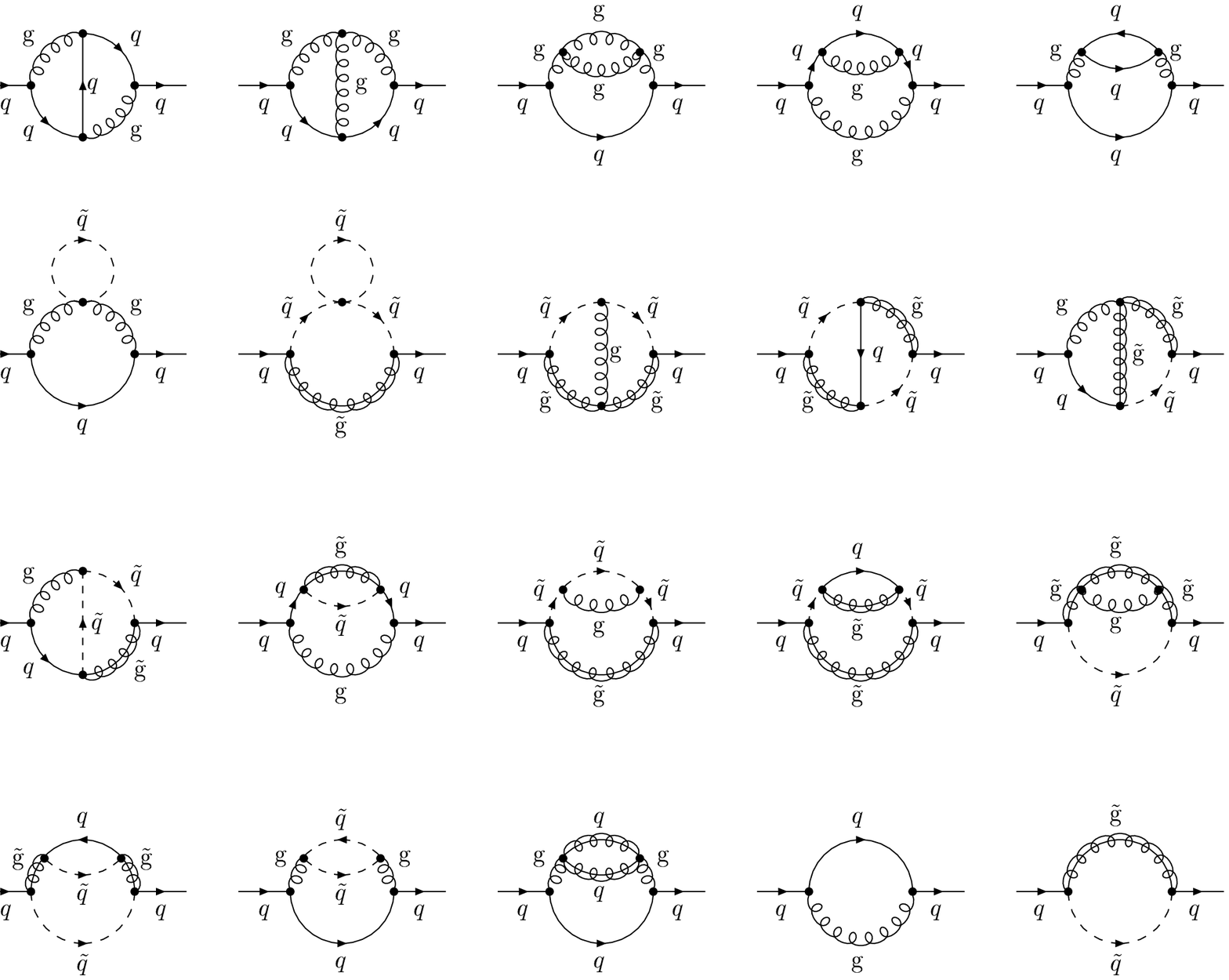}
\caption{Feynman diagrams for the ${\cal O}(\alpha_s^2)$ contribution
to quark self-energies
(last two diagrams at the one-loop level of the order ${\cal O}(\alpha_s)$).
The diagrams in the first line are the pure QCD contribution.}
\label{twoloopdiag}
\end{figure}
\end{center}

Making transition in eq.~(\ref{lagsqnd}) from ${\tilde q}_R$ and
${\tilde q}_L$ to ${\tilde q}_1$ and ${\tilde q}_2$ with the help
of the matrix from eq.~(\ref{squarkrotation}) we get for
quark-gluino-squark trilinear and four-squark vertices
complicated expressions due to squark mixing which we do not
present here. To produce input amplitudes of the diagrams needed
in our calculations we made use of the Mathematica package FeynArts
\cite{FeynArts}, which has SUSY QCD as part of MSSM model
\cite{FAMSSM}. The Feynman diagrams contributing to quarks
self-energies in the one and two loop order are shown in
Fig.~\ref{twoloopdiag}.

\vspace*{-12mm} Note again that doing these calculations we work
in the minimal subtraction scheme \MS adopted to supersymmetry ---
dimensional reduction with minimal subtraction \DR. To obtain a
relation between pole and \DR quark masses, we need to know  \DR
renormalization constants for quark, gluino, squark,
$\varepsilon$-scalar masses,
gauge and Yukawa charges%
    \footnote{By Yukawa charge we mean
              the Yukawa coupling of $\varepsilon$-scalar to fermions}
as well as renormalization of squark mixing angle. We will
describe evaluation of each of these quantities in the next
subsections.

\subsection{Renormalization of QCD sector}

First we have to renormalize the charge and the masses of quarks
and fields. The relations between bare and renormalized
parameters are given by
\begin{eqnarray}
  \alpha_{s,0} &=& \bar{\mu}^{2\,\varepsilon} Z_{\alpha_s} \alpha_s \,, \\
  m_{q,0} &=& Z_{m_q} m_q \,, \\
  G_0 &=& Z_G^{1/2} G \,, \\
  q_0 &=& Z_q^{1/2} q \,,
\end{eqnarray}
where $\bar{\mu}$ is a renormalization scale.

  The charge renormalization constant can be obtained from the
renormalization constants of the fields and vertex with the use of
the following relation:
\begin{eqnarray}
  Z_{\alpha_s} = Z^2_{qGq}\, Z^{-2}_{q} \, Z^{-1}_G \,, \label{renqgq}
\end{eqnarray}
where $Z_q$ and $Z_G$ correspond to renormalization 
of quark and
gluon fields and $Z_{qGq}$ renormalizes the quark-gluon
interaction vertex.

One can also renormalize the gauge parameter%
    \footnote{
              Gauge parameter $\xi$ is defined in such a way that $\xi=0$ and
              $\xi=1$ correspond to the Landau and Feynman gauge, respectively.
             }
with the help of the above-introduced constant $Z_G$
\begin{equation}
\xi_0= Z_G\xi \,.
\label{xirenorm}
\end{equation}
However, in the gauge invariant quantities (like, e.g., the pole
mass) the gauge parameter drops out already from the bare
expression and renormalization (\ref{xirenorm}) is not needed.

  Our results for renormalization constants
in the \DR scheme read
\begin{eqnarray}
Z_{m_q} &=& 1 - C_F\frac{\alpha_s}{2\pi}  \frac{1}{\varepsilon}
   - C_F\left(\frac{\alpha_s}{4\pi} \right)^2
\left( \left(6- 3C_A - 2C_F
       \right)
 \frac{1}{\varepsilon^2}
    + \left(-6 + 3C_A - 2C_F
      \right)
 \frac{1}{\varepsilon} \right) ,  \label{zm2} \\
Z_{q} & = & 1 - \frac{\alpha_s}{4\pi}C_F
\left( \xi + 1\right)\frac{1}{\varepsilon}\label{zquark} , \\
Z_{G} & = & 1 - \frac{\alpha_s}{4\pi}
\left( 6 - \frac{3-\xi}{2} C_A \right) \frac{1}{\varepsilon} , \label{zbg}\\
Z_{qGq} & = & 1 - \frac{\alpha_s}{4\pi}
       \left( (1 +\xi) C_F
    + \frac{3+\xi}{4} C_A \right) \frac{1}{\varepsilon} , \label{zqgq}\\
Z_{\alpha_s} &=&1-\frac{\alpha_s}{4\pi} \left( 3 C_A-6 \right)
    \frac{1}{\varepsilon} .  \label{zg}
\end{eqnarray}
For $SU(N_c)$ we have $C_F=(N_c^2-1)/(2N_c)$ and $C_A=N_c$. Here
we explicitly put the number of quark flavors to six.

There are several possibilities to check the correctness of our
results. The coupling renormalization constant (\ref{zg}) can be
compared with that obtained from the one-loop MSSM beta function
for $SU(3)$ group. We have also checked that in the case of
ordinary QCD we recover renormalization constants given in
\cite{Avdeev:1997sz}. The 2-loop \DR quark mass renormalization
constant (\ref{zm2}) has been obtained from eq. (\ref{2loop}) and
contains poles in $\varepsilon$ remaining after all other
renormalizations were done. Here we mean that one should perform
one-loop renormalizations of strong coupling constant and particle
masses. A subtle point is the renormalization of the Yukawa
interaction of $\eps$-scalars and quarks. In SUSY QCD discussed
here, one does not need to introduce a separate coupling constant $Y$
for this interaction, as supersymmetry insures that it is
renormalized  in the same way as a strong coupling
constant\footnote{See the section about the $\eps$-scalar sector
renormalization for details}. However, to make our renormalization
procedure as general as possible ( and also applicable in the case
of nonsupersymmetric QCD for checks), we introduce an additional
independent constant $Y$ and renormalize it separately. So, here
we follow the renormalization procedure described in
\cite{Avdeev:1997sz}. Renormalizing a physical quantity like a pole
mass, we ignore field renormalization constants which is a justified
procedure, provided we are evaluating the contribution 
of
counterterms for the whole sum of diagrams. The $1/\varepsilon^2$
poles in $Z_{m_q}$ can be checked with the use of the renormalization
group technique. Indeed, let us write
\begin{eqnarray}
m_0 &=& m \left(1+\frac{\alpha_s}{4\pi}\frac{Z^{(1,1)}}{\ep}
+ \left( \frac{\alpha_s}{4\pi} \right)^2 \frac{Z^{(2,1)}}{\ep}
+ \left( \frac{\alpha_s}{4\pi} \right)^2 \frac{Z^{(2,2)}}{\ep^2}
\right) \nonumber \\
&=& m \left(
1+\frac{z^{(1)}}{\ep}+\frac{z^{(2)}}{\ep^2} + \ldots
\right),
\end{eqnarray}
where
$
z^{(n)} = \sum_{l=1}^\infty(\alpha_s/4\pi)^l Z^{(l,n)} $. Now
using the renormalization group equation\footnote{In fact $m_0$ is not a
renormalization group invariant in our renormalization scheme.
However, to check the leading poles in $\eps$ we can treat it as an RG
invariant, as different prescriptions used to renormalize the
$\eps$-scalar sector give different results only in the  subleading
order in $\eps$. This point will be discussed in detail in the section
concerning the $\eps$-scalar sector renormalization.}
\begin{eqnarray}
0 = \bar{\mu}^2\frac{d}{d\bar{\mu}^2} m_0, \nonumber
\end{eqnarray}
we obtain
\begin{eqnarray}
\gamma &=& \frac{1}{2}g\frac{\partial}{\partial g}z^{(1)} \,, \nonumber \\
\left(\gamma + \beta_g\frac{\partial}{\partial g} +
\sum_i\gamma_i m_i \frac{\partial}{\partial m_i}
\right)z^{(n)} &=& \frac{1}{2}g\frac{\partial}{\partial g}z^{(n+1)} \,,
\end{eqnarray}
($g$ is related to $\alpha_s$ via $g^2/4\pi=\alpha_s$). The
formulae presented are general and are also applicable in a theory with
spontaneously broken symmetry \cite{Jegerlehner:2001fb}. In the
case of SUSY QCD, from the previous formula it follows that
\begin{eqnarray}
2Z^{(2,2)} = \left(Z^{(1,1)}\right)^2 +2\beta_g^{(1)}Z^{(1,1)}.
\end{eqnarray}
Substituting into this relation the expressions for  $Z^{(1,1)} =
-2C_F$ and $\beta_g^{(1)} = -\frac{1}{2}(3C_A-6)$, we get the same
result for the $1/\ep^2$ poles as in eq.~(\ref{zm2}).

\subsection{Gluino mass renormalization}

\begin{center}
\begin{figure}[t]
\hspace*{5cm}
\includegraphics[scale=1]{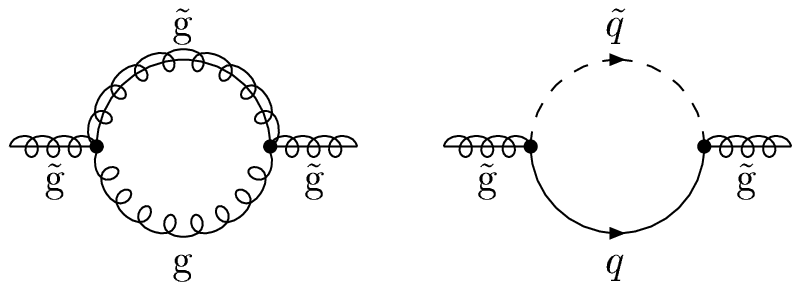}
\caption{One-loop diagrams  contributing to gluino mass counterterm.}
\label{diaggluino}
\end{figure}
\end{center}
\vspace*{-8mm}

The renormalization of the gluino mass is performed in a way similar to quark
mass renormalization. The relevant diagrams with gluino-gluon and
quark-squark loops are shown in Fig.\ref{diaggluino}.
Explicit calculations give the following results for the gluino
mass
$m_{\sg,0} = Z_{\msg} m_{\sg}$,
gluino wave function renormalization constants
$Z_{\sg}$
and
for gluino-gluon-gluino vertex renormalization constant $Z_{\sg G \sq}$
\begin{eqnarray}
Z_{\msg} & = & 1+ \frac{\alpha_s}{4\pi}
             \left( 6 - 3 C_A \right) \frac{1}{\varepsilon} , \\
Z_{\sg}  & = & 1 - \frac{\alpha_s}{4\pi}
   \left( \xi C_A + 6  \right) \frac{1}{\varepsilon} , \label{zgluino} \\
Z_{\sg G \sg} & = & 1 - \frac{\alpha_s}{4\pi} \left( 6
    + \frac{3+5\xi}{4} C_A \right) \frac{1}{\varepsilon} , \label{zglggl}.
\end{eqnarray}

   From eqs.~(\ref{zbg}), (\ref{zgluino}) and (\ref{zglggl})
one can extract the gauge coupling renormalization constant
\begin{equation}
     Z_{\alpha_s}   =   Z_{\sg G \sg}^2 Z_{\sg}^{-2} Z_{G}^{-1},
     \label{renglggl}
\end{equation}
which, of course, coincides with eq.~(\ref{zg}). This serves as an
additional check. The result for the gluino mass renormalization
constant can also be verified with help of the known gluino mass
beta-function.

\subsection{Squark sector renormalization}

To discuss the renormalization of the squark sector let us start with
the Lagrangian written in terms of physical states. The bare
Lagrangian in terms of bare parameters and bare fields is given by
\footnote{We do not include the bare Lagrangian for a
    four-squark interaction, because the
counterterms associated with the four-squark vertices
will give contribution of the ${\cal O}(\alpha_s^3)$ order.}
\begin{eqnarray}
{\cal L}^{\rm  bare}_{\sq} & =  & \demu \btq_i \demu \sq_i
- \msq{i}^2 \btq_i \sq_i + \Lag^{\rm bare}_{q \sg \sq} ,\label{lagbaresq}\\
{\cal L}^{\rm bare}_{q \sg \sq}  &= & - \sqrt{2} g_3
\bar{q} \left( P_R \cos \theta_{\sq} -
  P_L \sin \theta_{\sq} \right)T^a  \sg^a \sq_1 \nonumber\\
& +& \sqrt{2} g_3 \bar{q} \left(P_R \sin \theta_{\sq}
+ P_L \cos \theta_{\sq} \right)T^a \sg^a \sq_2 \nonumber\\
& - & \sqrt{2} g_3 \btq_1 T^a
\bar{\sg}^a \left( P_L \cos \theta_{\sq} -
  P_R \sin \theta_{\sq} \right) q  \nonumber\\
& + &\sqrt{2} g_3 \btq_2 T^a
\bar{\sg}^a \left(P_L \sin \theta_{\sq}
+ P_R \cos \theta_{\sq} \right) q \label{lagbareqglsq}
\end{eqnarray}
We can rewrite this Lagrangian in terms of renormalized physical
quantities and counterterms
\begin{eqnarray}
\label{lagrenormalized}
 {\cal L}^{\rm  bare}_{\sq} & =  & {\cal L}_{\sq} + \delta
{\cal
L}_{\sq} \\
\delta {\cal L}_{\sq} & =  & \delta Z_{\tilde{q}} (\demu \btq_i
\demu \sq_i - \msq{i}^2 \btq_i \sq_i)
 - \delta \msq{i}^2 \btq_i \sq_i
 -\delta { \tilde m_{12}}^2 (\btq_2 \sq_1+\btq_1 \sq_2)
 + \delta Z_{q \sg \sq} {\cal L}_{q \sg \sq}
  \label{ctlagbaresq}
\end{eqnarray}
where ${\cal L}_{\sq}$ and ${\cal L}_{q \sg \sq}$ have the same
form as (\ref{lagbaresq}) and (\ref{lagbareqglsq}). It should be
noticed that this set of counterterms is sufficient to cancel all
rising divergencies. The $\delta {\tilde m_{12}^2}$ counterterm is
used to render nondiagonal squark self-energies ${\tilde
\Sigma}_{\sq_i\sq_j}~(i\neq j)$ finite. In practical calculations,
it is convenient to trade $\delta {\tilde m_{12}^2}$ for the
counterterm for the squark mixing angle. Indeed, we can diagonalize
the Lagrangian (\ref{lagrenormalized}) with the following field
redefinitions:
\begin{equation}
\left(
\begin{array}{c}
{\tilde q'}_1\\[3mm]
{\tilde q'}_2
\end{array}
\right)= \left( \begin{array}{cc}
{1} & {\delta\theta_{\sq}} \\[3mm]
{-\delta\theta_{\sq}} & {1}
\end{array}\right)
\left(
\begin{array}{c}
{\tilde q}_1\\[3mm]
{\tilde q}_2
\end{array}
\right)
\end{equation}
where we have taken into account that $\delta\theta_{\sq}
\sim{\cal O}(\alpha_s)$, $\cos\left(\delta\theta_{\sq}\right)
=1+{\cal O}(\alpha_s^2)$ and
$\sin\left(\delta\theta_{\sq}\right)=\delta\theta_{\sq}+{\cal
O}(\alpha_s^3)$
Demanding that the new mass matrix of squarks
\begin{equation}
\left( \begin{array}{cc}
{1} & {\delta\theta_{\sq}} \\[3mm]
{-\delta\theta_{\sq}} & {1}
\end{array}\right)
\left( \begin{array}{cc} m_{\tilde{q}_1}^2+\delta Z_{\tilde{q}}
m_{\tilde{q}_1}^2 + \delta m_{\tilde{q}_1}^2 &
\delta {\tilde m_{12}^2} \\[3mm]
\delta {\tilde m_{12}^2} & m_{\tilde{q}_2}^2+\delta Z_{\tilde{q}}
m_{\tilde{q}_2}^2 + \delta m_{\tilde{q}_2}^2
\end{array}\right)
\left( \begin{array}{cc}
{1} & {-\delta\theta_{\sq}} \\[3mm]
{\delta\theta_{\sq}} & {1}
\end{array}\right) \,.
\label{diagmmmol}
\end{equation}
must be diagonal and expanding eq.~(\ref{diagmmmol}) in the gauge
coupling constant $\alpha_s$, one gets
\begin{equation}
\delta\theta_{\sq}=\frac{\delta {\tilde m_{12}^2}}
{m^2_{\sq_1}-m^2_{\sq_2}} =
\frac{({\tilde \Sigma}^{(1)}_{\sq_1\sq_2})_{\rm div}}
{m^2_{\sq_1}-m^2_{\sq_2}} ,
\end{equation}
where the subscript ``div'' stands for the $1/\varepsilon$ part
of the expression.
Rewriting eq.~(\ref{lagrenormalized}) in terms of the new fields
${\tilde q'}_i$, we come to the following renormalization
prescription for the squark sector:
\begin{eqnarray}
    m_{\sq,0}^2 &=& m_{\tilde{q}_i}^2+ \delta m_{\tilde{q}_i}^2 = Z_{m^2_{\tilde{q}_i}} m_{\tilde{q}_i}^2 ,\\
    \tilde{q}_{i,0} &=& (1+ {1\over 2} Z_{\tilde{q}} )\tilde{q}_i =  Z_{\tilde{q}}^{1/2} \tilde{q}_i  \\
    \theta_{\tilde{q},0} &=& \theta_{\tilde{q}} + \delta \theta_{\tilde{q}} = Z_{\theta_{\tilde{q}}} \theta_{\tilde{q}}
\end{eqnarray}

\begin{center}
\begin{figure}[t]
\hspace*{2cm}
\includegraphics[scale=1]{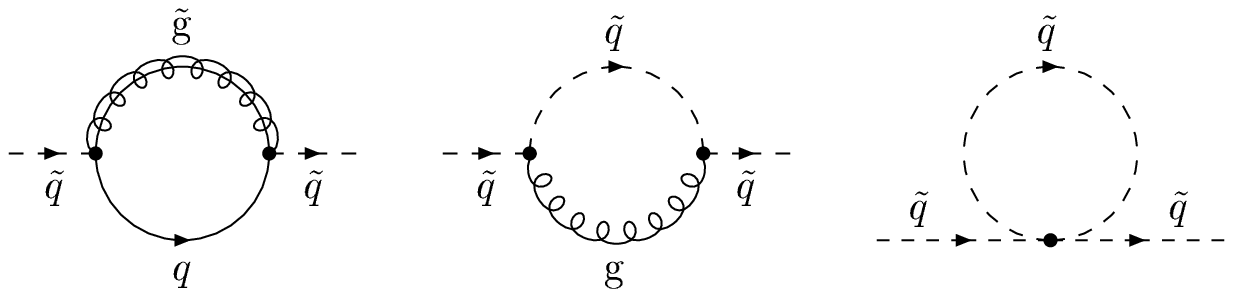}
\caption{One-loop diagrams contributing to the squark mass
and squark mixing angle counterterm.}
\label{squarksdiag}
\end{figure}
\end{center}

\vspace*{-8mm} Explicit calculation of squark self-energies gives
(see diagrams in Fig.~\ref{squarksdiag}) the following expression
for the renormalization of the squark mixing angle
\begin{equation}
\delta\theta_{\sq}=C_F\frac{\alpha_s}{4\pi}
\left( -\frac{\sin \left(4 \theta_{\sq}\right)}{2}
+\frac{4 m_q\msg\cos\left(2\theta_{\sq}\right)}{m_{\sq_1}^2-m_{\sq_2}^2}
\right) \frac{1}{\varepsilon} .
\label{deltatheta}
\end{equation}
Note that renormalization (\ref{deltatheta}) of the mixing angle
is minimal. It is the correct choice, if one wants to have the final
result expressed in terms of the running \DR sparticle masses.

  Summarizing the result of calculation, the renormalization constants
for the squark sector read
\begin{eqnarray}
Z_{\sq}&=&1-C_F\frac{\alpha_s}{4\pi}\left(\xi-1\right)\frac{1}{\varepsilon} ,
\label{zsq} \\
Z_{m^2_{\tilde{q}_1}} &=& 1 +  \frac{C_F}{m^2_{\tilde{q}_1}}
\frac{\alpha_s}{4\pi }
   \Biggl( \left(m_{\sq_2}^2-m_{\sq_1}^2\right) \sin^2 2 \theta_{\sq}
+4 \left(m_q \msg \sin 2 \theta_{\sq}- m_q^2 - \msg^2\right) \Biggr)
\frac{1}{\varepsilon} , \\
Z_{m^2_{\tilde{q}_2}}  & = &  1 + \frac{C_F}{m^2_{\tilde{q}_2}}
\frac{\alpha_s}{4\pi}
    \Biggl( \left(m_{\sq_1}^2-m_{\sq_2}^2\right) \sin^2 2 \theta_{\sq}
-4 \left(m_q \msg \sin 2 \theta_{\sq} + m_q^2 + \msg^2\right) \Biggr)
\frac{1}{\varepsilon} ,\\
Z_{q \sg \sq} & = & 1 - \frac{\alpha_s}{4\pi}
        \left(\xi C_F + \frac{3+\xi}{2} C_A \right)\frac{1}{\varepsilon} ,
\label{zqglsq} \\
Z_{\theta_{\tilde{q}}} &=& 1+
\frac{ \delta\theta_{\tilde{q}}}{\theta_{\tilde{q}}} .
\end{eqnarray}
Using eqs. (\ref{zquark}), (\ref{zgluino}), (\ref{zsq}) and (\ref{zqglsq}),
one can also get the gauge coupling renormalization constant
$Z_{\alpha_s}$ from quark-gluino-squark vertex
\begin{equation}
 Z_{\alpha_s}   =   Z_{q\sg \sq}^2 Z_{q}^{-1} Z_{\sg}^{-1} Z_{\sq}^{-1}
\end{equation}
which, due to supersymmetry, of course coincides with the results
of eqs.~(\ref{renqgq}) and (\ref{renglggl}).

\begin{center}
\begin{figure}[t]
\hspace*{0.8cm}
\includegraphics[scale=0.75]{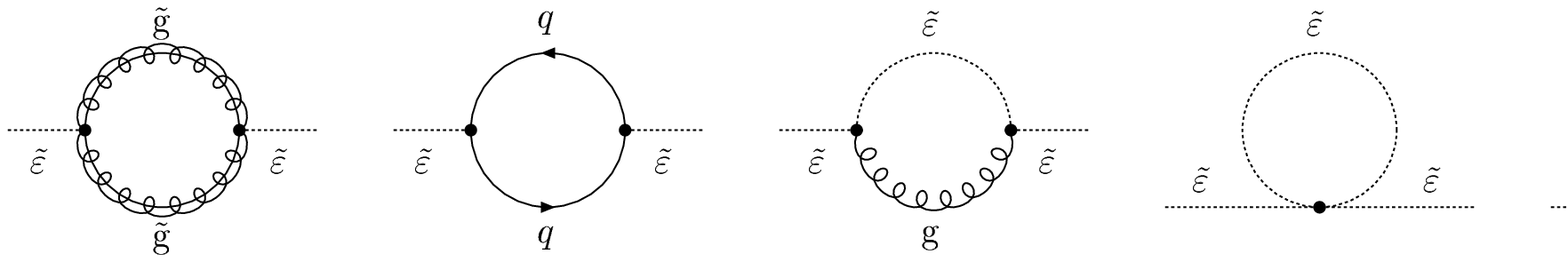}
\caption{One-loop diagrams contributing to the
$\varepsilon$-scalar wave function and mass counterterms}
\label{escalardiag2}
\end{figure}
\end{center}
\vspace*{-12mm}

\subsection{The $\eps$-scalar sector renormalization}

The renormalization of the $\eps$-scalar sector in SUSY QCD goes along
the same lines as for the ordinary QCD in dimensional reduction.
Here we will remind the tricks used to obtain the corresponding
renormalization constants and write down their values in SUSY QCD.
The part of the SUSY QCD Lagrangian containing $\eps$-scalar fields
could be obtained through the dimensional reduction of the Lagrangian
from eq.(\ref{lagsqnd}). As a result we have
\begin{eqnarray}
{\cal L}_{\eps} &=& \frac{1}{2}\left(D^{\mu}G_{\sigma}\right)^\dag
D_{\mu}G_{\sigma} - g_3\bar q\gamma_{\sigma}T^aG^a_{\sigma}q
-\frac{1}{4}g_3^2f^{a b c}f^{a d e}G^b_{\sigma}G^c_{\sigma '}
G^d_{\sigma}G^e_{\sigma '} \nonumber \\
&&-\frac{1}{2}g_3\bar{\sg}\gamma_{\sigma}T^aG^a_{\sigma}{\sg}
+g_3^2{\btq} T^aT^bG^a_{\sigma}G^b_{\sigma}{\sq}.
\end{eqnarray}
Here $G_{\sigma (\sigma ')}$ denote $\eps$-scalar fields, and
indices $\sigma$, $\sigma '$ belong to $2\eps$ subspace.

\begin{center}
\begin{figure}[t]
\hspace*{2cm}
\includegraphics[scale=1]{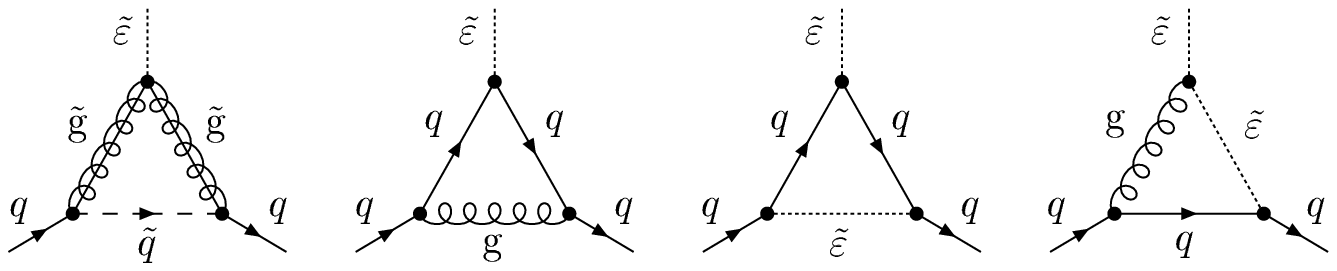}
\caption{One-loop diagrams for $\eps$-scalar Yukawa coupling
renormalization constant.} \label{escalardiag3}
\end{figure}
\end{center}

A standard way is to use the Feynman rules derived from this
Lagrangian to account for the contribution of $\eps$-scalars. However,
it turns out that for some problems one can use far 
simpler
arguments to get a desired solution. For example, we are
interested in both full and separate contributions of vectors and
$\eps$-scalars to the quantity, like a Green function or amplitude,
without external vector indices. So, vector and $\eps$-scalar
fields are located on internal lines only. Then to solve this
problem one should
\begin{enumerate}
\item perform $d$-dimensional vector and spinor algebras \item for
full contribution - replace $d$ in the numerator of scalarized
expression with 4; \\
for $\eps$-scalar contribution - replace $d^n$ with
$4^n-(4-2\eps)^n$.
\end{enumerate}
When we deal with a Green function containing external
vector indices and want to extract a contribution of either
external vector or $\eps$-scalar fields, the  corresponding projector
onto $4-2\varepsilon$ or $2\varepsilon$ subspaces should be
constructed. Internal $\eps$-scalars and vectors are treated 
in the same way, as described above. At this point one should keep in
mind that particle momenta are always orthogonal to
$2\varepsilon$ subspace. For example, in the case of the
$\varepsilon$-scalar propagator (see Fig.\ref{escalardiag2}) we
use $\frac{1}{2\eps}g^{\mu\nu}$ (the $\eps$-scalar
propagator is always diagonal, see the notice above)
as a projector and after
contraction of external indices replace $d^n$ with
$4^n-(4-2\eps)^n$. At the same moment, evaluating corrections to
the interaction between the $\eps$-scalar and quarks (see
Fig.\ref{escalardiag3}), required for the determination of the Yukawa
charge renormalization, one should follow the same steps, as in the
case of $\eps$-scalar propagator with the only difference that now
the projector is given by
\begin{eqnarray}
{\rm Tr}\left( \frac{1}{8C_FC_A\eps}T^a\gamma^{\mu}
 \Gamma_{\mu}
\right),
\end{eqnarray}
where $\Gamma^{\mu}$ is an interaction vertex of the gluon field with
quarks computed in SUSY QCD without separating $\epsilon$-scalar
contributions,  but using the 4-dimensional vector and spinor algebras.
\vspace*{-12mm}
In the one-loop order the $\eps$-scalar field renormalization
constant is given by
\begin{eqnarray}
Z_s = 1 - \frac{\alpha_s}{4\pi} \Bigl( 6-C_A(2-\xi)  \Bigr)
    \frac{1}{\varepsilon} .
\end{eqnarray}
For the one-loop Yukawa charge renormalization constant,
describing the interaction  of the $\epsilon$-scalar with fermions in
SUSY QCD, we have
\begin{eqnarray}
Z_Y = Z^2_{qsq}Z_q^{-2}Z_s^{-1} =
1 - \frac{\alpha_s}{4\pi}\left(3C_A-6\right)\frac{1}{\varepsilon} .
\end{eqnarray}

  Note that in SUSY QCD, contrary to the case of nonsupersymmetric QCD,
it coincides with the gauge charge renormalization constant. It is due
to supersymmetry, which now protects a tree-level coincidence of
the $\varepsilon$-scalar and vector coupling constants.

The one-loop nonminimal mass counterterm for $\varepsilon$-scalars reads
\begin{eqnarray}
\delta m_s^2 = -\frac{\alpha_s}{4\pi}\left(
\sum_{n_f}\left(\frac{m_f^2}{\bar{\mu}^2}\right)^{\!\!-\varepsilon}
\!\!\!\!m_f^2\left[\frac{2}{\varepsilon}+2\right]
-\sum_{n_{\tilde f}}\left(\frac{m_{\tilde f}^2}{\bar{\mu}^2}\right)^{\!\!-\varepsilon}
\!\!\!\!m_{\tilde f}^2\left[\frac{1}{\varepsilon}+1\right]
+C_A\left(\frac{m_{\sg}^2}{\bar{\mu}^2}\right)^{\!\!-\varepsilon}
\!\!\!\!m_{\sg}^2\left[\frac{2}{\varepsilon}+2\right]
\right) .
\end{eqnarray}
It is necessary to insure that the pole mass of the
$\varepsilon$-scalar is zero. Here $n_f$ denotes the sum over
different quark flavours and $n_{\tilde f}$ is used to denote the sum
over different squarks. It is just this counterterm that cancels
$m_{{\rm hard}}^2$ terms occurring in the two-loop diagrams and
ensures decoupling of large scale physics for a physical quantity
like a quark pole mass. A way of renormalizing the $\eps$-scalar mass
nonminimally was first proposed in \cite{Avdeev:1997sz} and was shown
to be a consistent renormalization procedure. In fact, there are
two theoretically admissible prescriptions to deal with the
$\eps$-scalar mass renormalization. One is to renormalize the
$\eps$-scalar mass minimally \cite{jones} and the other is the
total subtraction of loop corrections to the $\eps$-scalar mass
\cite{Avdeev:1997sz}. In the first approach, the $\eps$-scalar mass
becomes a running quantity and physical quantities, like pole mass
in our case, will in general depend on it together with other
physical renormalized parameters. However, the $\eps$-scalar mass is
not a physical quantity and it is more correct to treat it as an
artifact of our regularization and renormalization schemes. From
this point of view the second approach is more natural, if  we are
going to follow as close as possible the original idea of
dimensional reduction suggesting the zero tree-level mass for both
$\eps$-scalars and gauge bosons. Of course, all renormalization
schemes are equivalent and could be related via a final
recalculation of parameters. Introduction of nonminimal
subtraction for the $\eps$-scalar mass leads to the fact that the bare
quark mass is not anymore a renormalization group invariant. The
point is that the physical quark mass depends on two mutually
correlated bare masses: quark mass itself and $\eps$-scalar mass.
Both these quantities are $\bar{\mu}$ dependent and only the physical
pole mass is RG invariant. For precisely this reason the RG
equations for the bare quark mass should be written with great
care.


\section{Results of calculation}

Here we present the results of our calculation. Corrections
to the pole masses are parameterized as follows:
\begin{equation}
\frac{M_{\rm pole}}{m} = 1 + \left( \frac{\Delta m}{m} \right),
\label{PM}
\end{equation}
where $\Delta m$ starts from the one-loop order.

Through
calculations we reproduced all known results about radiative
corrections to the pole masses of the bottom and top quarks: one-loop
MSSM
\cite{Hempfling:1993kv,Hall:1993gn,SUSY94&Wright,Donini95,Pierce:1996zz}
and two loop QCD \cite{Gray:1990yh,Fleischer:1998dw} (in \MS) and
\cite{Avdeev:1997sz} (in \DR) corrections.
As an additional check, the renormalization group analysis can be
applied to confirm independently the $1/\varepsilon$ and $1/\varepsilon^2$
terms. We get complete agreement between our pole terms
and those from RG for the \DR quark mass renormalization constant. And
the last but not least: calculations were performed in the linear
gauge with a free QCD gauge parameter. The correction to the pole
mass (\ref{PM}) however appears to be gauge independent.

\subsection{One-loop result}
The corrections to the bottom and top quark pole masses 
in the one-loop order
have been well known for a long time for QCD as well as for MSSM.
Here we reproduce them only up to terms of the order
${\cal O}({m_{\rm soft}^2}/{m_{\rm hard}^2})$
in a large mass expansion
procedure. For pure QCD we have
\begin{equation}
\left({\Delta m_b\over m_b}\right)^{\rm QCD}=
C_F\frac{\alpha_s}{4\pi}\left[5 -3 \ln
\left(\frac{m^2_b}{\bar{\mu}^2}\right)\right]
\end{equation}
and the squark-gluino MSSM contribution is given by
\begin{eqnarray}
\left({\Delta m_b\over m_b}\right)^{\tilde b {\tilde {\mathrm
g}}}&=& C_F\frac{\alpha_s}{8\pi}\left[-3 + \frac{m^2_{\tilde
b_1}}{ m^2_{\tilde b_1}-m^2_{\tilde {\mathrm g}}} +
\frac{m^2_{\tilde b_2}}{ m^2_{\tilde b_2}-m^2_{\tilde {\mathrm
g}}} + 2 \ln \left(\frac{m^2_{\tilde {\mathrm
g}}}{\bar{\mu}^2}\right)\right . \nonumber \\
&+&\left . \frac{m^2_{\tilde b_1}}{ m^2_{\tilde b_1}-m^2_{\tilde
{\mathrm g}}}
  \left(2 - \frac{{m^2_{\tilde b_1}}}
   {{m^2_{\tilde b_1}-m^2_{\tilde {\mathrm g}}}}  -
 2\sin (2\theta_{\tilde b})\frac{m_{\tilde {\mathrm g}}}{m_{b} } \right)
\,\ln \left(\frac{m^2_{\tilde b_1}}{m^2_{\tilde {\mathrm
g}}}\right) \right . \nonumber \\ &+& \left .
  \frac{m^2_{\tilde b_2}}{m^2_{\tilde b_2}-m^2_{\tilde {\mathrm g}}}
  \left(2- \frac{{m^2_{\tilde b_2}}}
{{m^2_{\tilde b_2}-m^2_{\tilde {\mathrm g}} }} + 2\sin
(2\theta_{\tilde b}) \frac{m_{\tilde {\mathrm g}}} {m_{b}} \right)
\,\ln\left(\frac{m^2_{\tilde b_2}}{m^2_{\tilde {\mathrm
g}}}\right)\right]
\end{eqnarray}
which coincide with the results of
\cite{Hempfling:1993kv,Hall:1993gn,SUSY94&Wright,Donini95,Pierce:1996zz}.
To obtain similar expressions for the case of the top one should
perform an obvious substitution $b\rightarrow t$. The results for
squark-chargino and higgs-quarks loops also known, but we are
interested only in ${\cal O}(\alpha_s)$ and ${\cal
O}(\alpha_s^2)$ corrections in this paper.

\subsection{Two-loop result}

The two-loop QCD results in the $\overline{\rm DR}$-scheme for the $b$-quark
(with four light quarks)
\begin{eqnarray}
&&\left(\frac{\Delta m_b}{m_b}\right)^{\rm QCD}= C_F
  \left(\frac{\alpha_s}{4\pi} \right)^2
\left \{ - \frac{623}{18}   -
       8\,\zeta_2 + {\ln^2 \left(\frac{m_b^2}{m_t^2} \right) } -
       \frac{13}{3}\ln \left(\frac{m_b^2}{m_t^2}
\right) \right . \nonumber\\ &&+
       C_F\,\left[ - \frac{59}{8}   +
          30\,\zeta_2 - 48\,\ln (2)\,\zeta_2 +
          12\,\zeta_3 +
          \frac{3}{2} \ln \left(\frac{m_b^2}{\bar{\mu}^2} \right)
            + \frac{9}{2}{ \ln^2 \left(\frac{m_b^2}{\bar{\mu}^2}
\right) } \right] \nonumber\\ &&+       C_A\,\left[
\frac{1093}{24} - 8\,\zeta_2 +
          24\,\ln (2)\,\zeta_2 - 6\,\zeta_3 -
          \frac{179}{6} \ln \left(\frac{m_b^2}{\bar{\mu}^2} \right)
           + \frac{11}{2}{ \ln^2 \left(\frac{m_b^2}{\bar{\mu}^2}
\right) } \right] \nonumber\\ &&+
      \left . 26\, \ln \left(\frac{m_b^2}{\bar{\mu}^2} \right)  -
       6\,{ \ln^2 \left(\frac{m_b^2}{\bar{\mu}^2} \right) } \right \}
\end{eqnarray}
and for the $t$-quark (with five light quarks)
\begin{eqnarray}
&&\left(\frac{\Delta m_t}{m_t}\right)^{\rm QCD}=
   C_F \left( \frac{\alpha_s}{4\pi} \right)^2
   \left \{ - 43   -
       12\,\zeta_2  +
       26\, \ln \left(\frac{m_t^2}{\bar{\mu}^2} \right)  -
       6\,{ \ln^2 \left(\frac{m_t^2}{\bar{\mu}^2} \right) }
\right . \nonumber\\ &&+
       C_F\,\left[ - \frac{59}{8}   +
          30\,\zeta_2 - 48\,\ln (2)\,\zeta_2 +
          12\,\zeta_3 +
          \frac{3}{2} \ln \left(\frac{m_t^2}{\bar{\mu}^2} \right)
            + \frac{9}{2}{ \ln^2 \left(\frac{m_t^2}{\bar{\mu}^2}
\right) } \right] \nonumber\\ &&+ \left .      C_A\,\left[
\frac{1093}{24} - 8\,\zeta_2 +
          24\,\ln (2)\,\zeta_2 - 6\,\zeta_3 -
          \frac{179}{6} \ln \left(\frac{m_t^2}{\bar{\mu}^2} \right)
           + \frac{11}{2}{ \ln^2 \left(\frac{m_t^2}{\bar{\mu}^2}
\right) } \right]  \right \}
\end{eqnarray}
coincide with the first terms of the expansion in $m_b/m_t$ from
\cite{Avdeev:1997sz}

We find very a simple answer to one particular limit. Namely, in the case, when
$m_{{\tilde t}_L}=m_{{\tilde t}_R}=m_{{\tilde b}_L}=m_{{\tilde b}_R}=
m_{{\tilde \gl}}=m_{{\tilde u}_{1,2}}=m_{{\tilde d}_{1,2}}= m_{{\rm SUSY}}=M$,
the expression looks like
(see eq.~(\ref{lagmsq}) for the definition of $a_q$)
\begin{eqnarray}
&&\left(\frac{\Delta m_q}{m_q}\right)^{\rm MSSM}=
C_F \left( \frac{\alpha_s}{4\pi} \right)^2
    \left\{ 
      \frac{47}{3} 
     + 20\ln\left(\frac{M^2}{\bar{\mu}^2}\right) 
     + 6\ln\left(\frac{M^2}{\bar{\mu}^2}\right) \ln\left(\frac{M^2}{m_q^2}\right)
      \right. \nonumber\\ 
&& +
 C_F \left[ \frac{23}{24} 
           - \frac{13}{6}\ln\left(\frac{M^2}{\bar{\mu}^2}\right) 
           + \frac{1}{2}\ln^2\left(\frac{M^2}{\bar{\mu}^2}\right)
       - 3\ln\left(\frac{M^2}{\bar{\mu}^2}\right) 
                     \ln\left(\frac{m_q^2}{\bar{\mu}^2}\right)
      \right] \nonumber\\ 
&& +
   C_A \left[ 
      \frac{175}{72} + \frac{41}{6}\ln\left(\frac{M^2}{\bar{\mu}^2}\right) 
     - \frac{1}{2}\ln^2\left(\frac{M^2}{\bar{\mu}^2}\right) 
     - 2\ln\left(\frac{M^2}{\bar{\mu}^2}\right)\ln\left(\frac{m_q^2}{\bar{\mu}^2}\right)
       \right] \nonumber\\
&& + 
  \frac{
  a_q}{M} \left[
    - 4 - 8\ln\left( \frac{M^2}{\bar{\mu}^2}\right) \right] 
+ C_F \, \frac{
  a_q}{M} \left[
    \frac73 - \frac{11}{3}\ln\left( \frac{M^2}{\bar{\mu}^2}\right) 
       + 3\ln\left( \frac{m^2_q}{\bar{\mu}^2}\right)  \right]\nonumber\\
&& +\left.
  C_A \, \frac{
  a_q
}{M} \left[
    - \frac83 
     + 4\ln\left( \frac{M^2}{\bar{\mu}^2}\right)  \right]
 \right\} \label{limit}
\end{eqnarray}
Note, that we put equal to each other the left and right squark masses
$m^2_{{\tilde q}_R}=m^2_{{\tilde q}_L}$ but not its physical mass
eigenstates $m^2_{{\tilde q}_1}=m^2_{{\tilde q}_2}$. The latest is true only
when $a_q=0$ because a minimal mixing takes place when $\theta_q=\pi/4$
and
$m^2_{{\tilde q}_{1,2}}=M^2\pm a_q m_q$ (for $a_q<0$).

Since the complete formulae\footnote{They could be found in the source archive
file for this paper submitted to the HEP database }
are too large to be presented here,
we present only the influence of our results
on the masses of $b$- and $t$-quarks and the spectra of
supersymmetric particles obtained as solutions of RG equations.

We incorporated our formulae into the SoftSUSY code \cite{SoftSUSY}
and analyzed the difference in the predictions for particle masses
obtained with and without our corrections to the $b$- and $t$-quark
pole masses. To measure quantitatively influence on the spectra let us
introduce for each species of particle the quantity
\begin{equation}
\Delta_p=
\frac{m_p^{\mathrm
{CODE}}-m_p^{\mathrm {CODE+Corrections}}} {m_p^{\mathrm
{CODE}}\;\;\;\;\;\;\;\;\;\;\;\;\;\;\;},
\label{deltap}
\end{equation}
where $p$ stands for quarks or different supersymmetric particles.
Below we put in our analysis the renormalization scale $\bar{\mu}$ to be
equal to $m_Z$.

\begin{figure}[h]
\psfig{figure=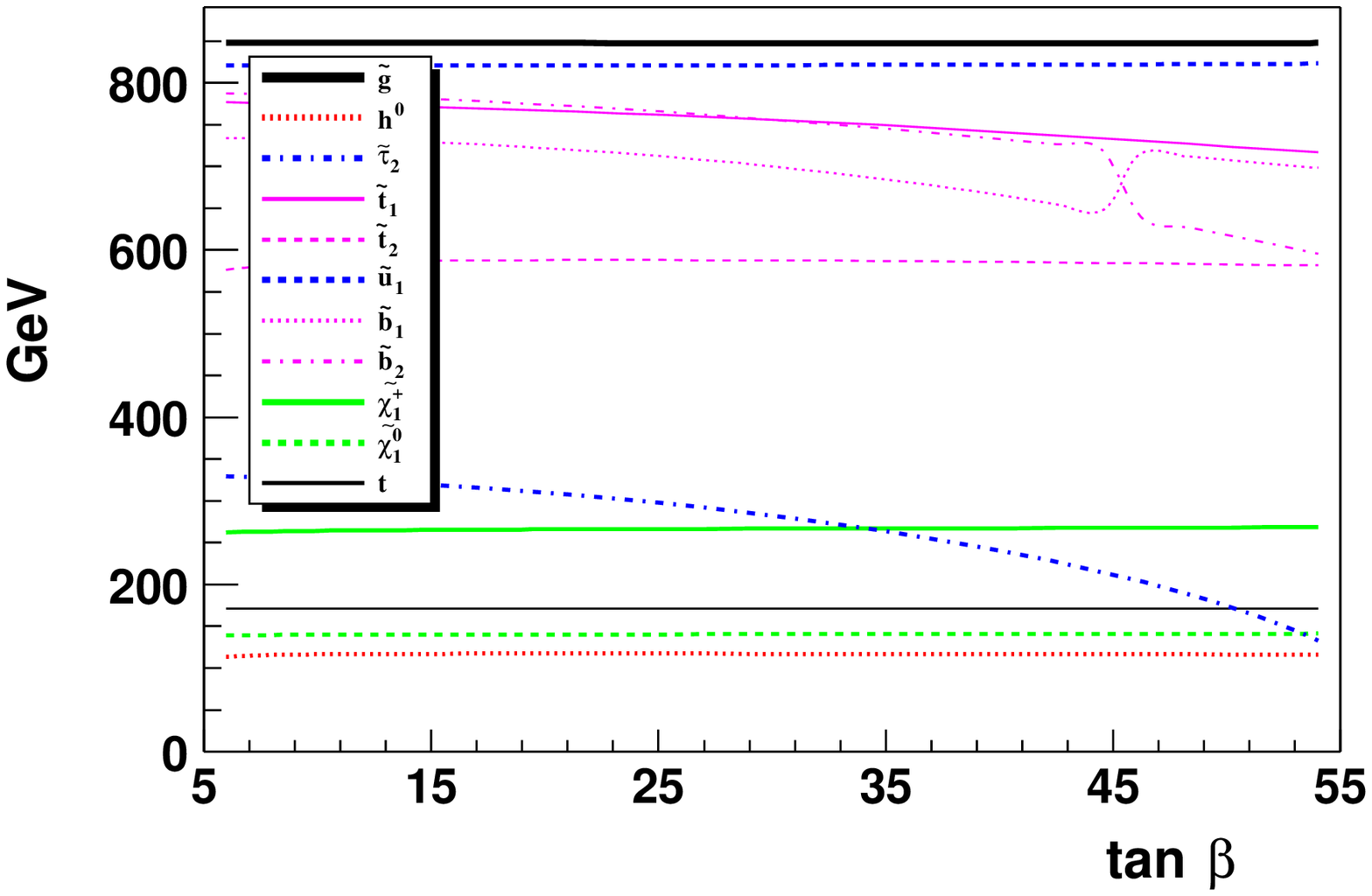,height=4.5cm,width=8.5cm}
\psfig{figure=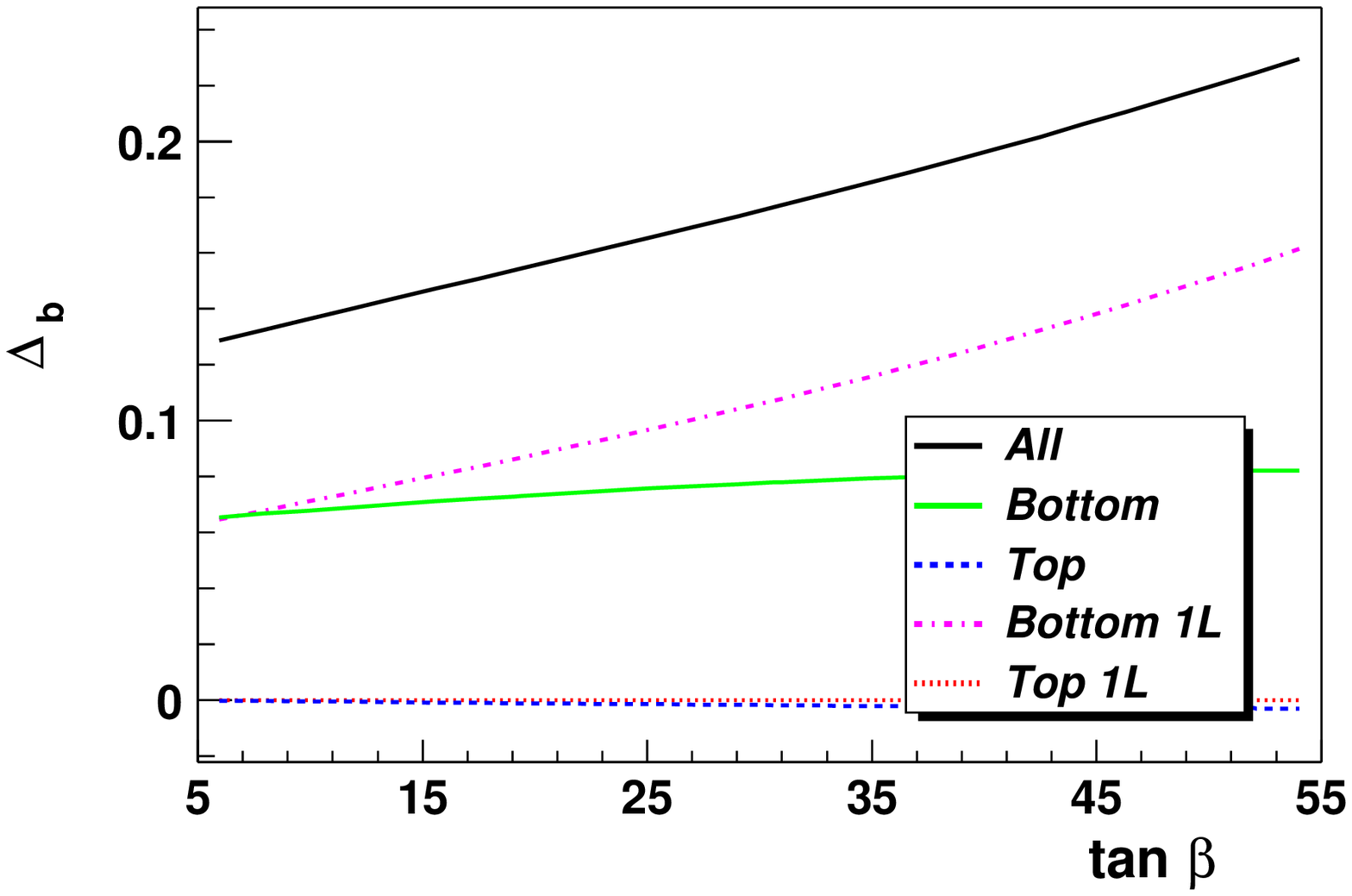,height=4.5cm,width=8.5cm}
\psfig{figure=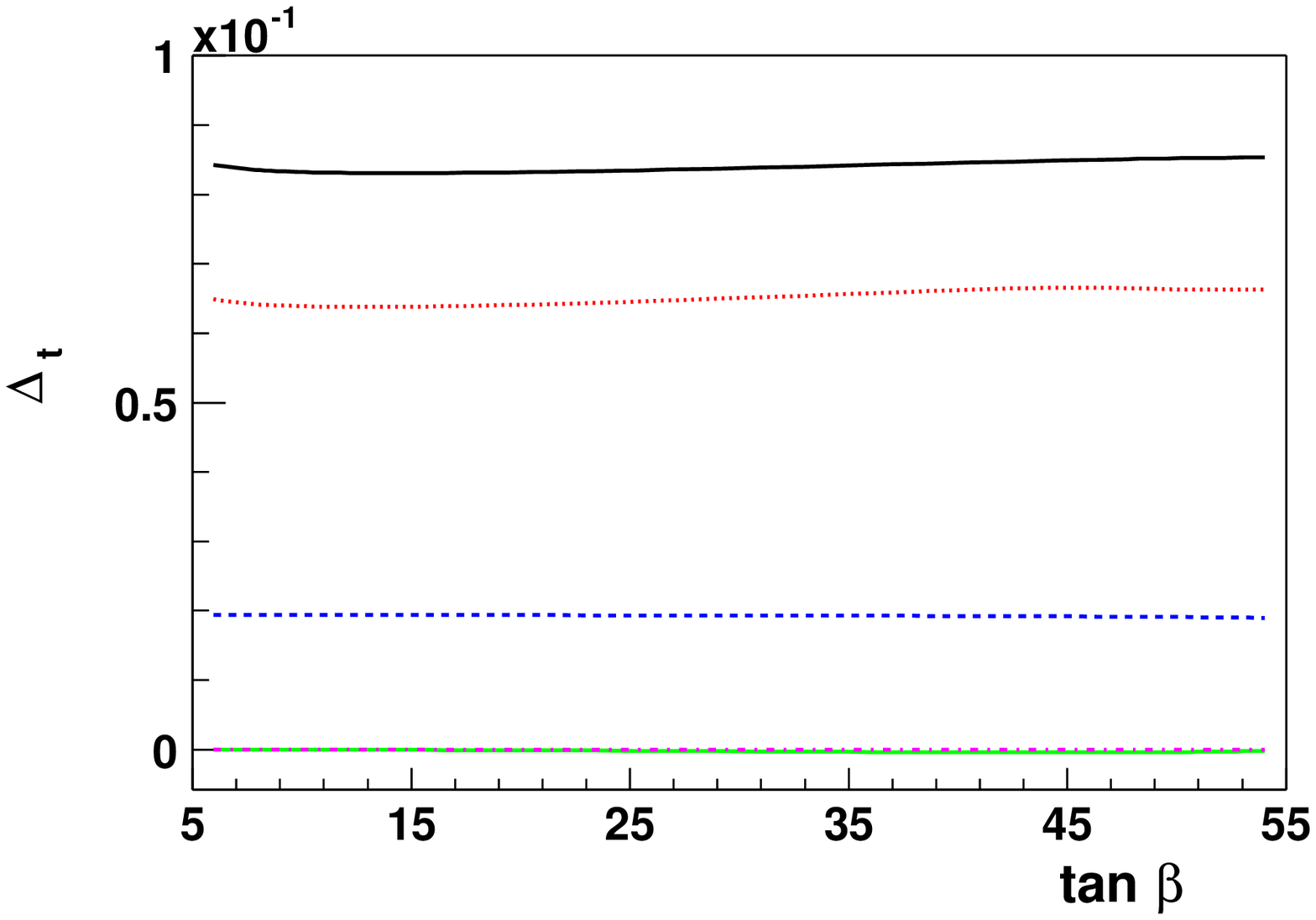,height=4.5cm,width=8.5cm}
\psfig{figure=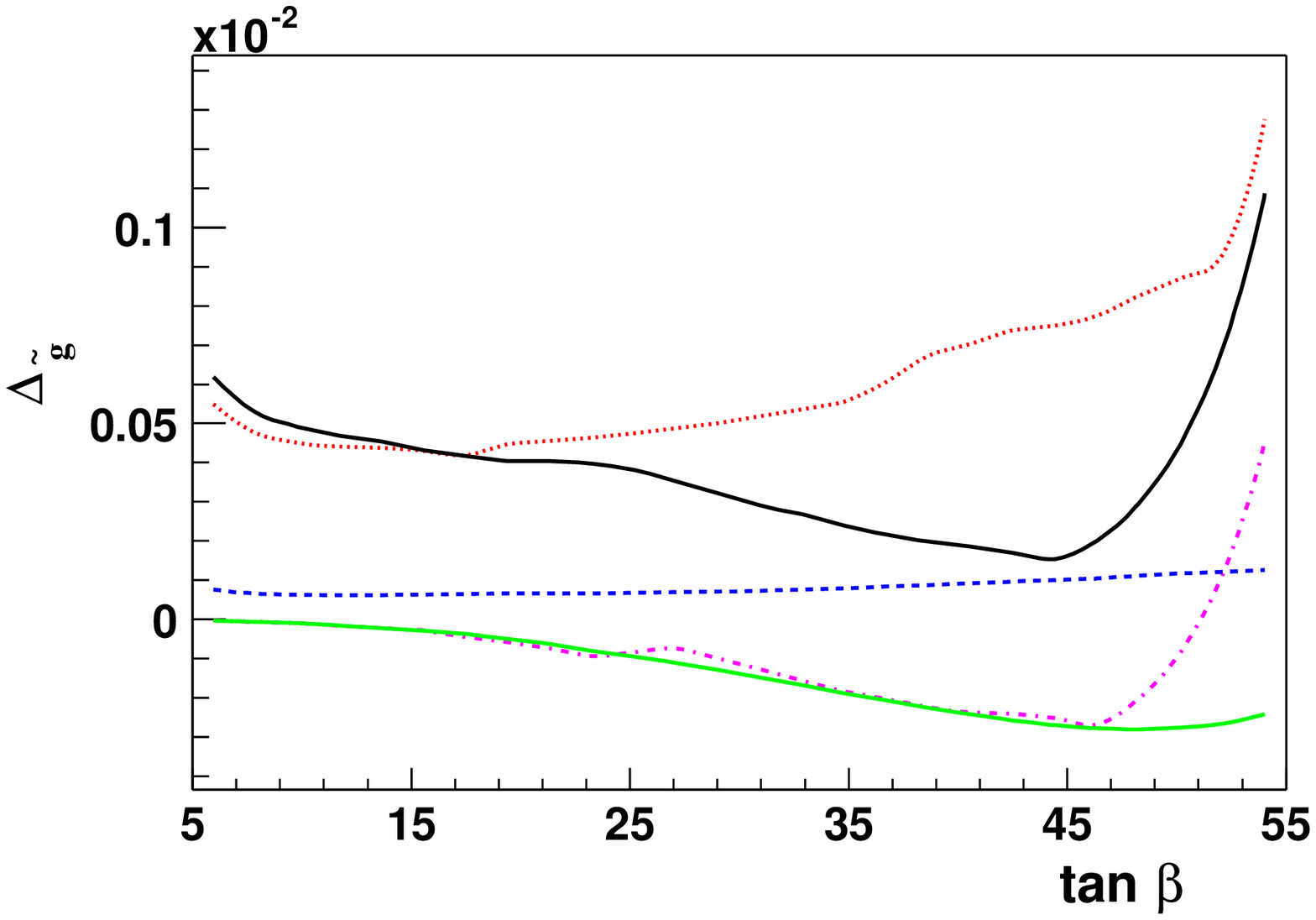,height=4.5cm,width=8.5cm}
\psfig{figure=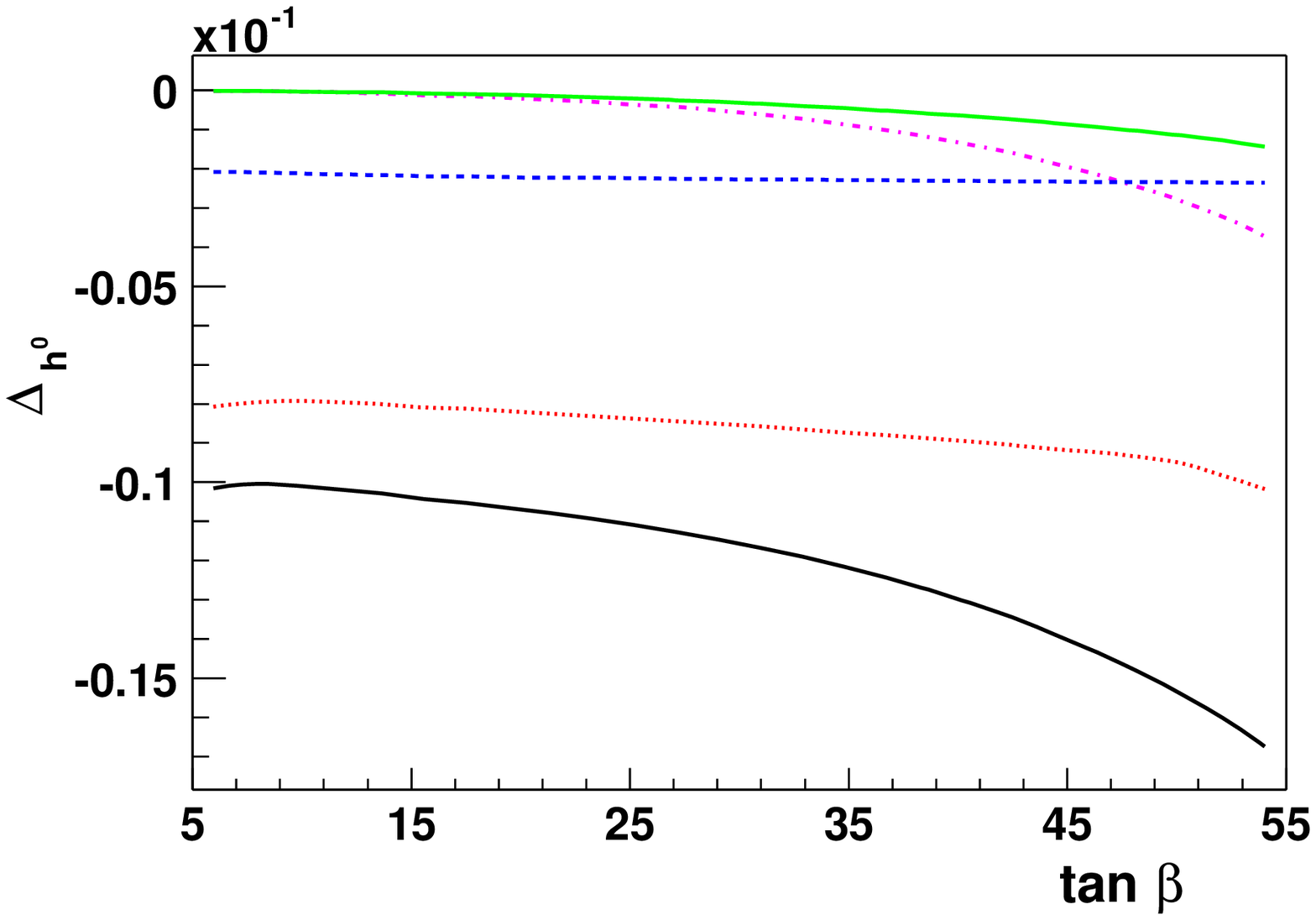,height=4.5cm,width=8.5cm}
\psfig{figure=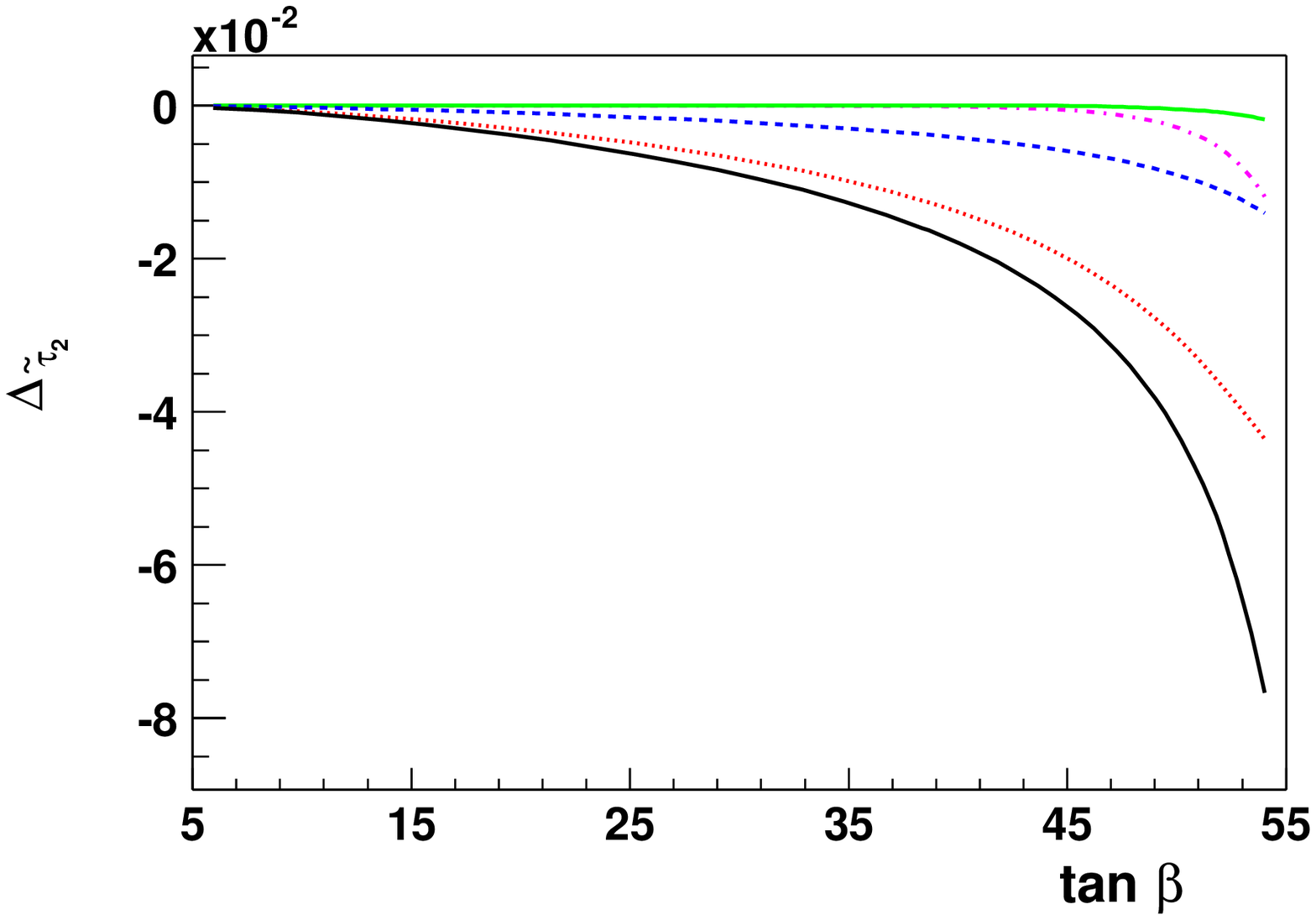,height=4.5cm,width=8.5cm}
\psfig{figure=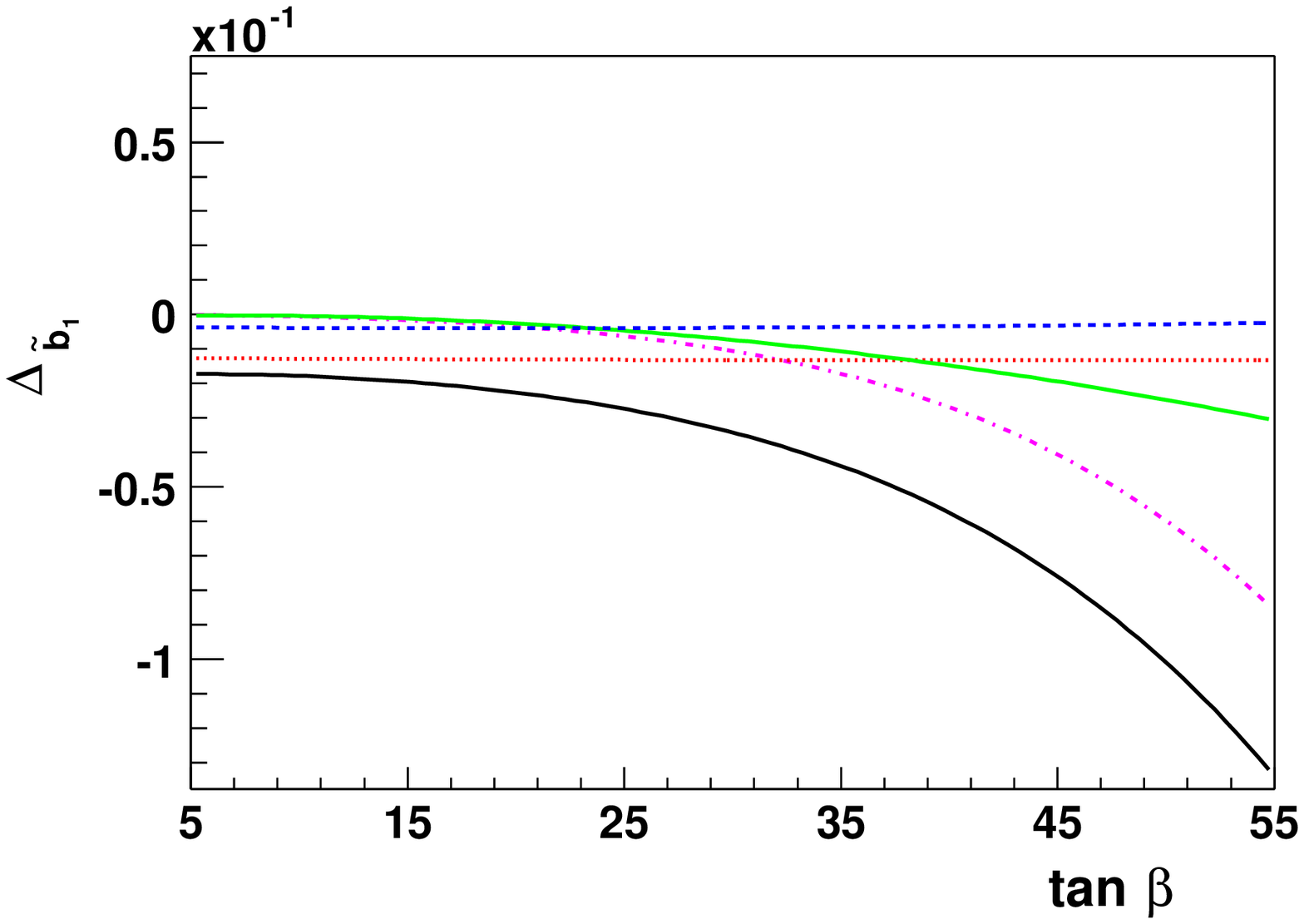,height=4.5cm,width=8.5cm}
\psfig{figure=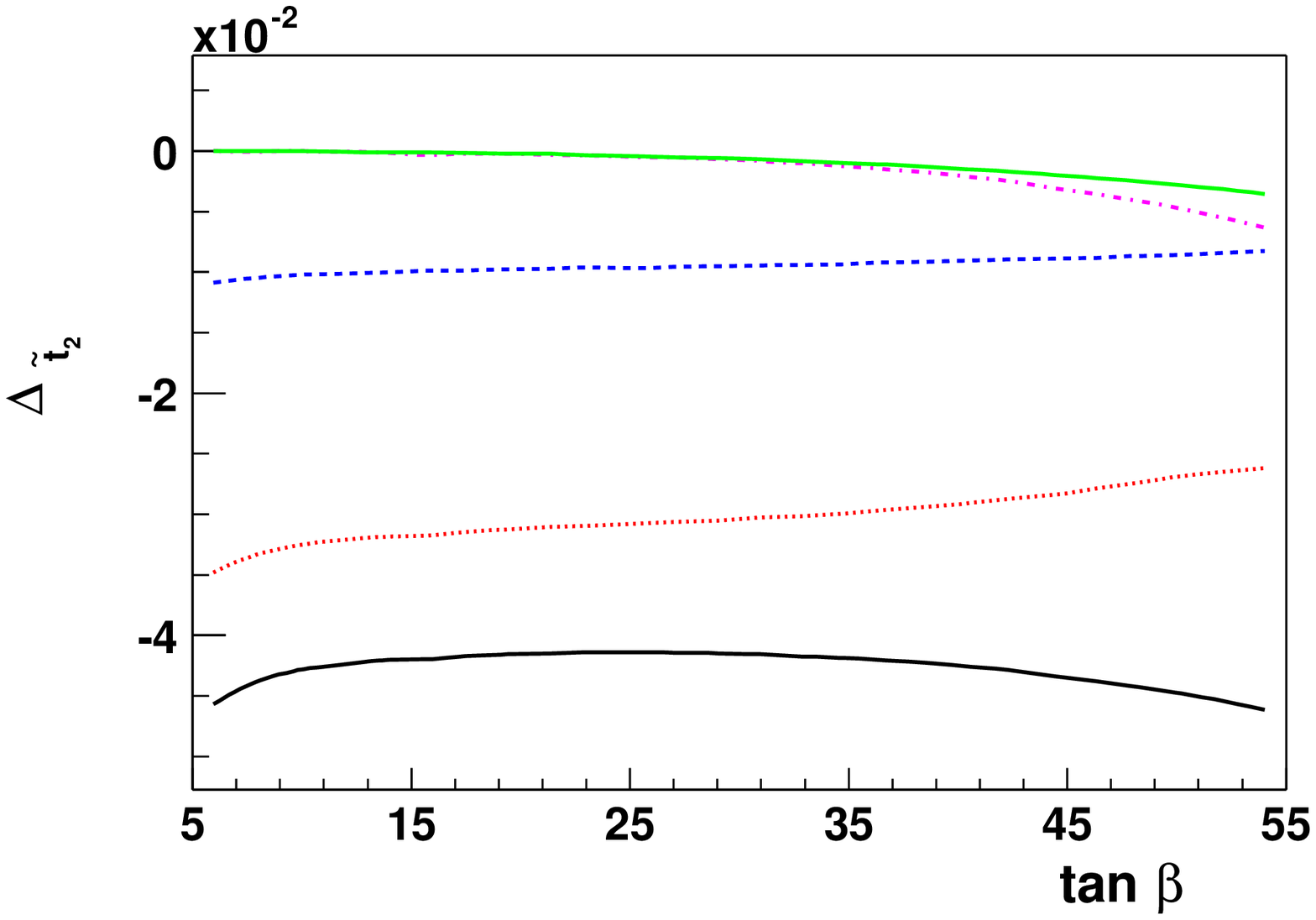,height=4.5cm,width=8.5cm}
\psfig{figure=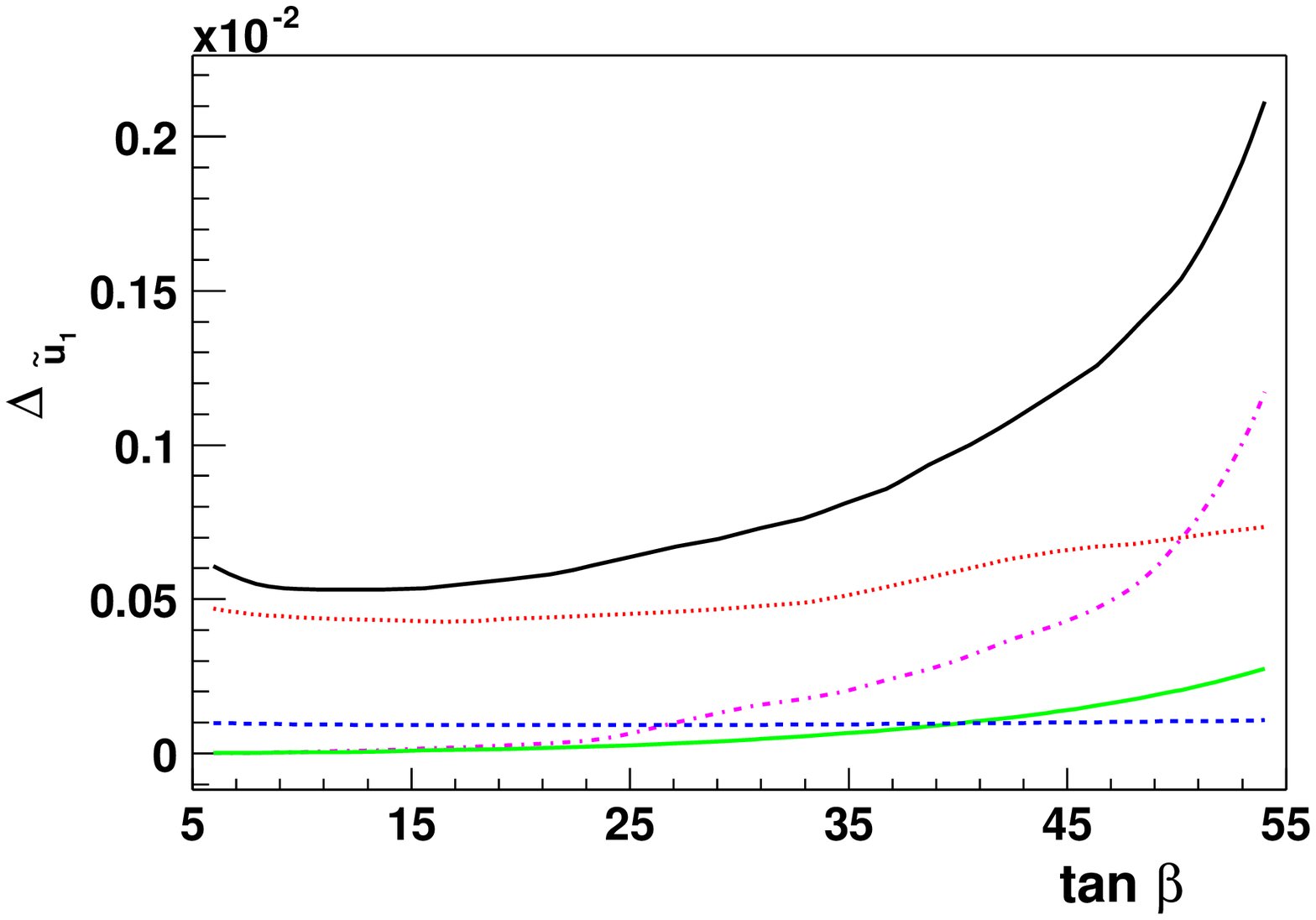,height=4.5cm,width=8.5cm}~~~~~~~
\psfig{figure=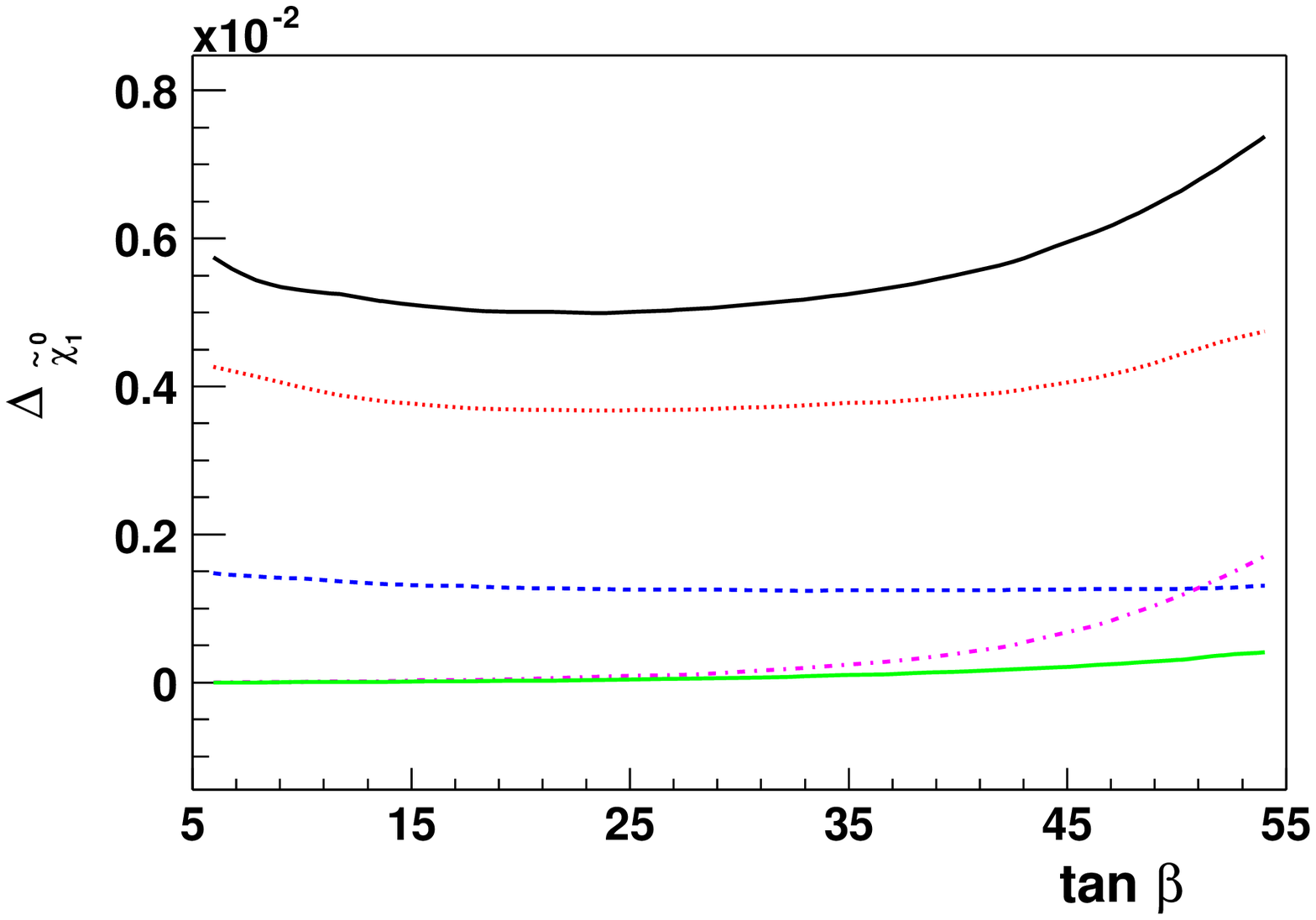,height=4.5cm,width=8.5cm}
\caption{$\Delta_p$ from the formula (\ref{deltap})
as functions of $\tan \beta$. In the first
picture we plotted a spectrum of masses (see also explanation in the text).} \label{fig:tanb}
\end{figure}

In Fig.\ref{fig:tanb}, we present the variation of $\Delta_p$ from the
above formula as a
function of $\tan \beta$ for $m_0=400\GeV$, $m_{1/2}=400\GeV$,
$A_0=0$ and $M_{GUT}=1.9\times 10^{16}\GeV$. We plotted $\Delta_p$
for five different cases:

$\bullet$ {\bf Bottom 1L} --- difference between the results obtained
with code including and excluding MSSM one-loop corrections to the
pole mass of the $b$-quark  

$\bullet$ {\bf Top 1L} --- the same as above for the $t$-quark

$\bullet$ {\bf Bottom} --- difference between the results obtained with
the code containing two-loop 
SUSY QCD
correction to
the $b$-quark together with MSSM one-loop corrections to the pole
masses of $b$- and $t$-quarks and results obtained with the code
containing MSSM one-loop corrections to the pole masses of $b$- and
$t$-quarks

$\bullet$ {\bf Top} --- the same as above for the $t$-quark

$\bullet$ {\bf All} --- difference between the results obtained with the
code containing  two-loop 
SUSY QCD correction
together with MSSM one-loop corrections to the pole masses of $b$-
and $t$-quarks and the results obtained with the code without any MSSM
(one-loop and two-loop) corrections to the pole masses of $b$- and
$t$-quarks.

From this figure one can see that two-loop correction to the
relation between the $b$-quark pole and \DR masses gives a sizeable
contribution only to the solution obtained with the SoftSUSY program
for the \DR $b$-quark mass at $m_Z$ and almost does not influence
spectra of SUSY particles. However, if one considers large
$\tan\beta$ it becomes important for other particle spectra and
thus should be taken into account in an accurate RG analysis. Note
that our two-loop SUSY QCD correction for a wide range of
parameter space is always less than a similar one-loop contribution.
The two-loop MSSM correction to the $t$-quark pole mass, on the other
hand, contributes to almost all SUSY particle masses obtained as
a solution of renormalization group equations with universal
boundary conditions at the GUT scale. This 2-loop correction is smaller
than 1-loop MSSM contribution, however its effect on particle
spectra is greater then corresponding 1-loop and 2-loop MSSM contributions
to the $b$-quark pole mass almost everywhere in MSSM parameter
space. This could be attributed to potentially large corrections
coming from higher terms in the ${m_t}/{M_{SUSY}}$
expansion.

To see the dependence of $\Delta_p$ on $m_0$, $m_{1/2}$ and $A_0$,
we use so called
a set of benchmark points and parameter lines in 
MSSM parameter space from \cite{BDEGMOPW,aix,modelline,SPS} which
corresponds to different scenarios in the search for Supersymmetry
at present and future colliders. While a detailed scan over the
more-than-hundred-dimensional parameter space of MSSM is
clearly not practicable, even a sampling of the three-
(four-)dimensional parameter space of $m_0$, $m_{1/2}$ and $A_0$
($\tan \beta$) is beyond the present capabilities 
of
phenomenological studies, especially when one tries to simulate
experimental signatures of supersymmetric particles within
detectors. For this reason, one often resorts to specific benchmark
scenarios, i.e.,\ one studies only specific parameter points or at
best samples the one-dimensional parameter space (the latter is
sometimes called a model line~\cite{modelline}), which exhibit
specific characteristics of 
MSSM parameter space.

Some recent proposals for SUSY benchmark scenarios may be found in
\cite{BDEGMOPW,aix,SPS}. We refer to the ``Snowmass Points and
Slopes'' (SPS)~\cite{modelline,SPS} which  consists of model lines
(``slopes''), i.e., continuous sets of parameters depending on the
one dimensionful parameter and specific benchmark points, where each
model line goes through one of the benchmark points.

\begin{figure}[h]
\psfig{figure=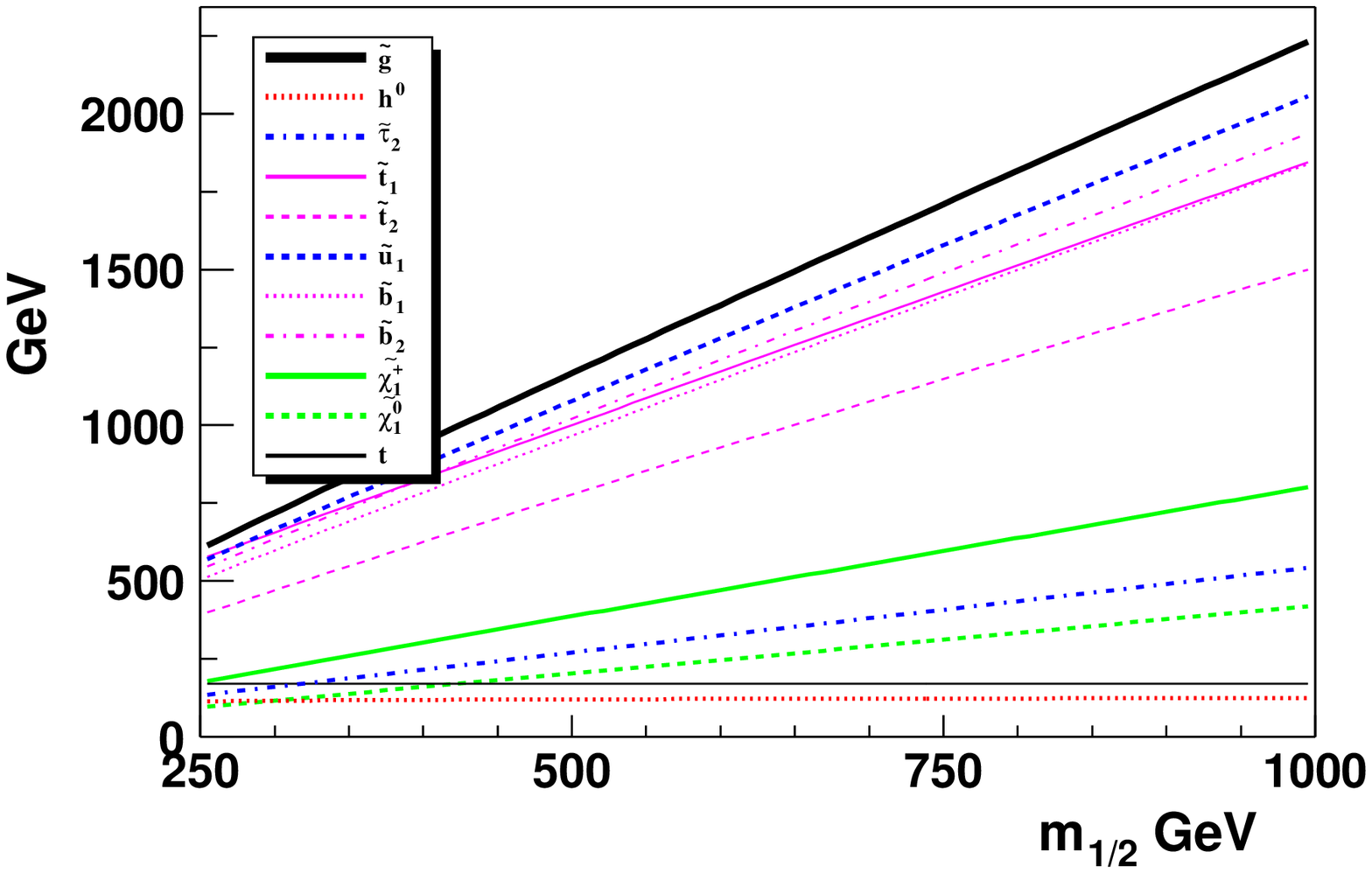,height=4.5cm,width=8.5cm}
\psfig{figure=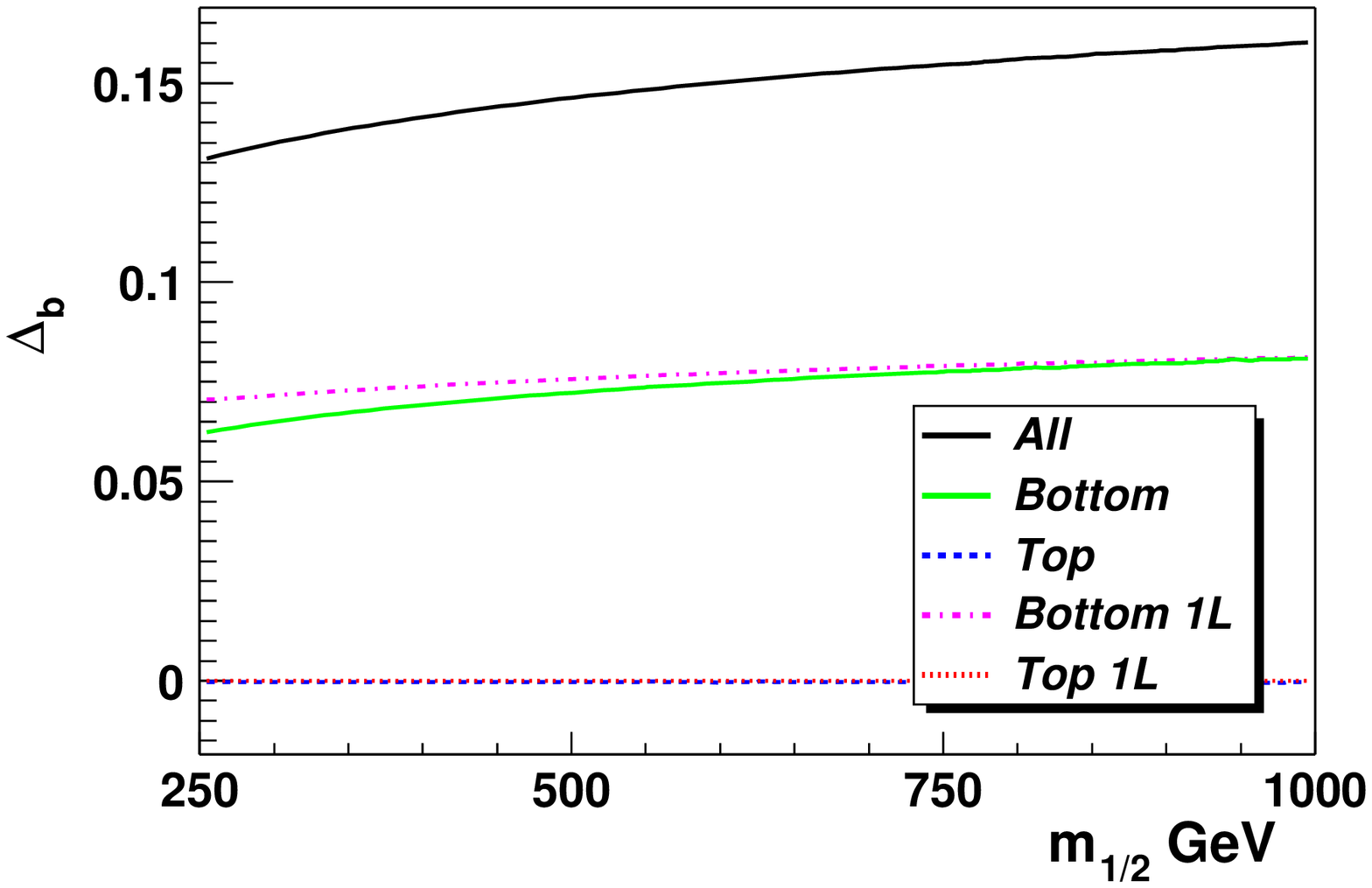,height=4.5cm,width=8.5cm}
\psfig{figure=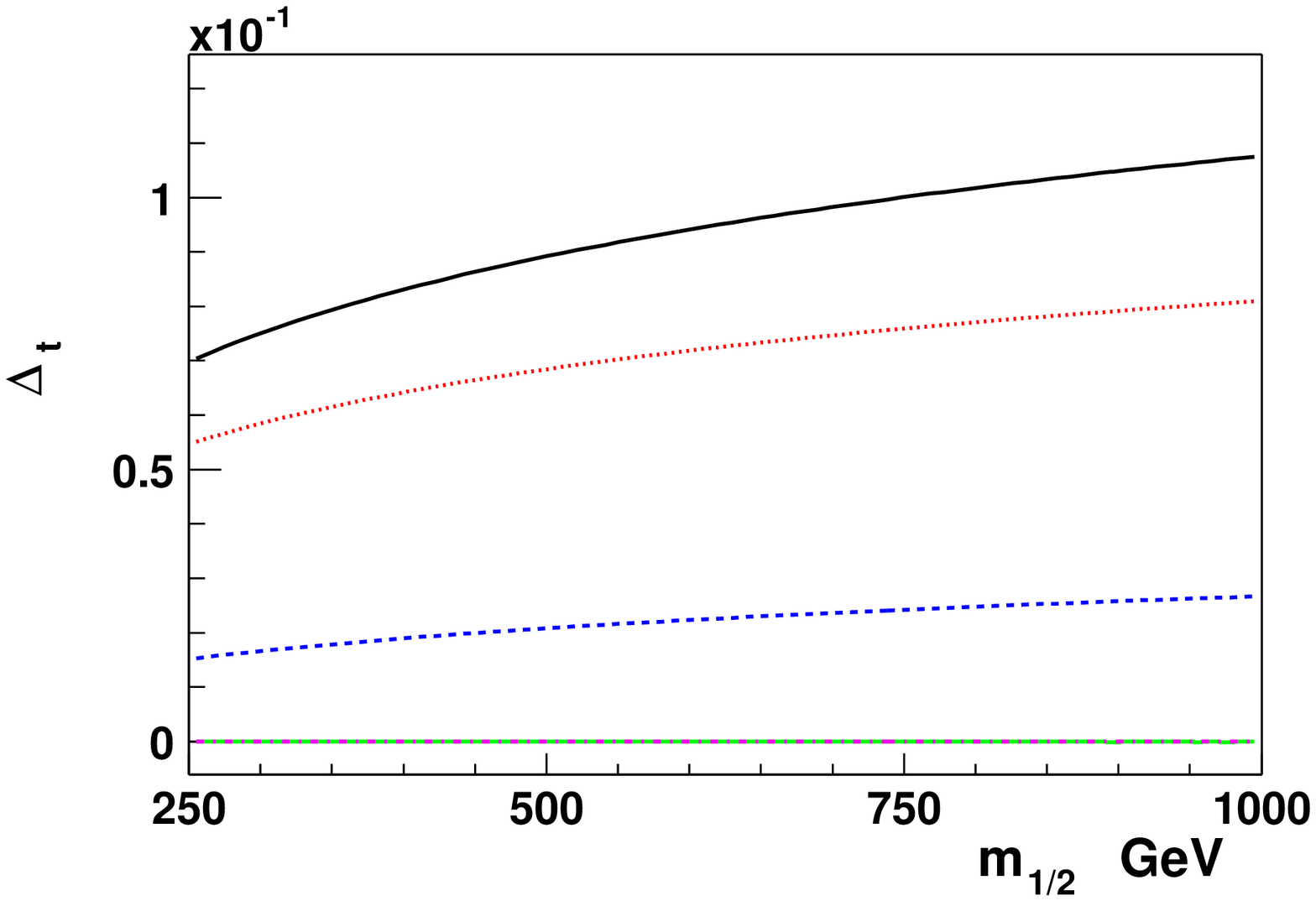,height=4.5cm,width=8.5cm}
\psfig{figure=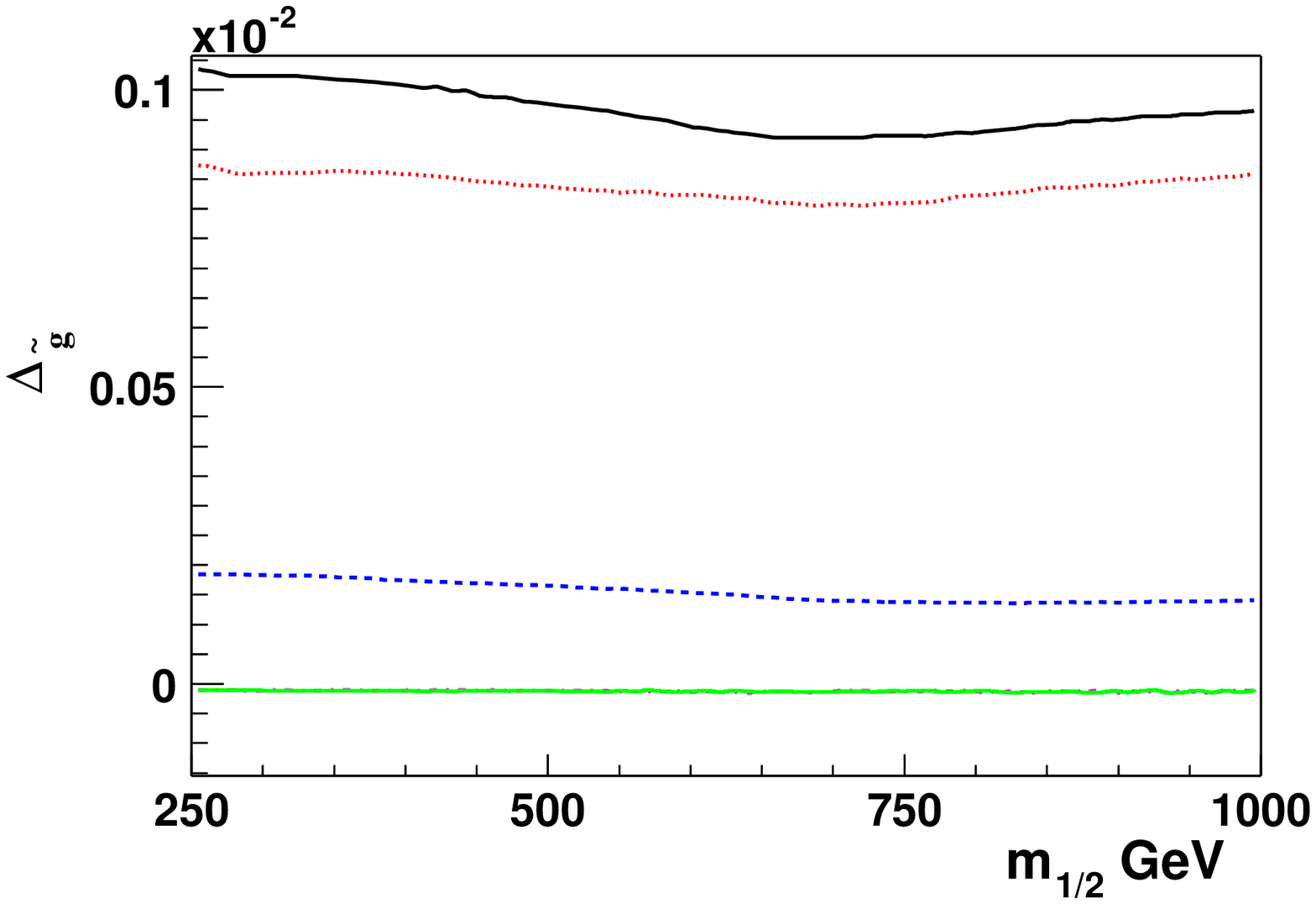,height=4.5cm,width=8.5cm}
\psfig{figure=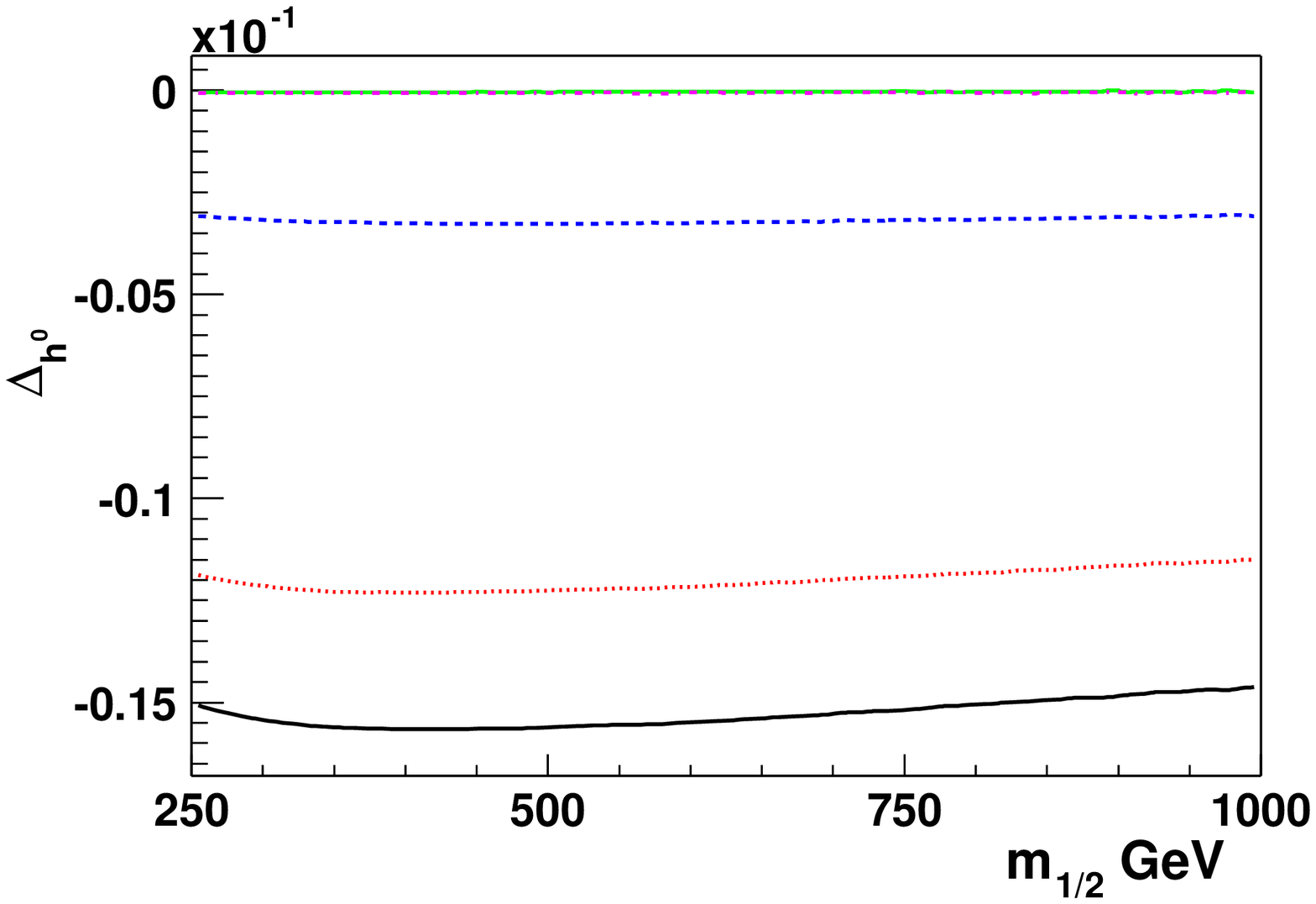,height=4.5cm,width=8.5cm}
\psfig{figure=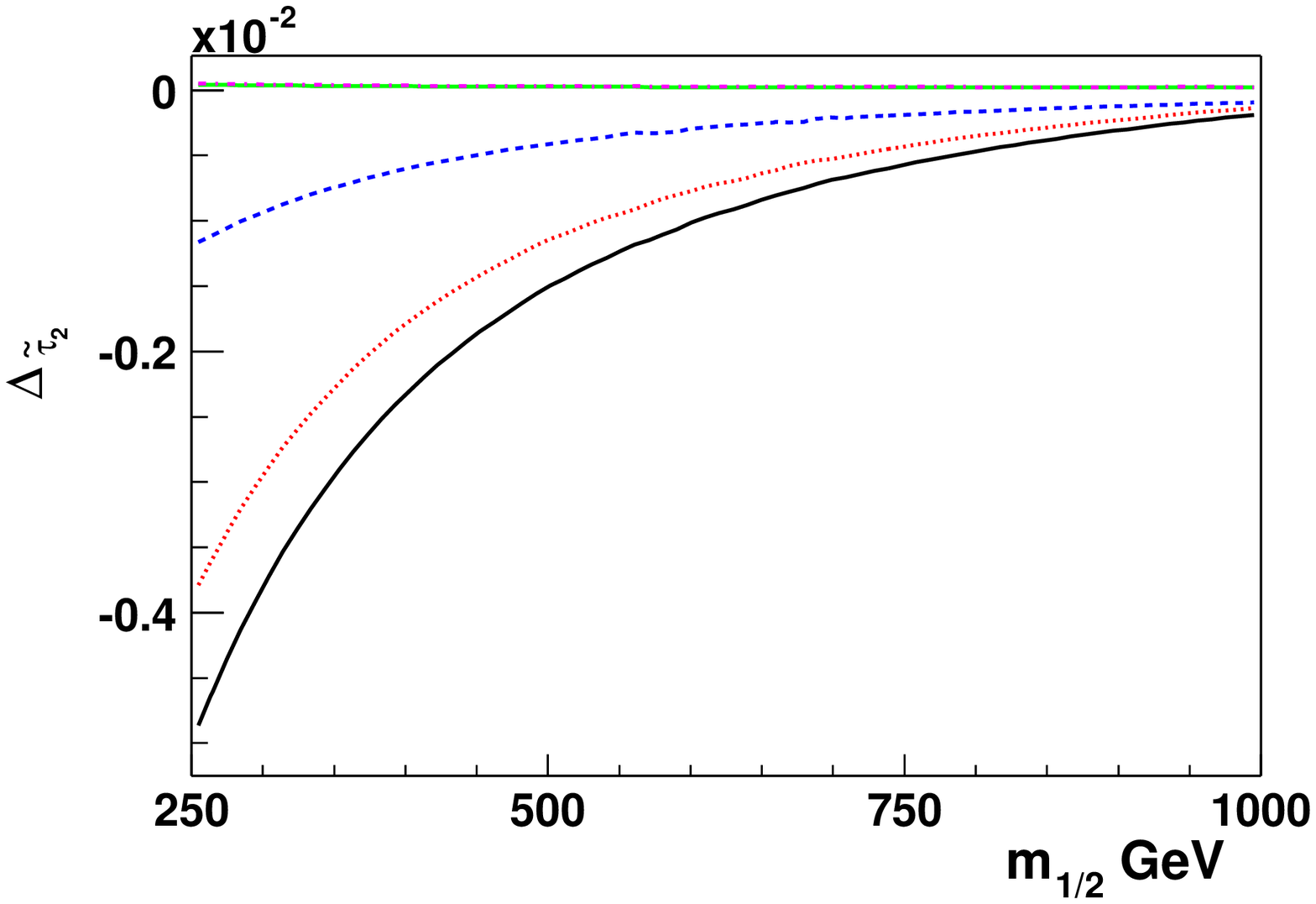,height=4.5cm,width=8.5cm}
\psfig{figure=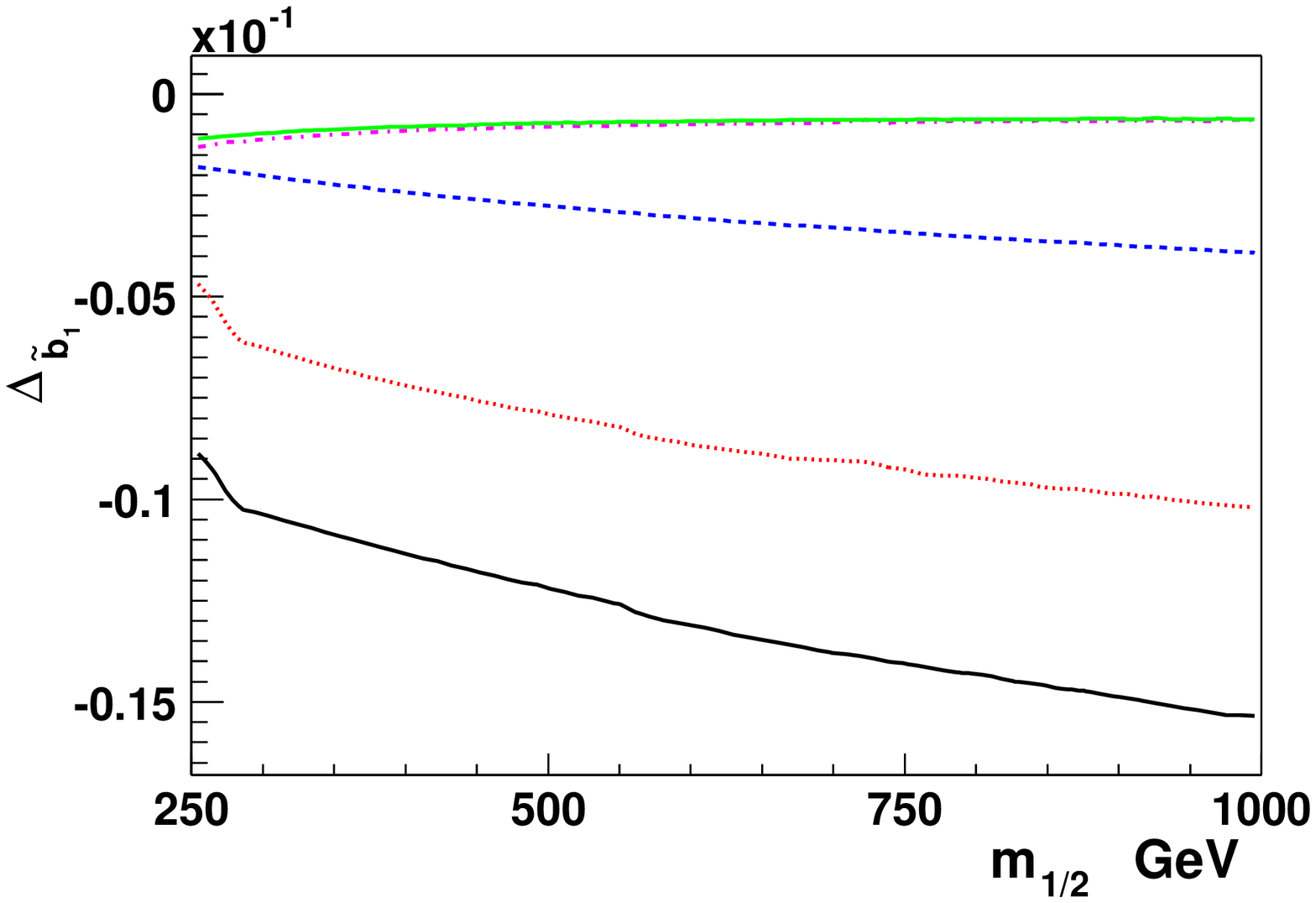,height=4.5cm,width=8.5cm}
\psfig{figure=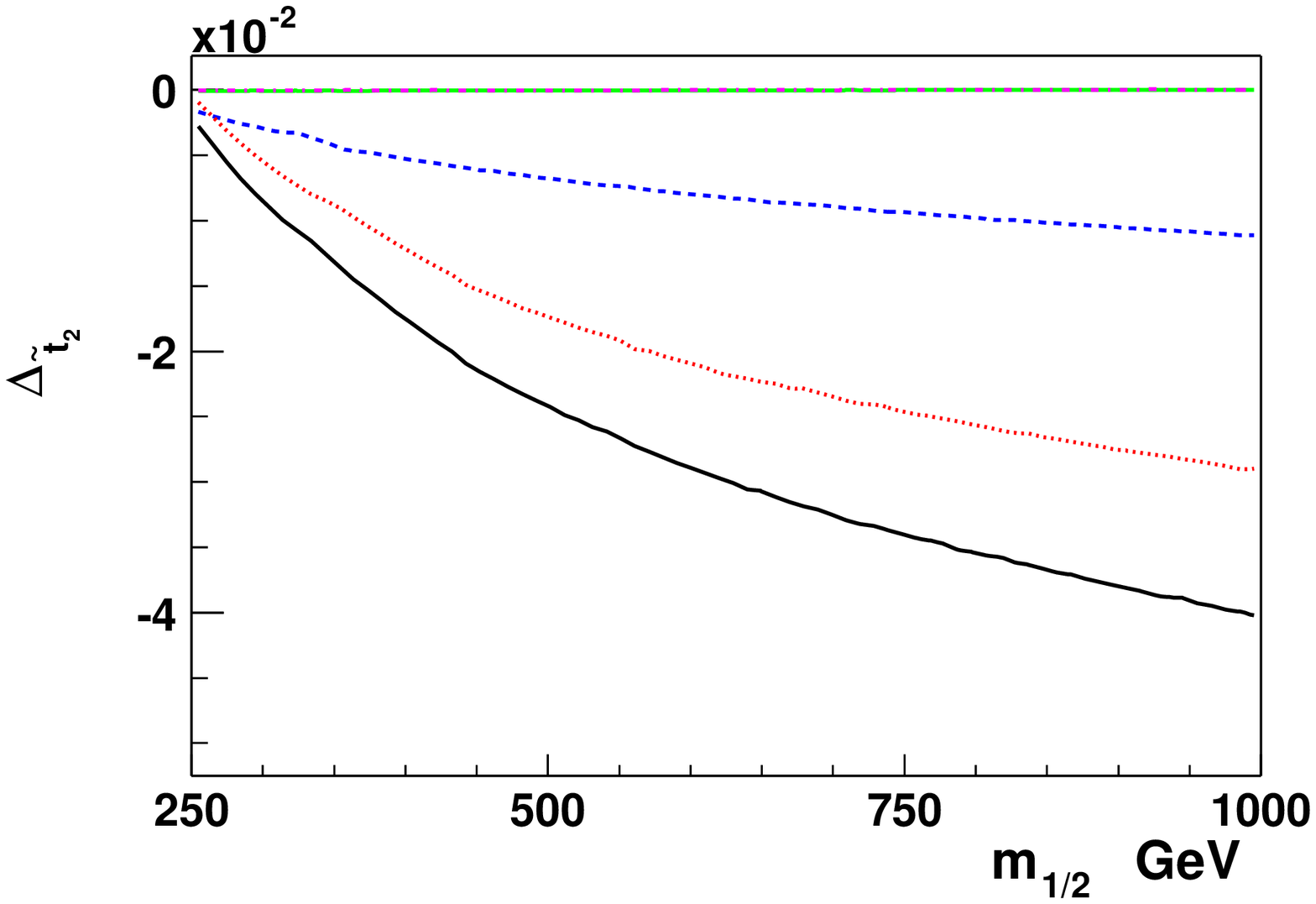,height=4.5cm,width=8.5cm}
\psfig{figure=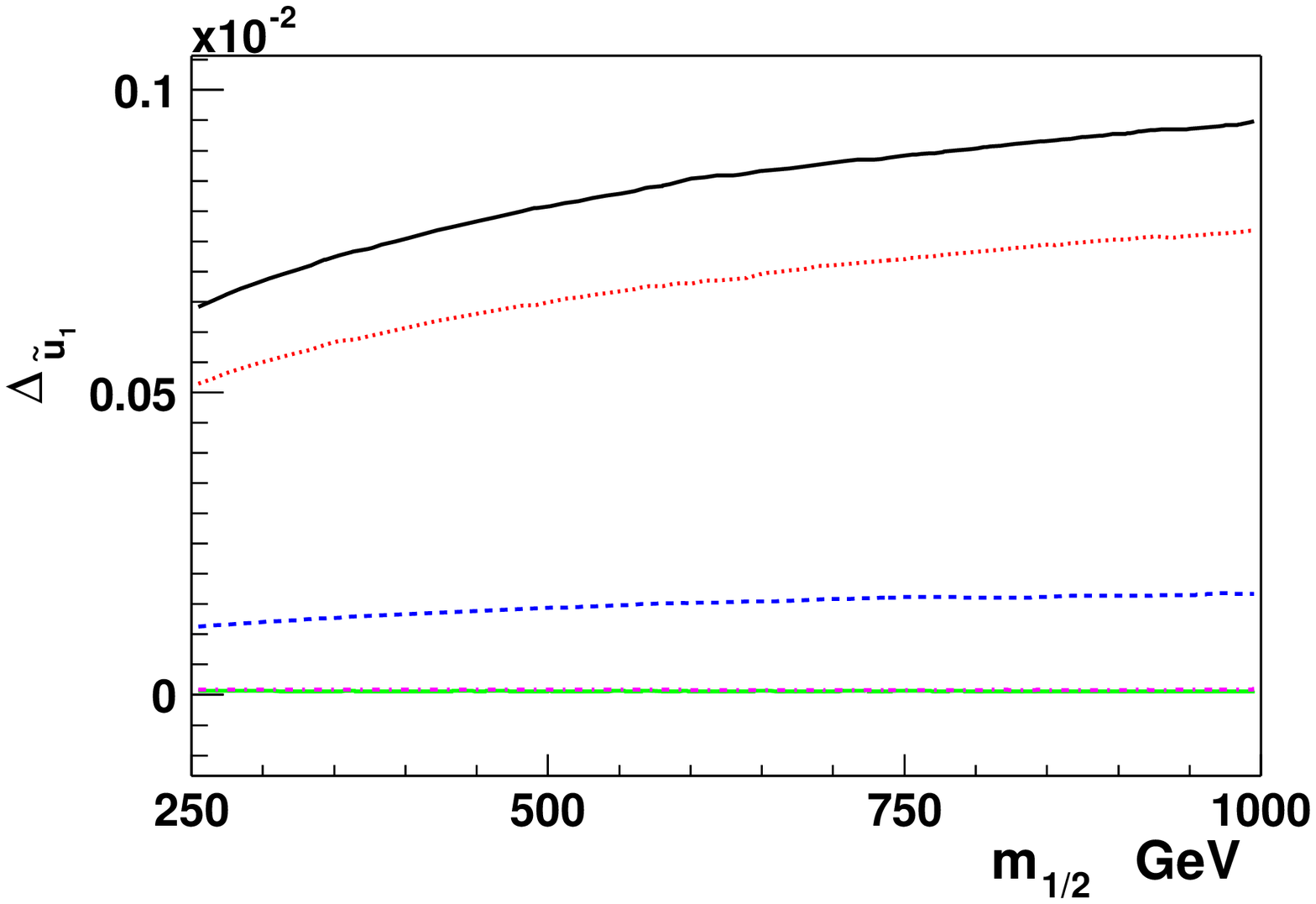,height=4.5cm,width=8.5cm}~~~~~~~
\psfig{figure=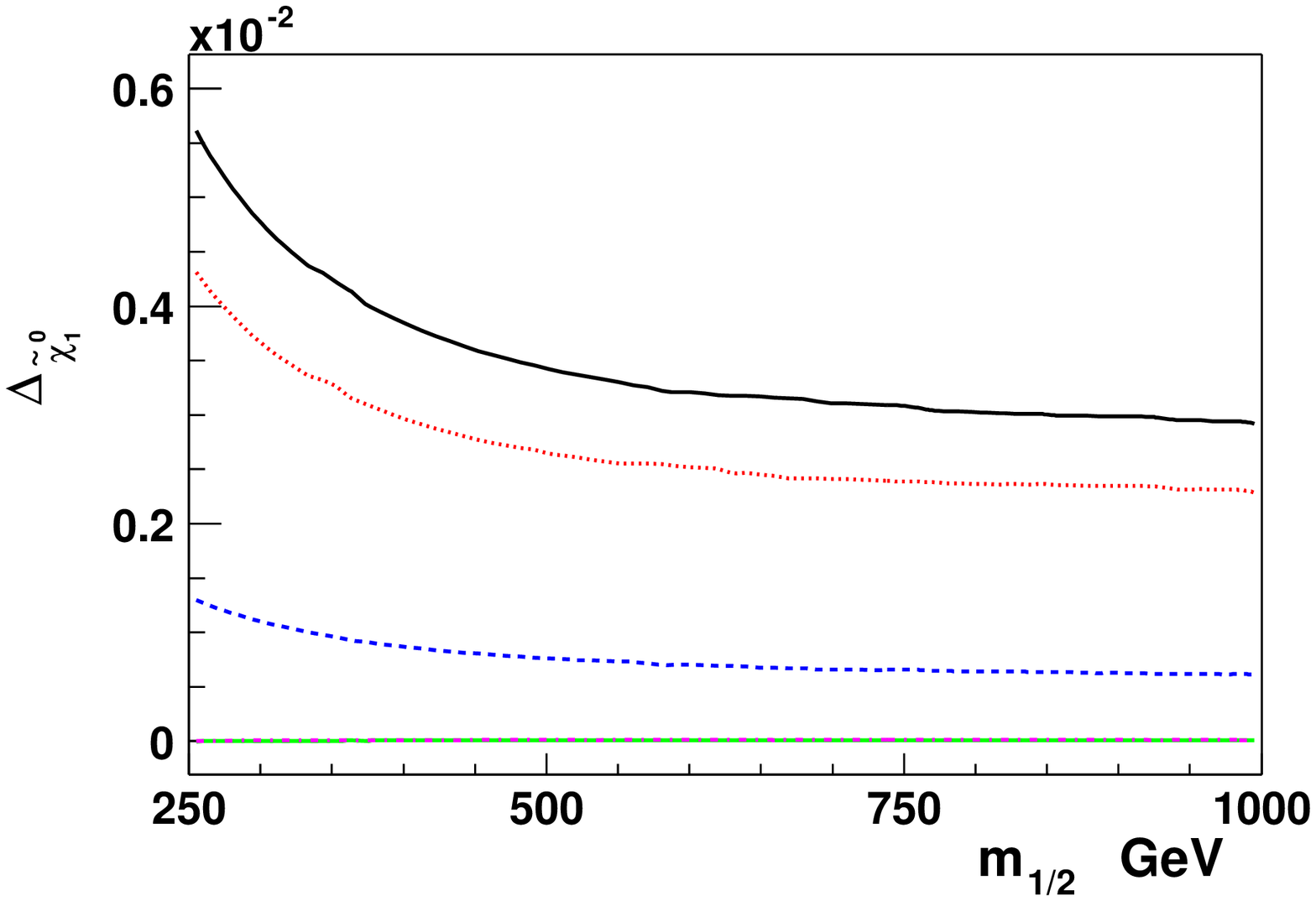,height=4.5cm,width=8.5cm}
\caption{$\Delta_p$ from the formula (\ref{deltap})
as functions defined along {\bf Model Line A}
from corrections depending on $m_{1/2}$. In the first picture we plotted a
spectrum of masses } \label{fig:linea}
\end{figure}

In Fig.\ref{fig:linea}, we present the results for $\Delta_p$ as
functions defined along {\bf Model Line A} $$ m_0 = -A_0 = 0.4 \,
m_{1/2}, \quad m_{1/2} \mbox{ varies}, \quad \tan\beta = 10, \quad
\mu > 0 . $$ where the benchmark point is $$ m_0 = 100 \GeV, \quad
m_{1/2} = 250 \GeV, \quad A_0 = -100 \GeV, \quad \tan\beta = 10,
\quad \mu > 0 .
$$
Here and below $\mu$ is the supersymmetric Higgs mass parameter

As one can easily see from Fig.\ref{fig:linea}, varying
$m_{1/2}$ at fixed $\tan\beta$ 2-loop SUSY QCD correction to the
$b$-quark pole mass almost does not contribute to heavy
supersymmetric particle spectra. On the other hand, our correction
to the $t$-quark pole mass is comparable to the corresponding 1-loop
contribution.

\begin{figure}[h]
\psfig{figure=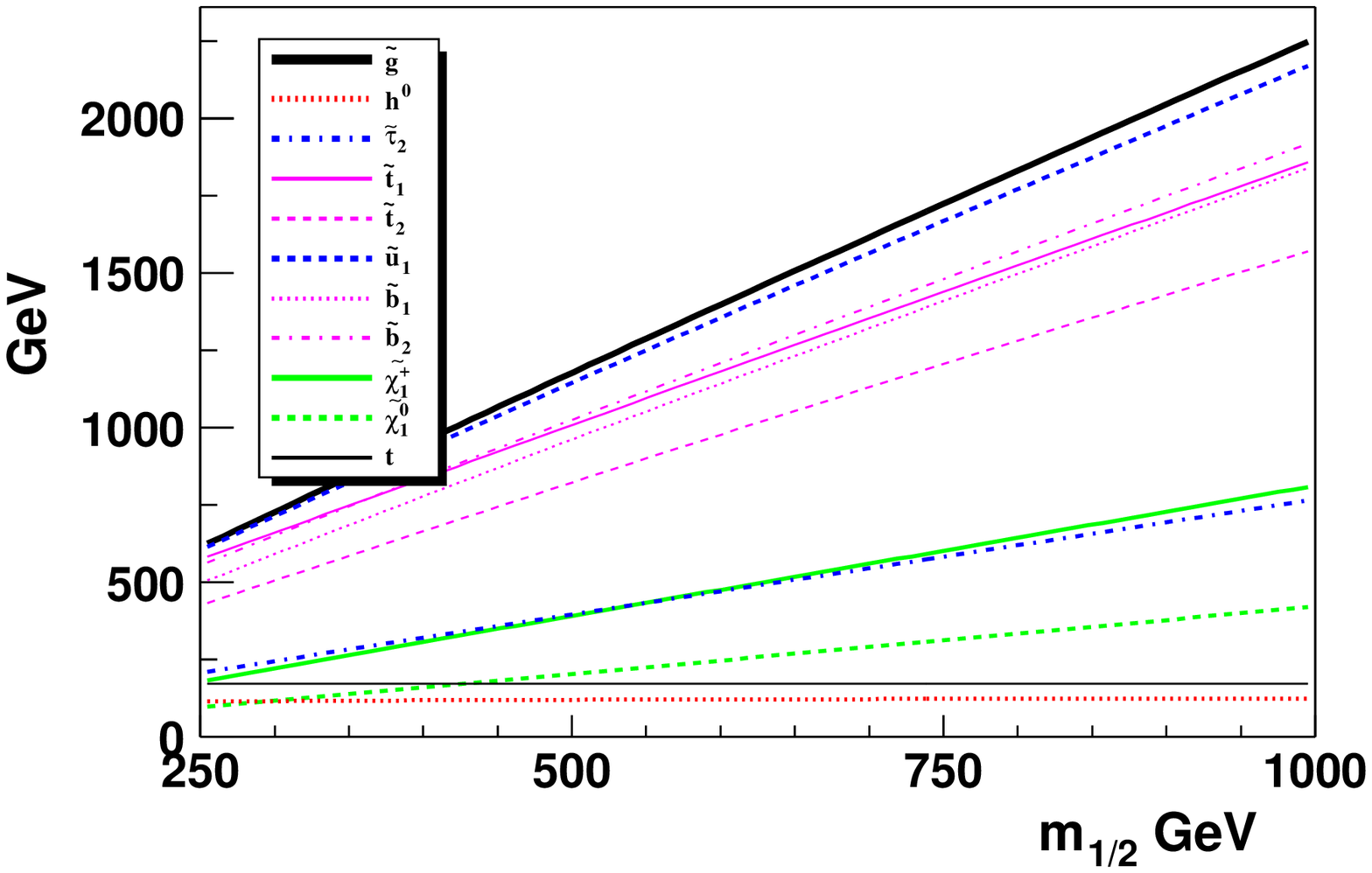,height=4.5cm,width=8.5cm}
\psfig{figure=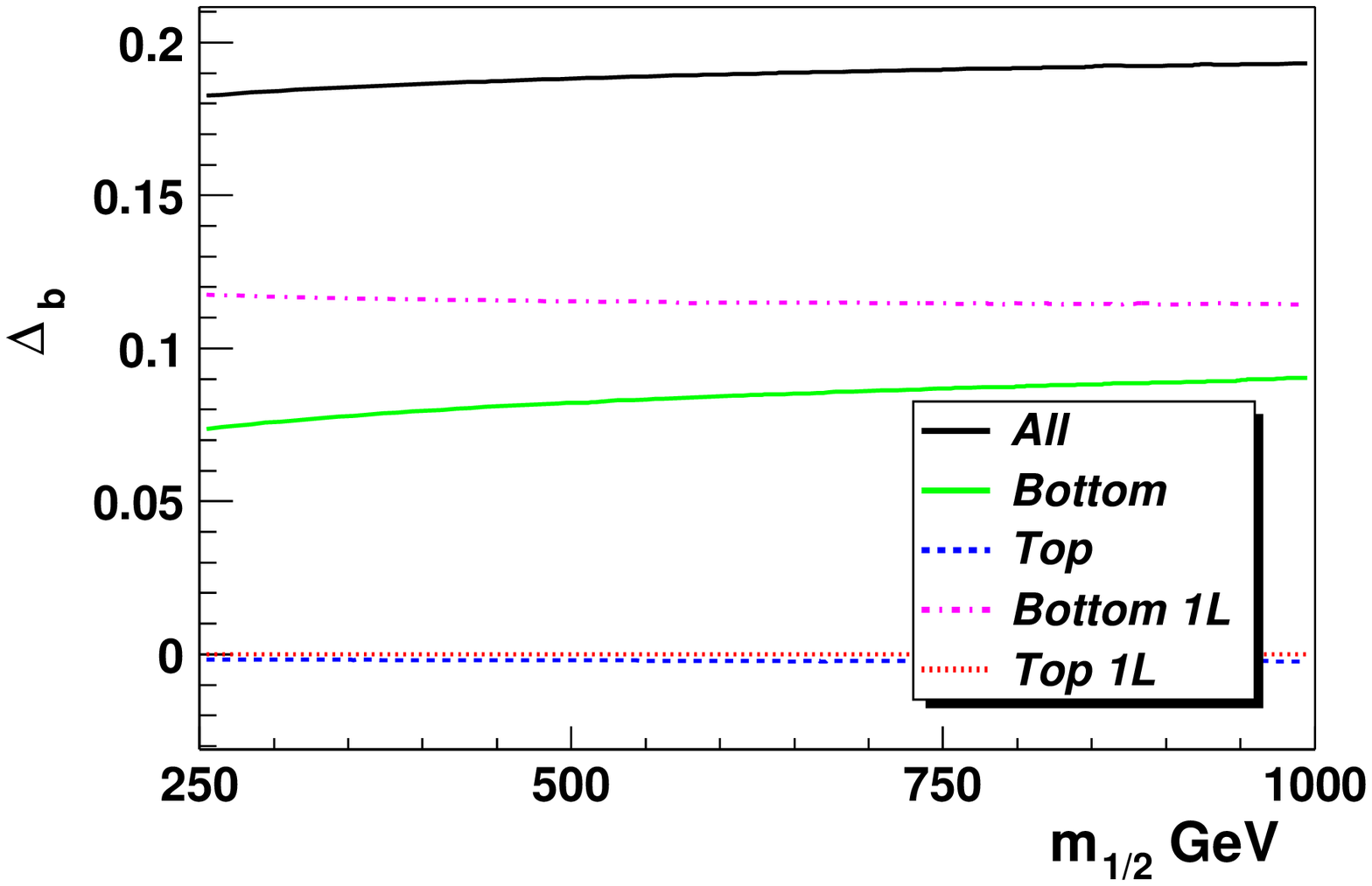,height=4.5cm,width=8.5cm}
\psfig{figure=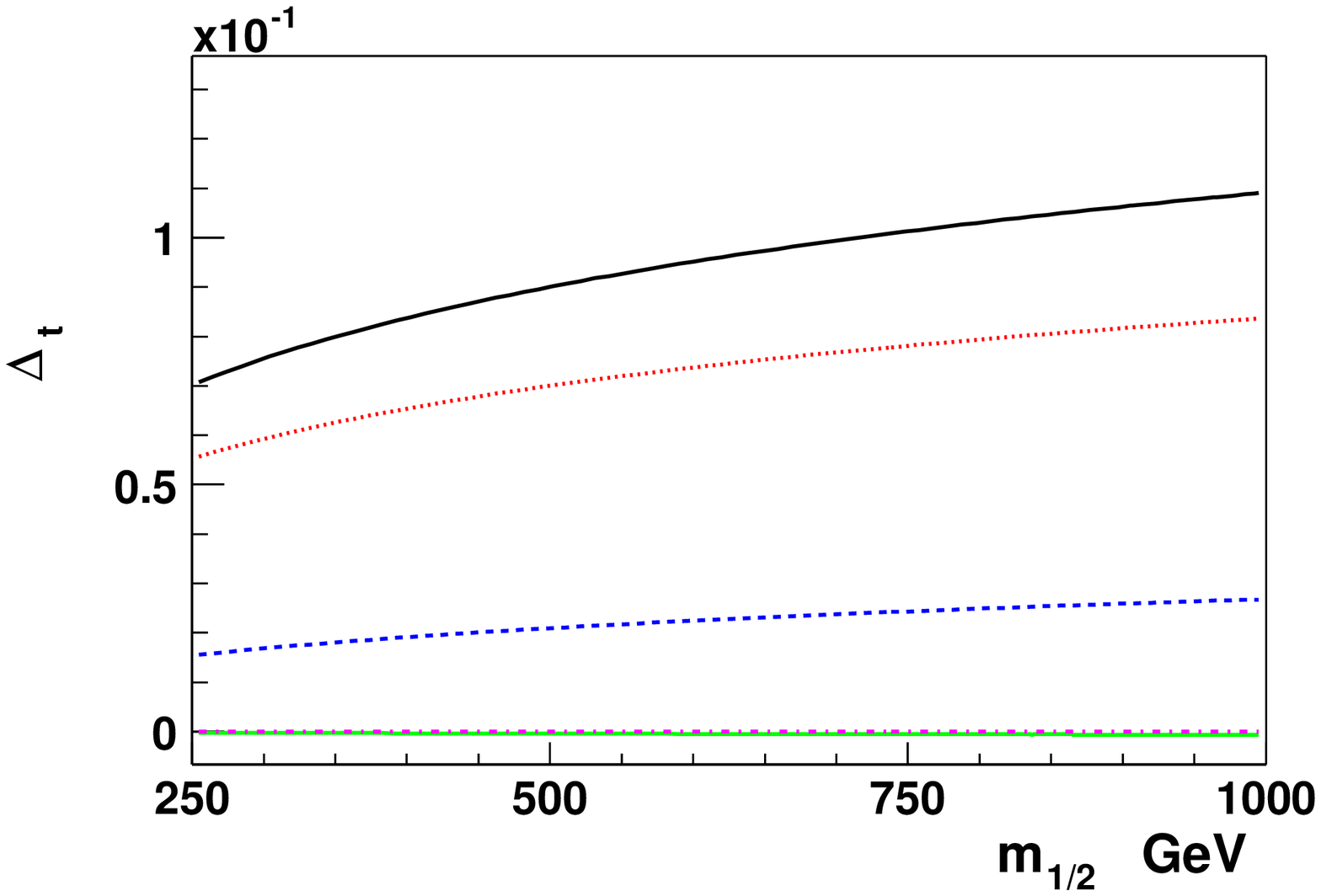,height=4.5cm,width=8.5cm}
\psfig{figure=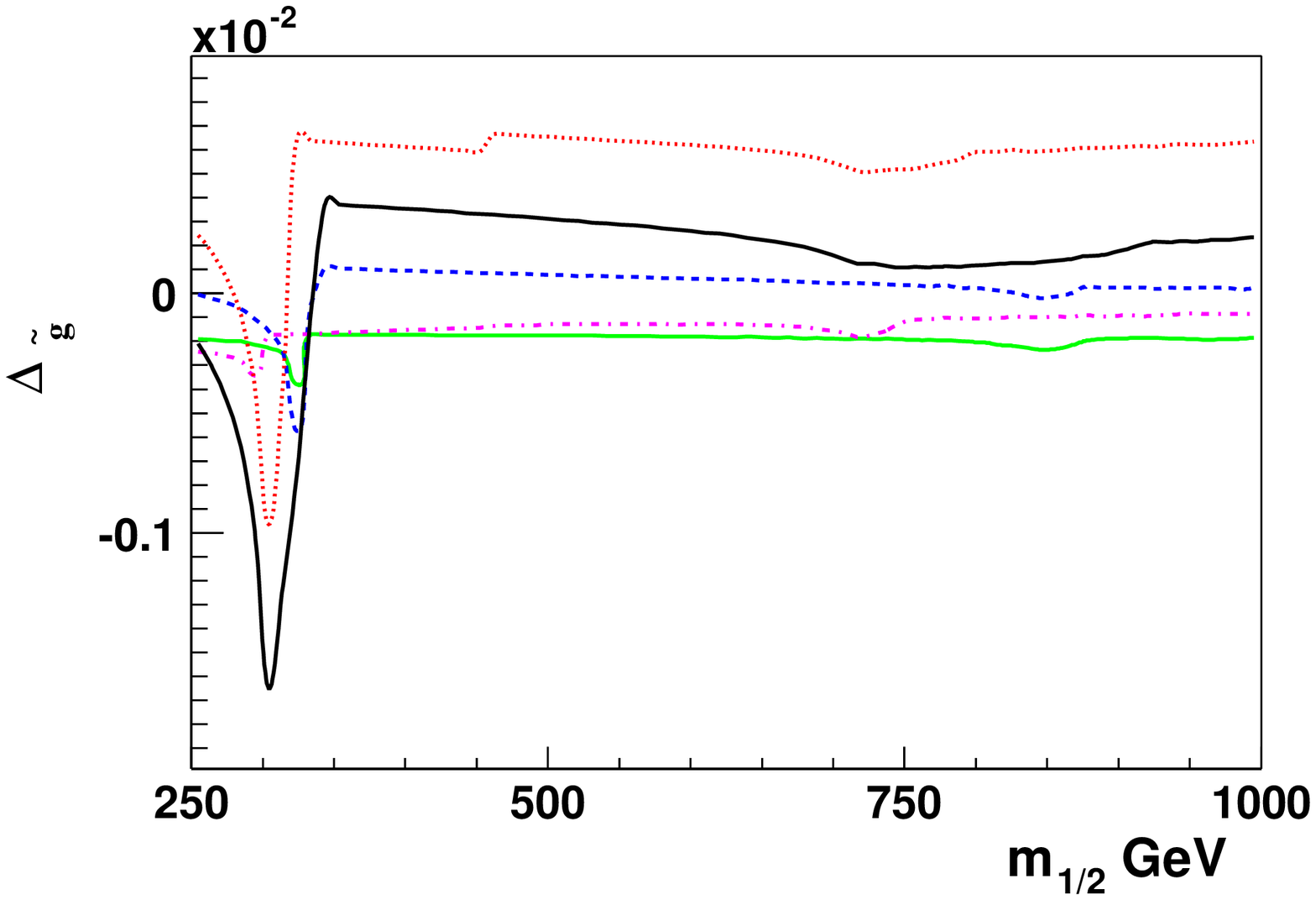,height=4.5cm,width=8.5cm}
\psfig{figure=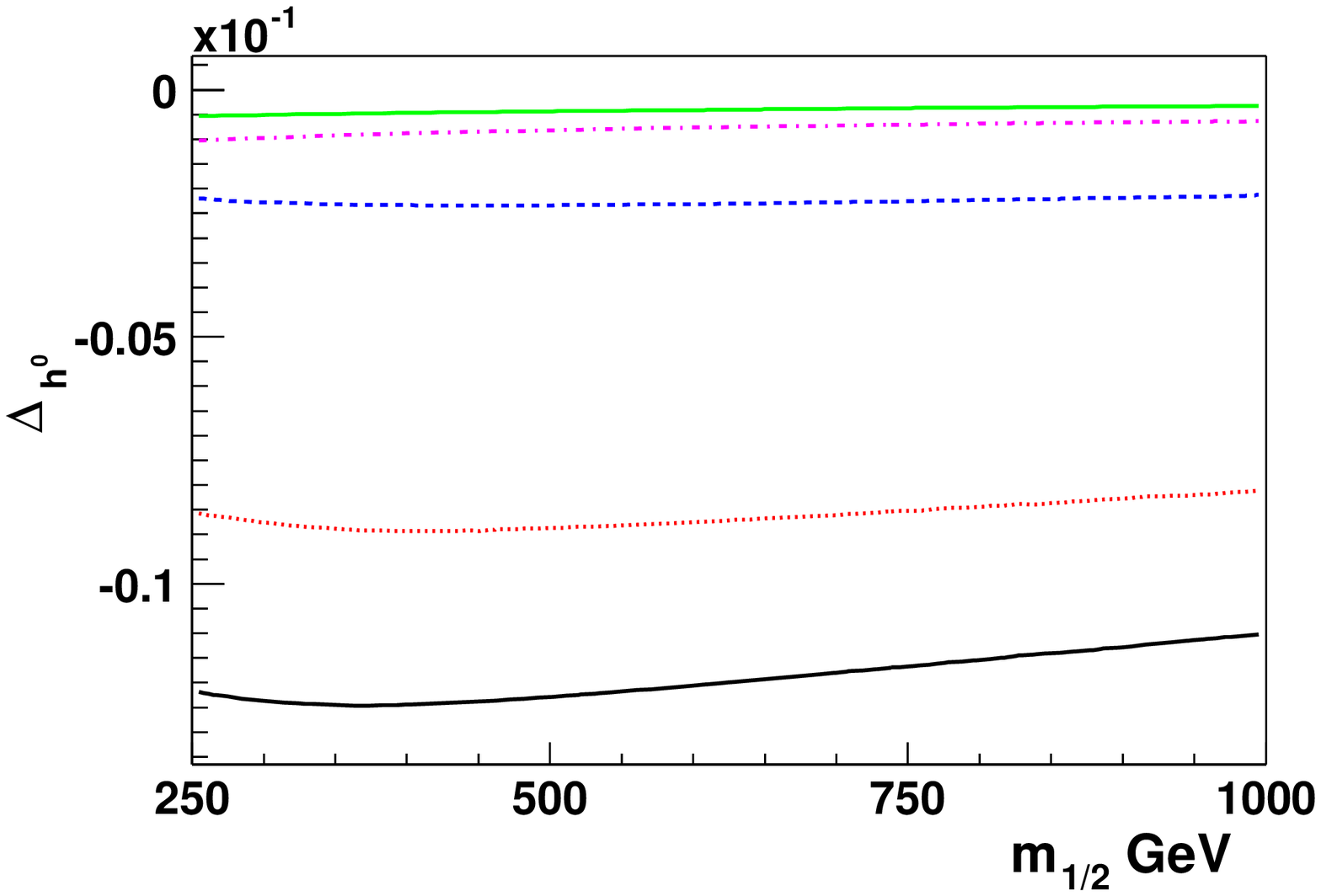,height=4.5cm,width=8.5cm}
\psfig{figure=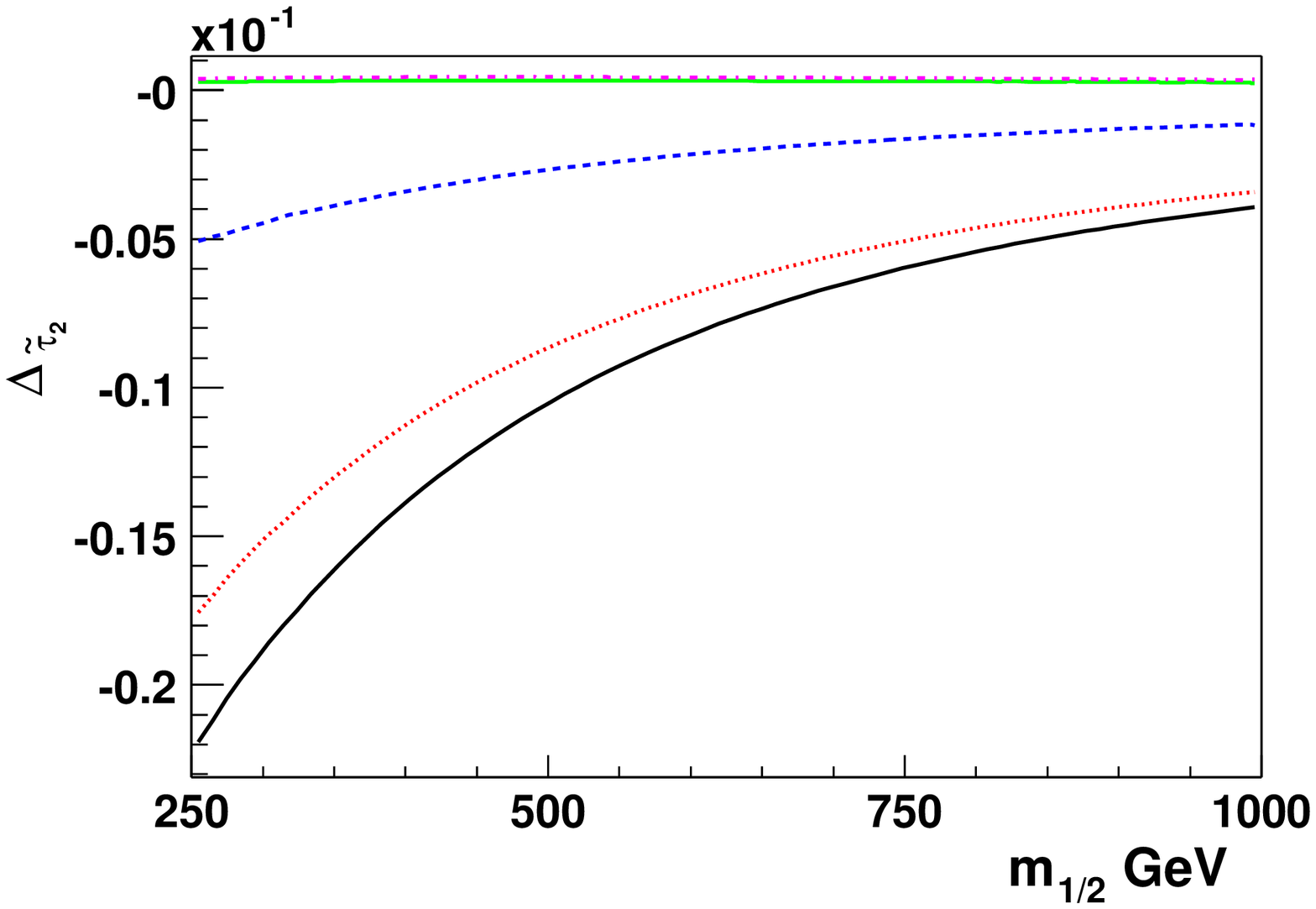,height=4.5cm,width=8.5cm}
\psfig{figure=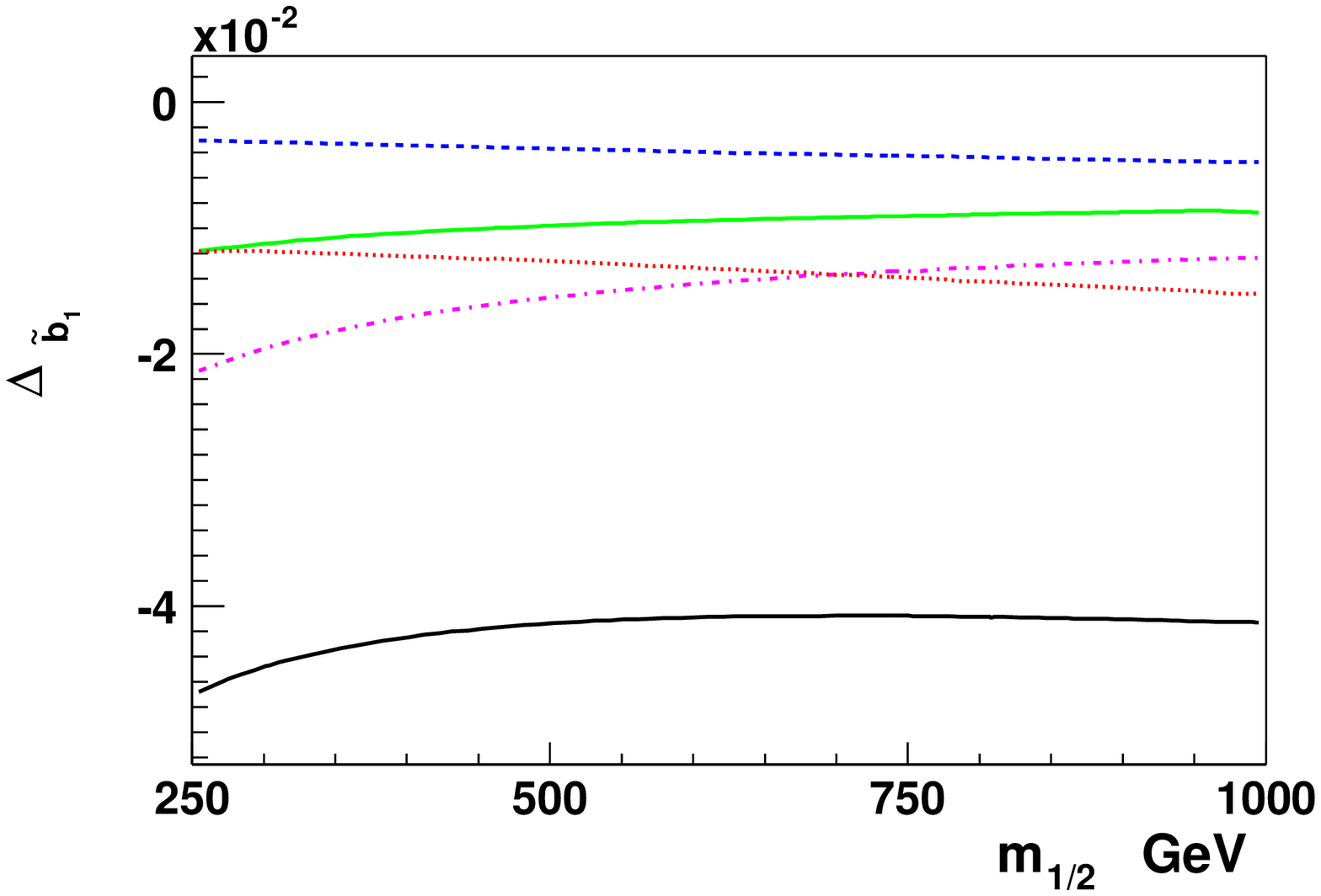,height=4.5cm,width=8.5cm}
\psfig{figure=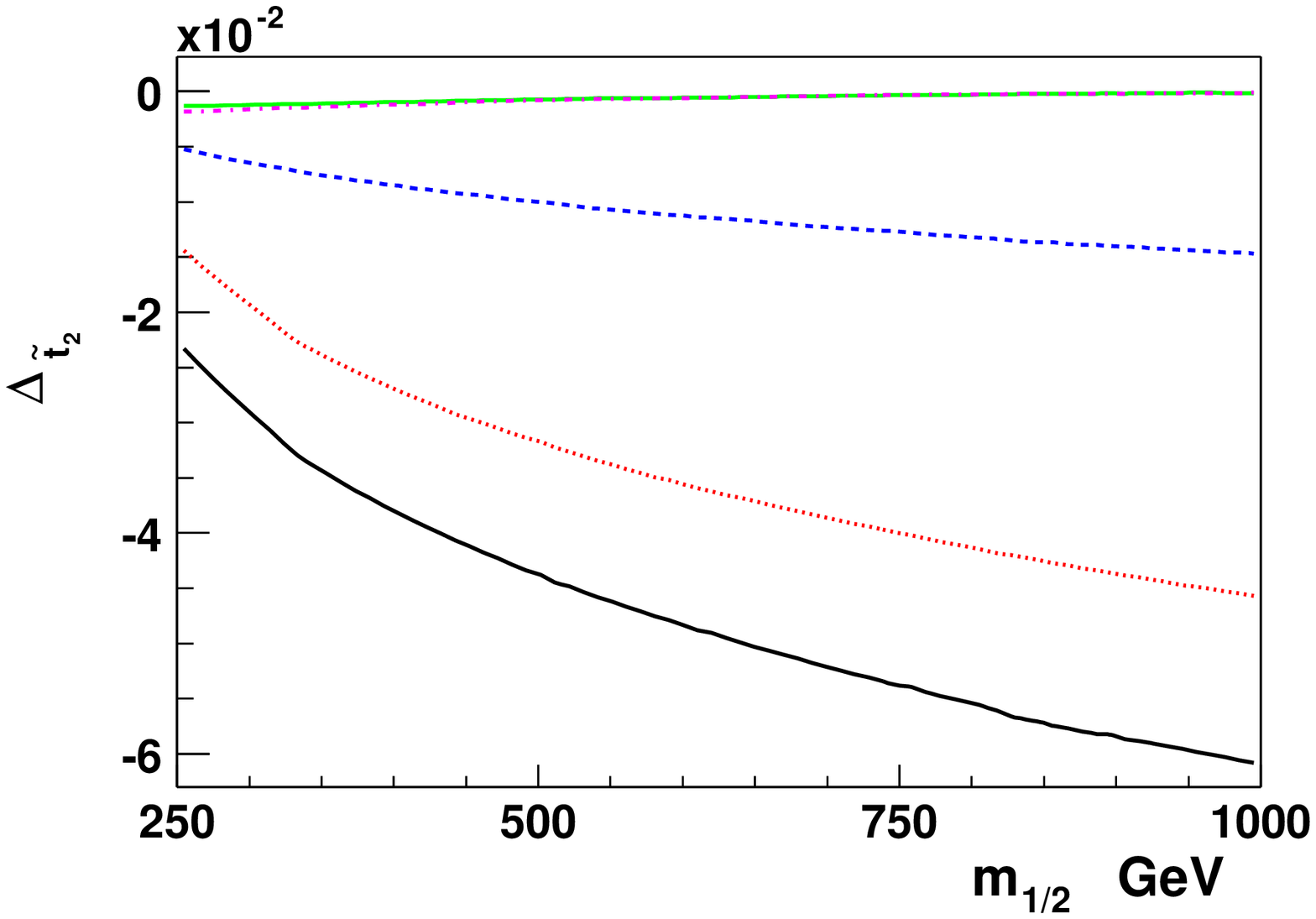,height=4.5cm,width=8.5cm}
\psfig{figure=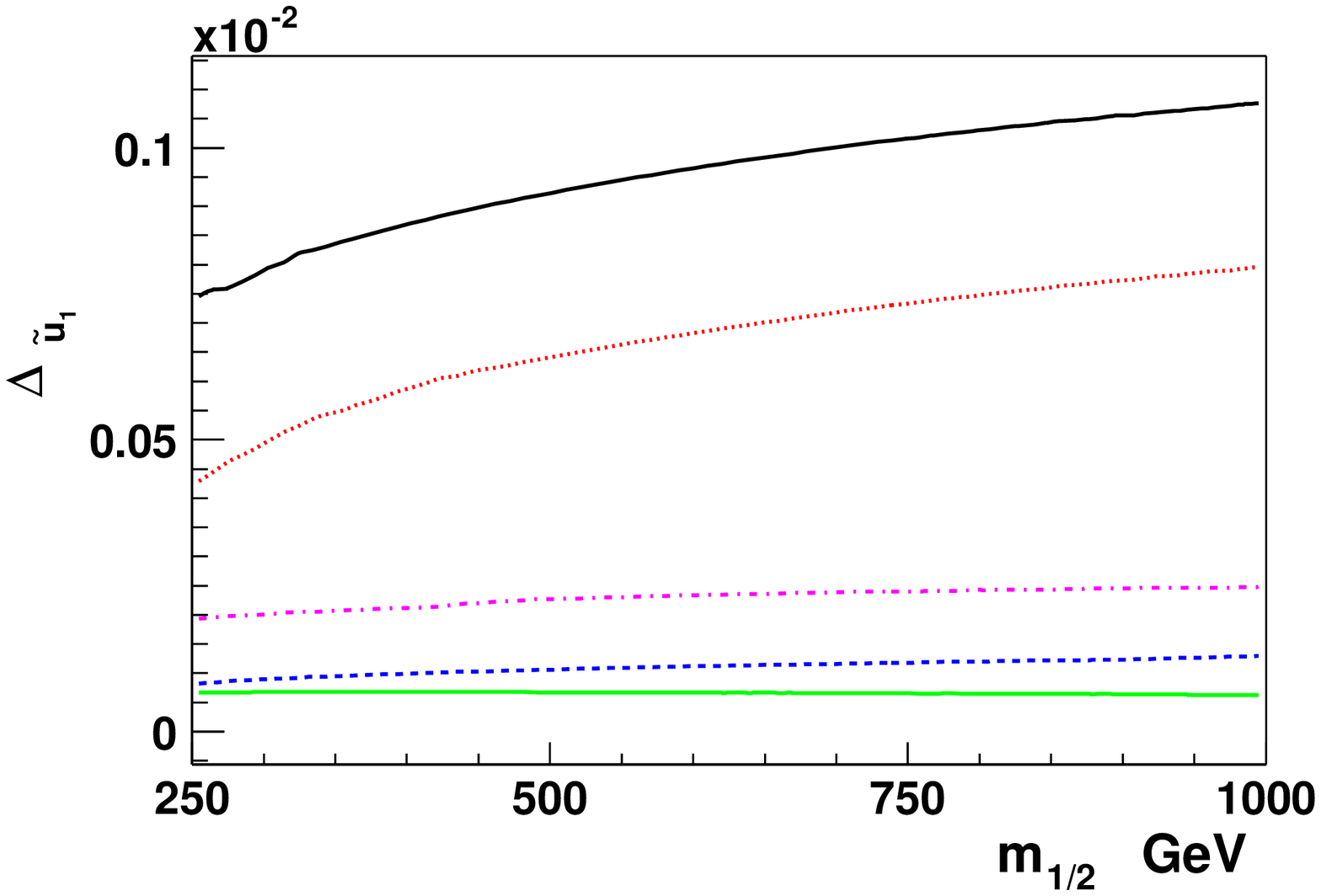,height=4.5cm,width=8.5cm}~~~~~~~
\psfig{figure=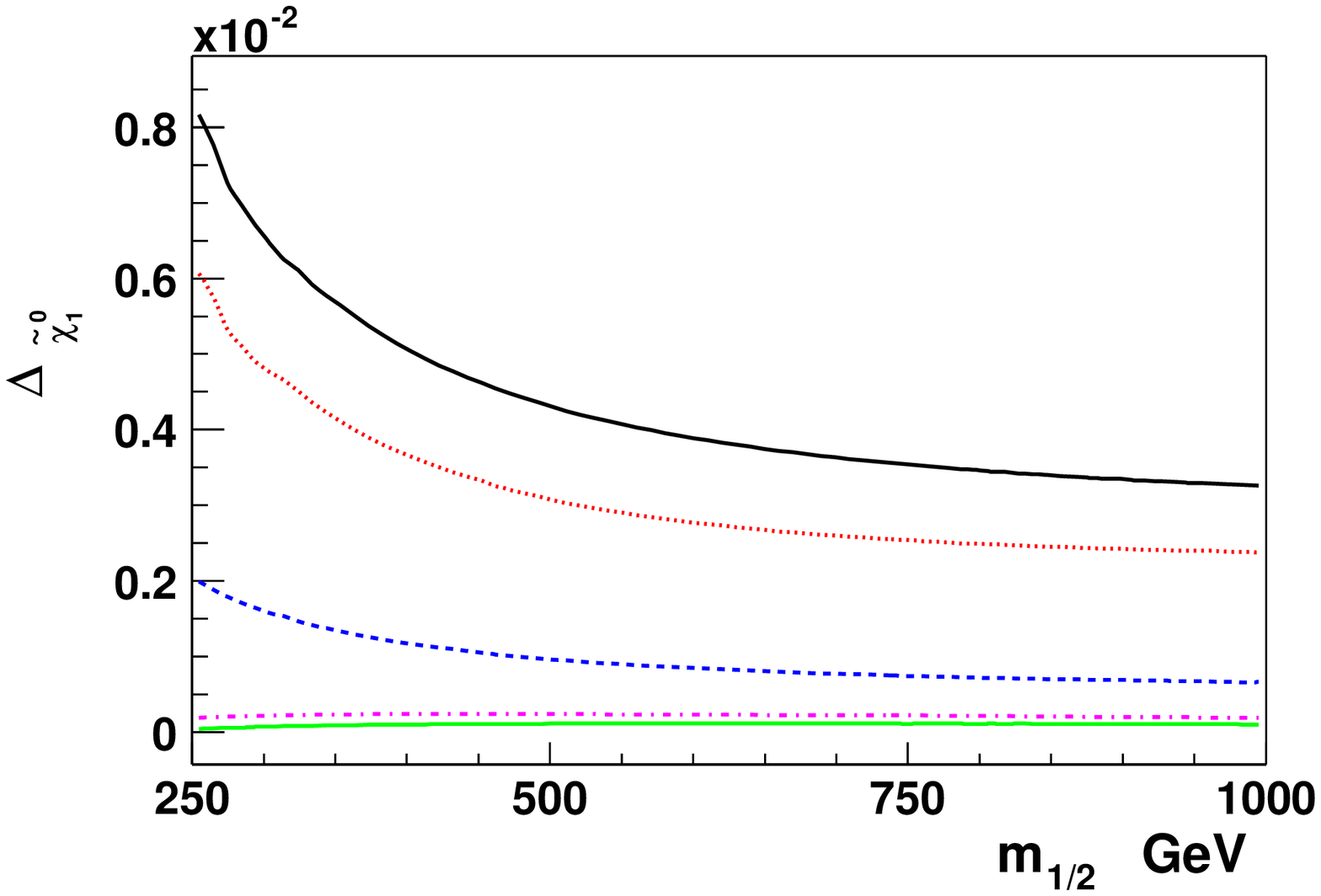,height=4.5cm,width=8.5cm}
\caption{$\Delta_p$ from the formula (\ref{deltap})
as functions defined along {\bf Model Line C}
from corrections depending on $m_{1/2}$. In the first picture we plotted a
spectrum of masses} \label{fig:linec}
\end{figure}

In Fig.\ref{fig:linec}, we present the results for $\Delta_p$
defined as functions along {\bf Model Line C} $$ m_0 = m_{1/2},
\quad m_{1/2} \mbox{ varies}, \quad A_0=0, \quad \tan\beta = 35,
\quad \mu > 0 . $$ where the benchmark point is $$ m_0 = 300 \GeV,
\quad m_{1/2} = 300 \GeV, \quad A_0 = 0 \GeV, \quad \tan\beta =
35, \quad \mu > 0 . $$

Situation with Fig.\ref{fig:linec} is analogous to that of
Fig.\ref{fig:linea}, even if we increase $\tan\beta$ from 10
up to 35. Importance of our corrections increases only at very
large values of $\tan\beta$ (50 and more), as one can see from
Fig.\ref{fig:tanb} .

\begin{figure}[h]
\psfig{figure=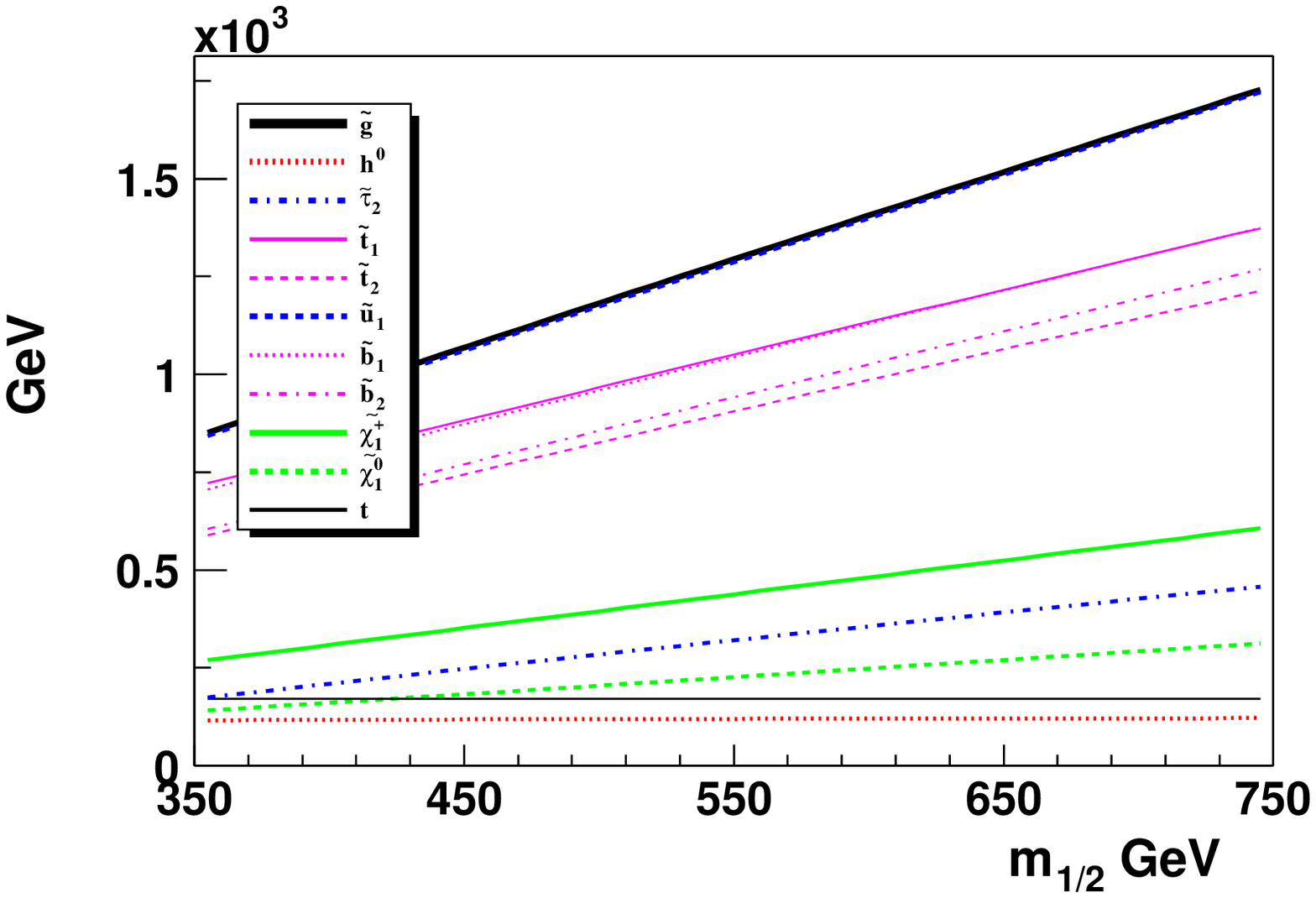,height=4.5cm,width=8.5cm}
\psfig{figure=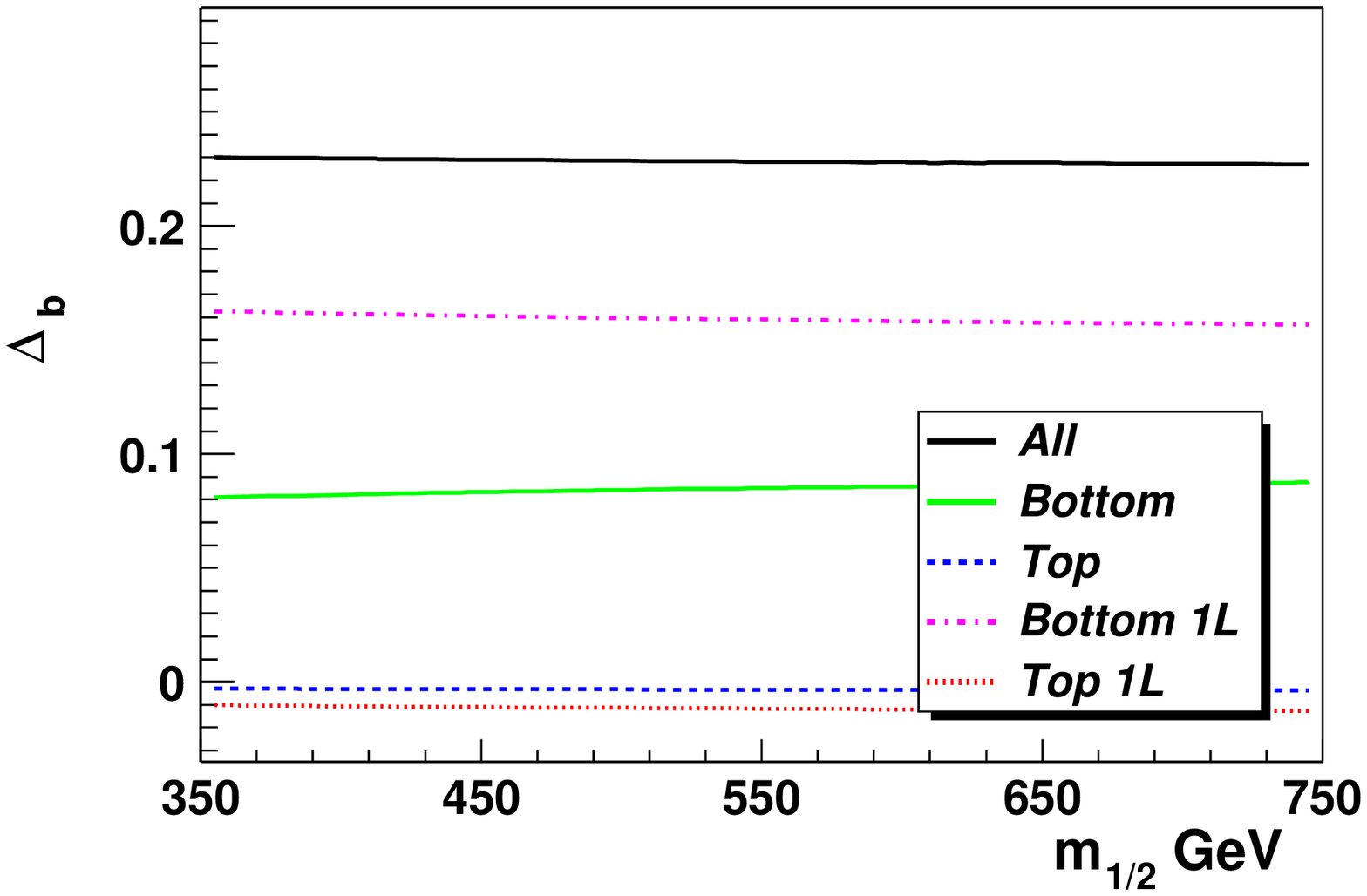,height=4.5cm,width=8.5cm}
\psfig{figure=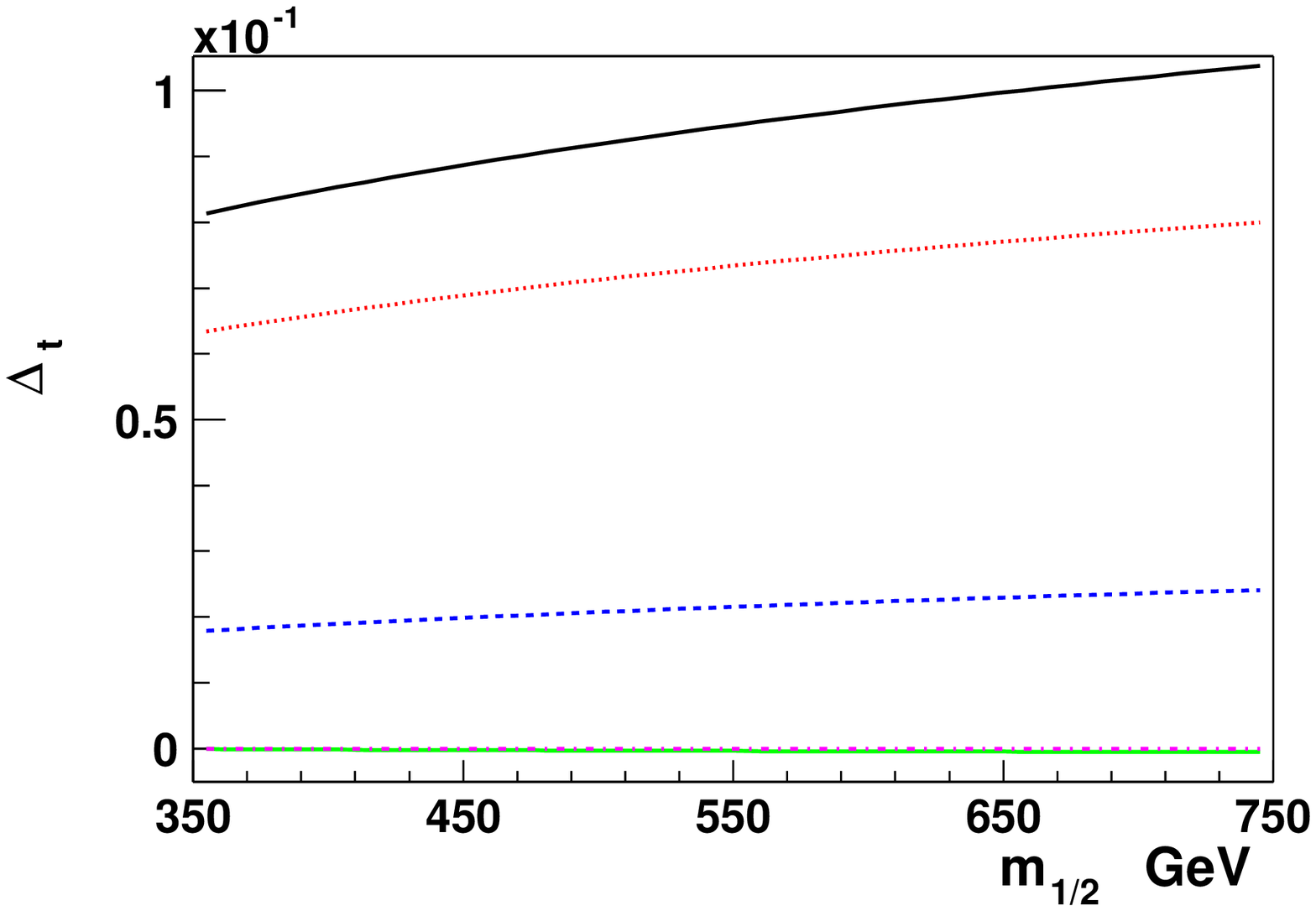,height=4.5cm,width=8.5cm}
\psfig{figure=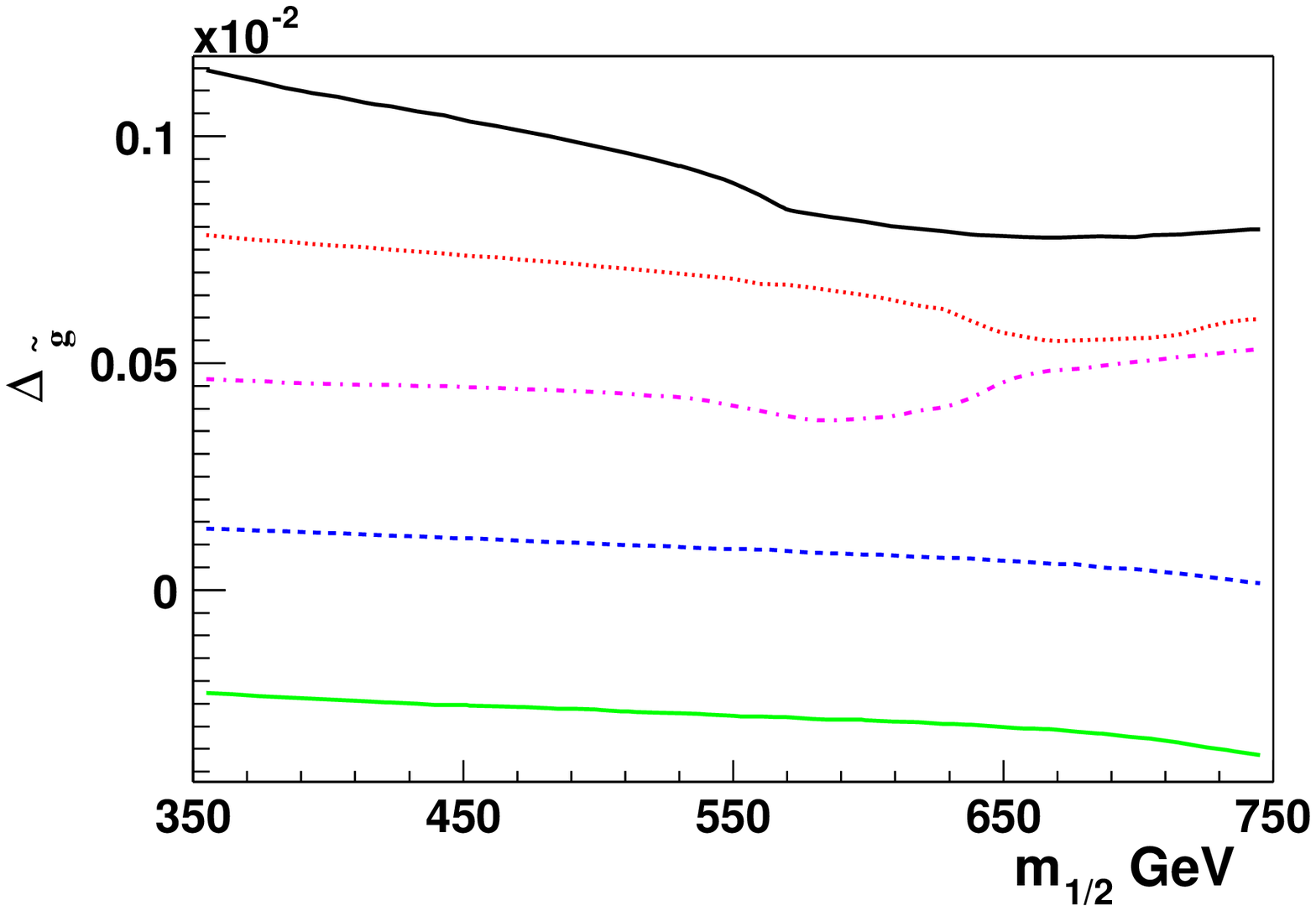,height=4.5cm,width=8.5cm}
\psfig{figure=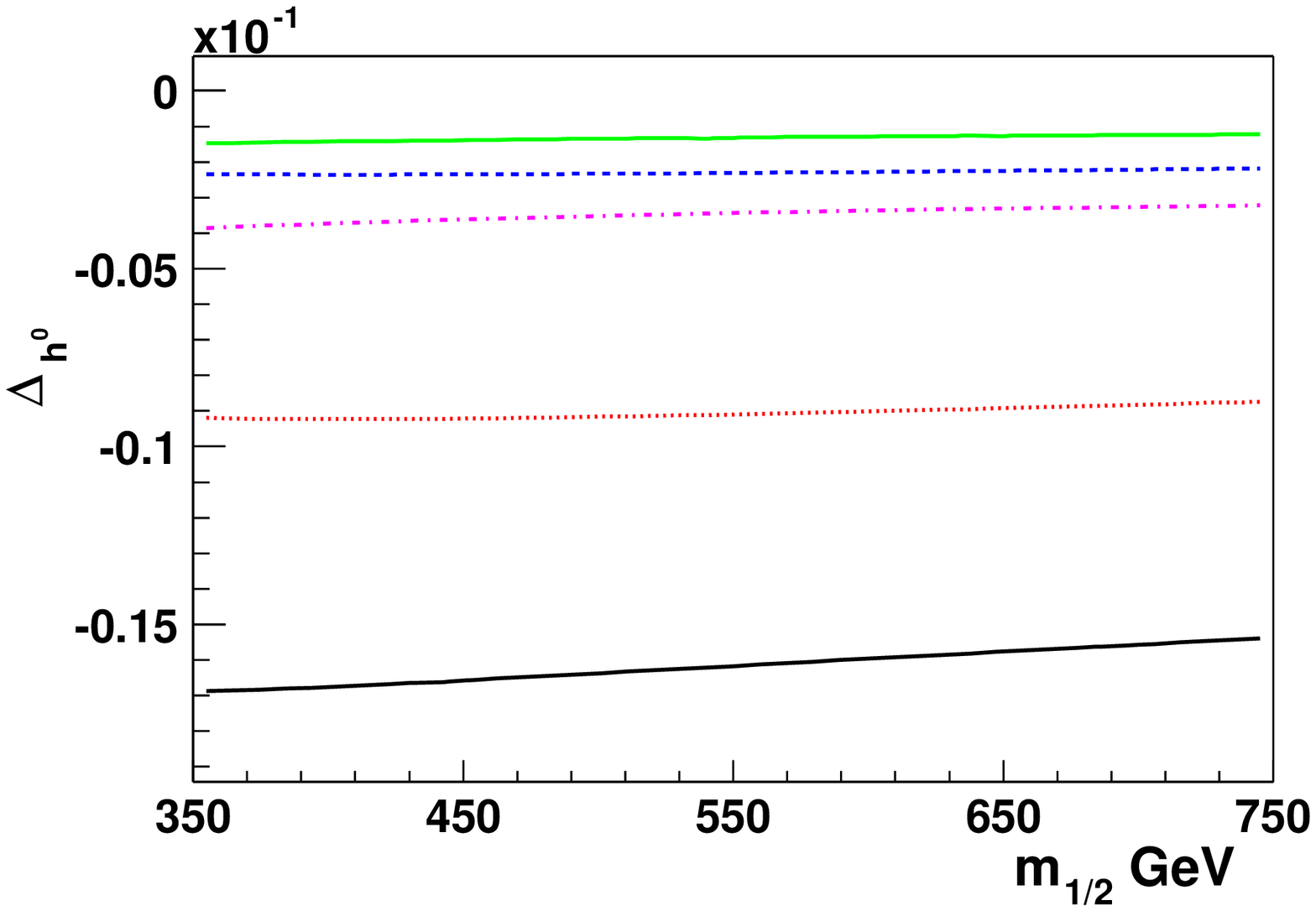,height=4.5cm,width=8.5cm}
\psfig{figure=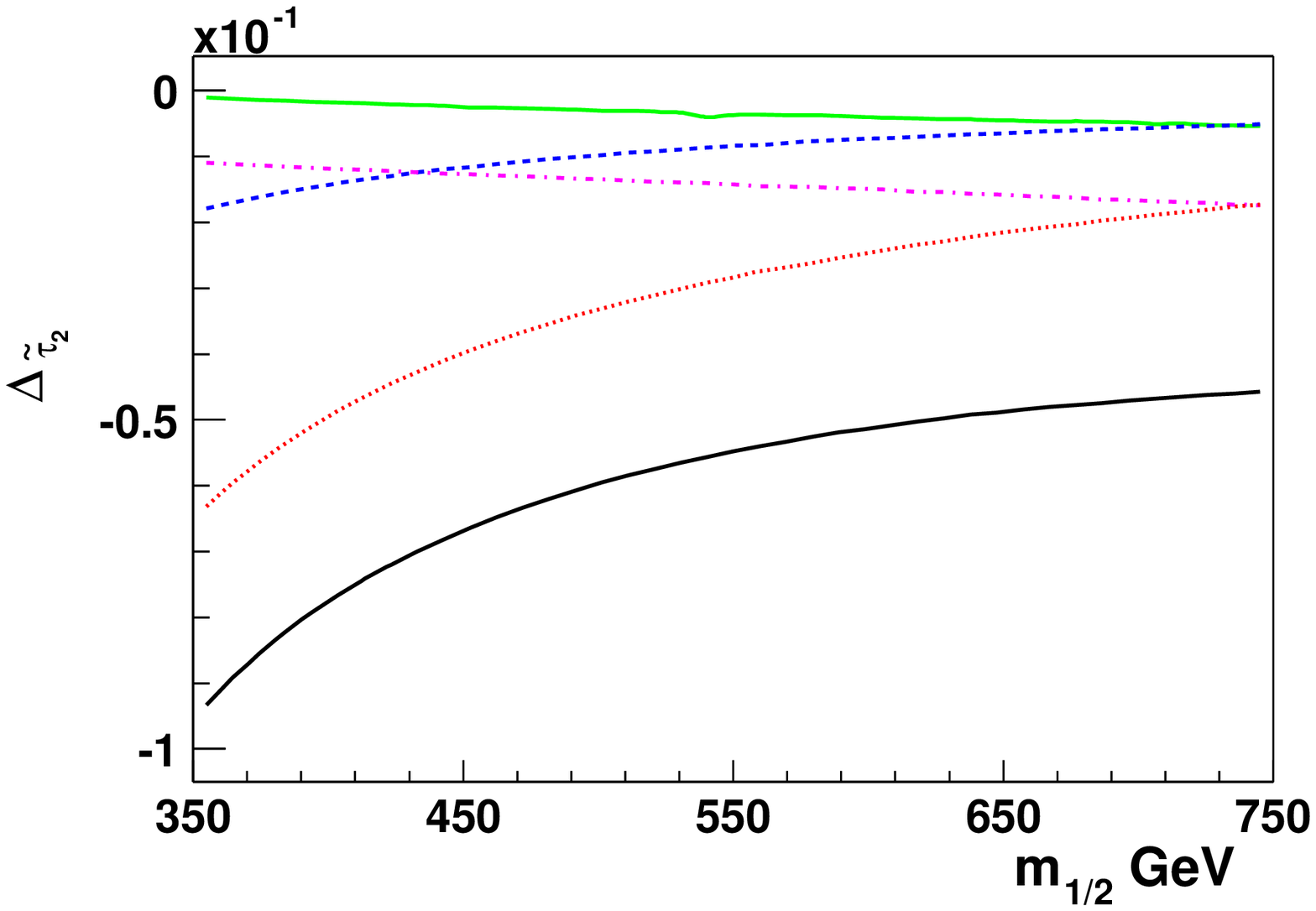,height=4.5cm,width=8.5cm}
\psfig{figure=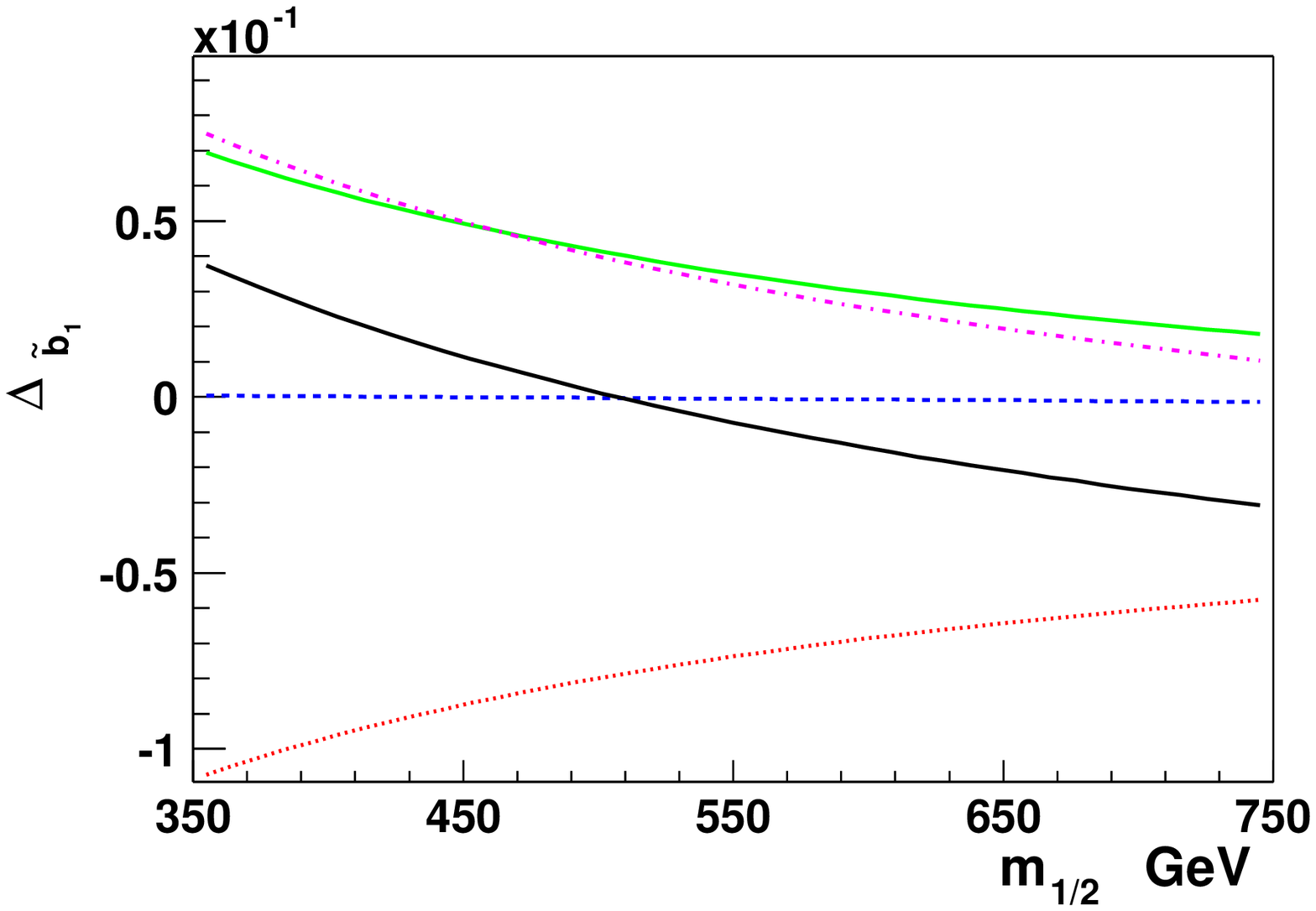,height=4.5cm,width=8.5cm}
\psfig{figure=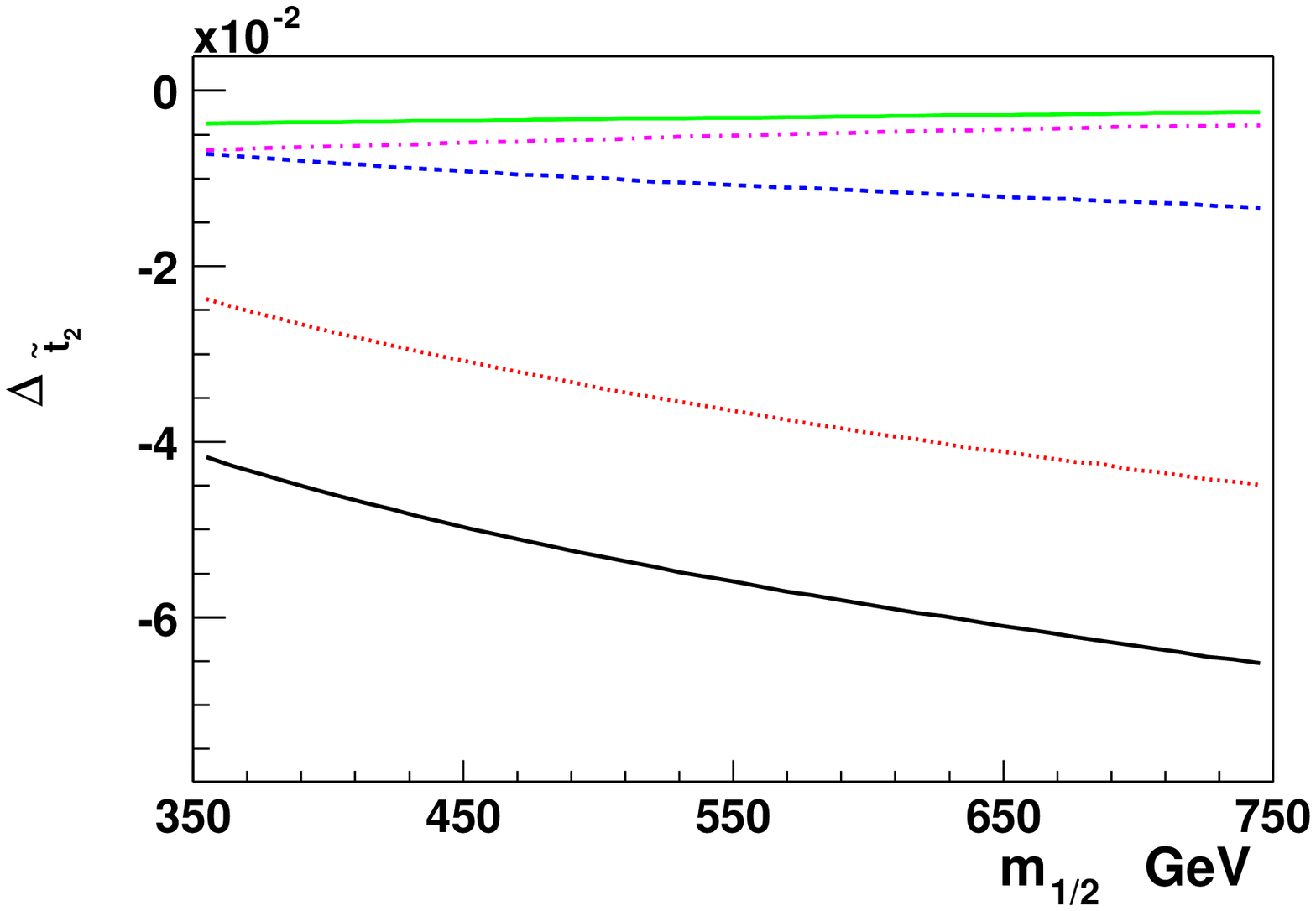,height=4.5cm,width=8.5cm}
\psfig{figure=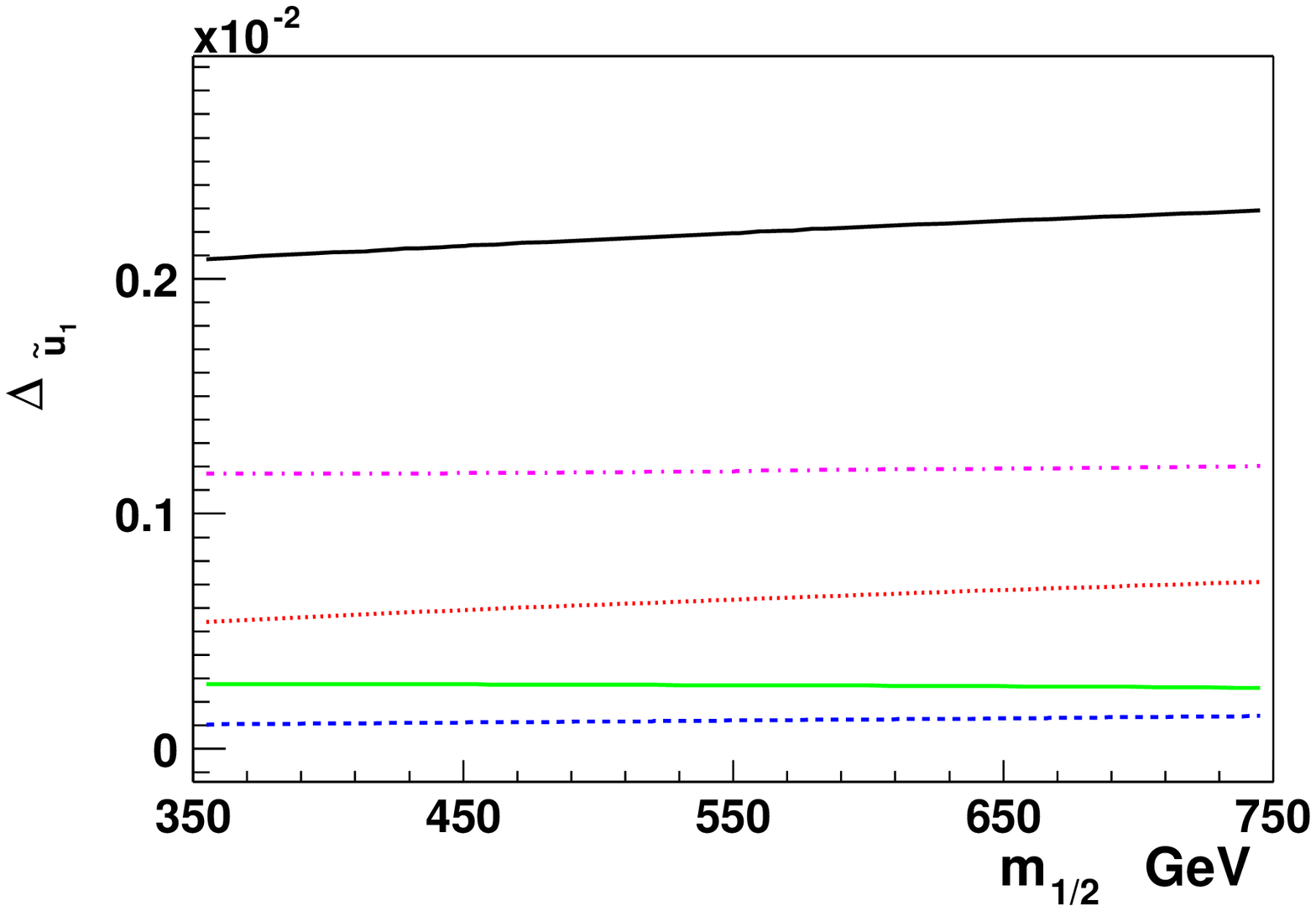,height=4.5cm,width=8.5cm}~~~~~~~
\psfig{figure=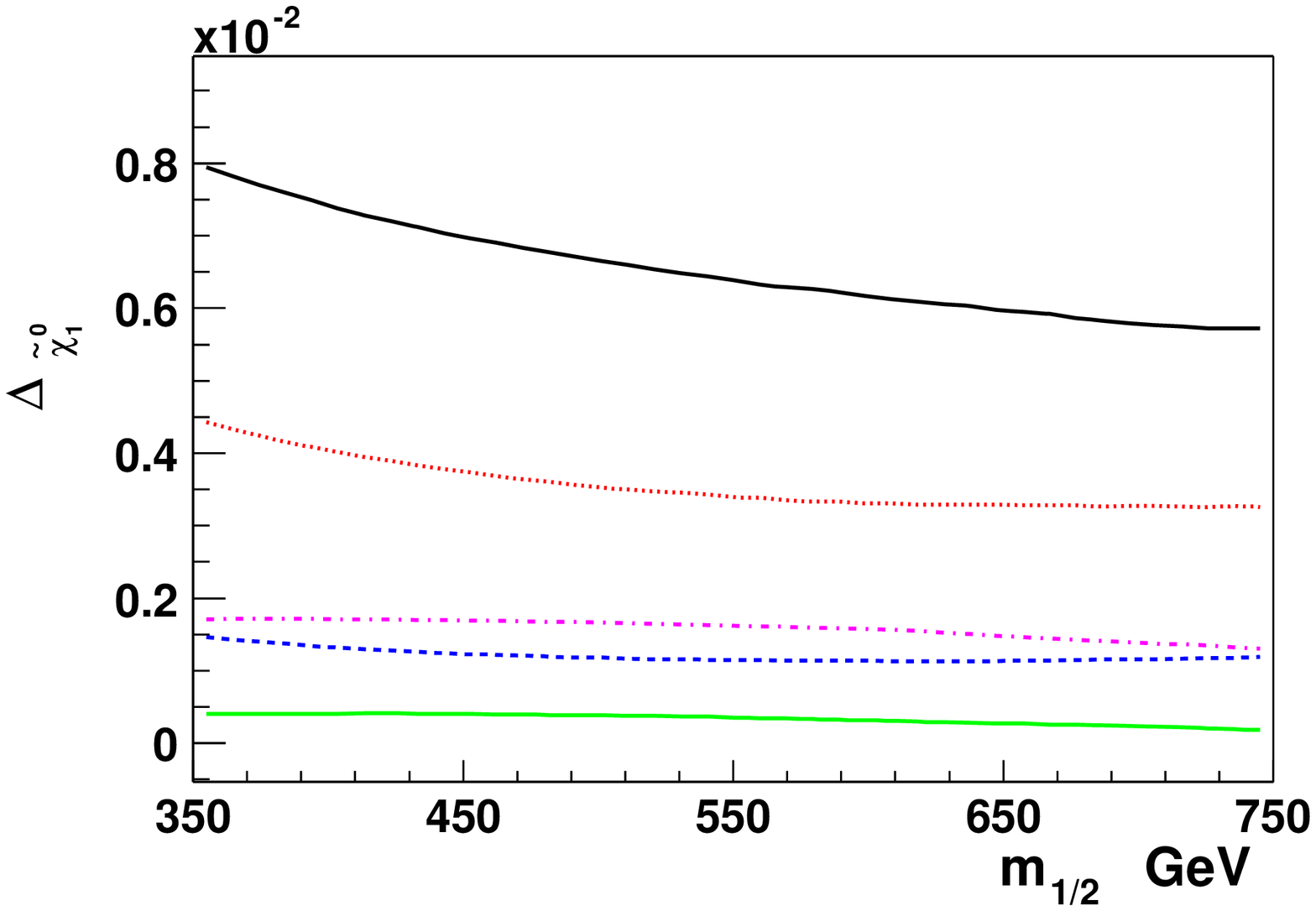,height=4.5cm,width=8.5cm}
\caption{$\Delta_p$ from the formula (\ref{deltap})
at large value of $\tan\beta=55$ as functions
of $m_{1/2}$. In the first picture we plotted a spectrum of masses}
\label{fig:linem}
\end{figure}

For such a large $\tan \beta$ we just used some particular values
$m_0$ and $m_{1/2}$ from the allowed region from~\cite{Ellis:2001ms}.
We use the following parameterization: $$ m_0 = m_{1/2}, \quad m_{1/2} \mbox{
varies}, \quad A_0=0, \quad \tan\beta = 55, \quad \mu > 0 .
$$ The results for $\Delta_p$ are shown in Fig.\ref{fig:linem}
and we can see that in
this case our 2-loop correction to the $b$-quark pole mass gives a
sizeable contribution to the masses of gluino and sbottoms.

\begin{center}
\begin{figure*}[t]
\psfig{figure= 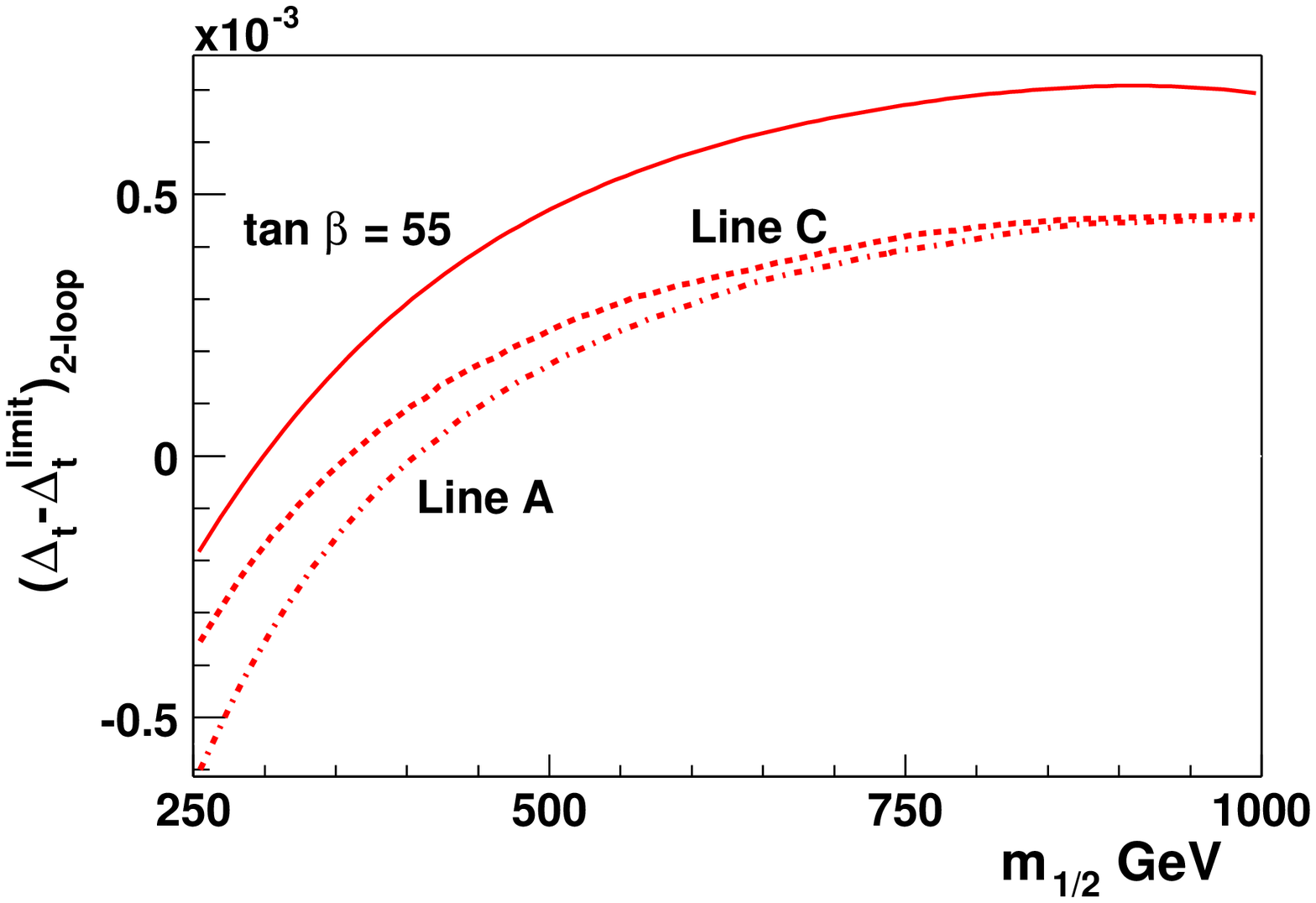,height=5.5cm,width=8.5cm}
\psfig{figure=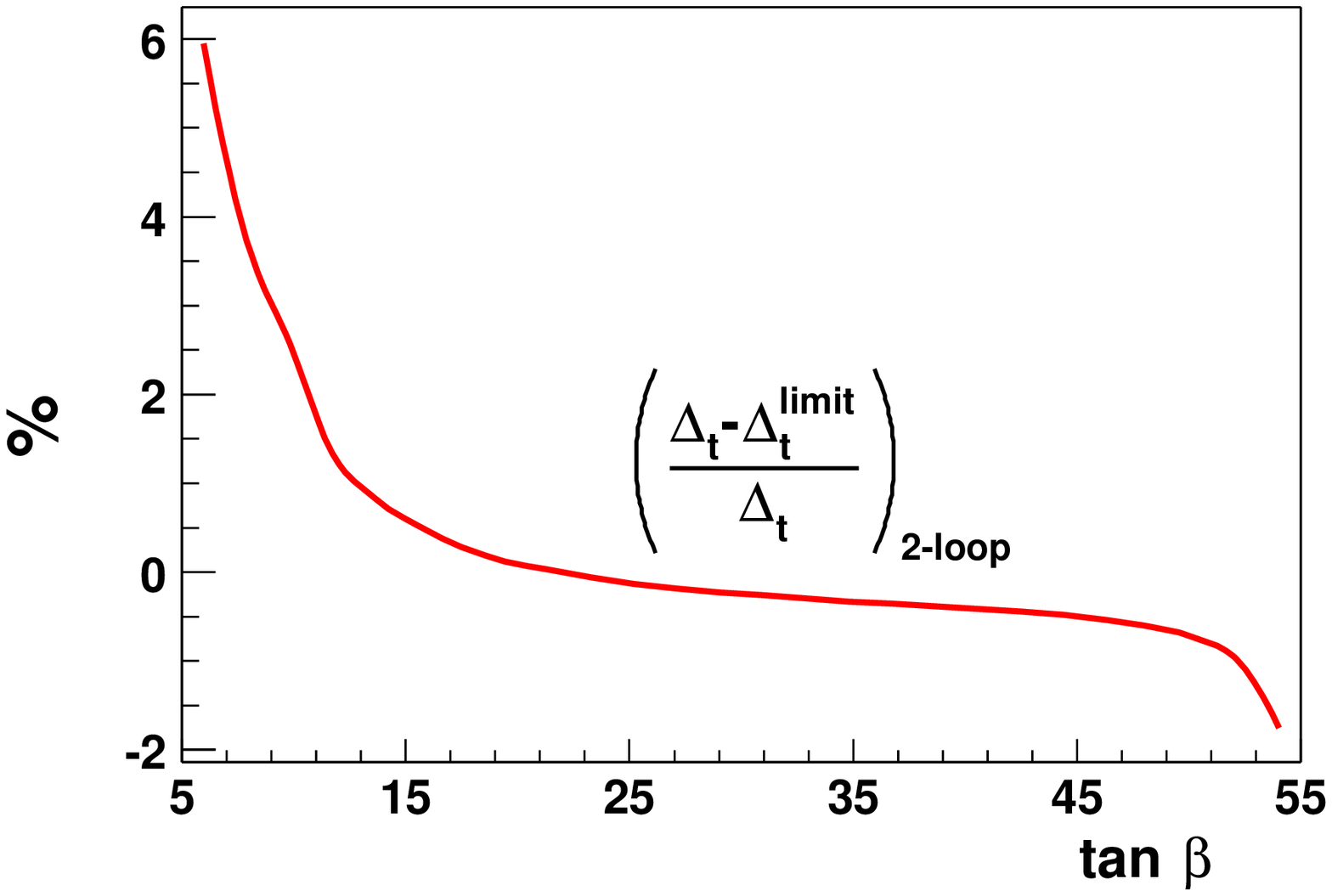,height=5.5cm,width=8.5cm}\\
\psfig{figure= 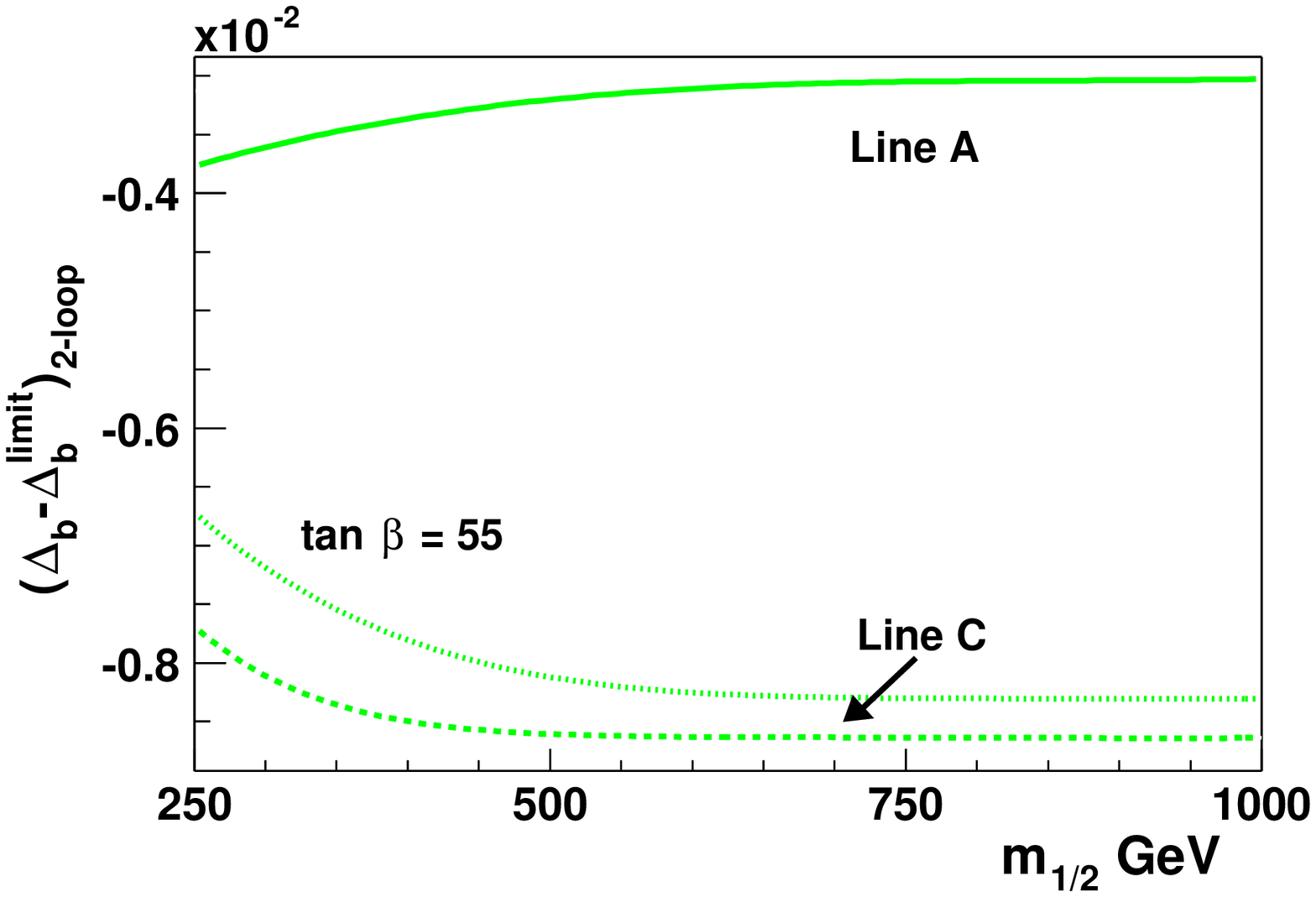,height=5.5cm,width=8.5cm}
\psfig{figure=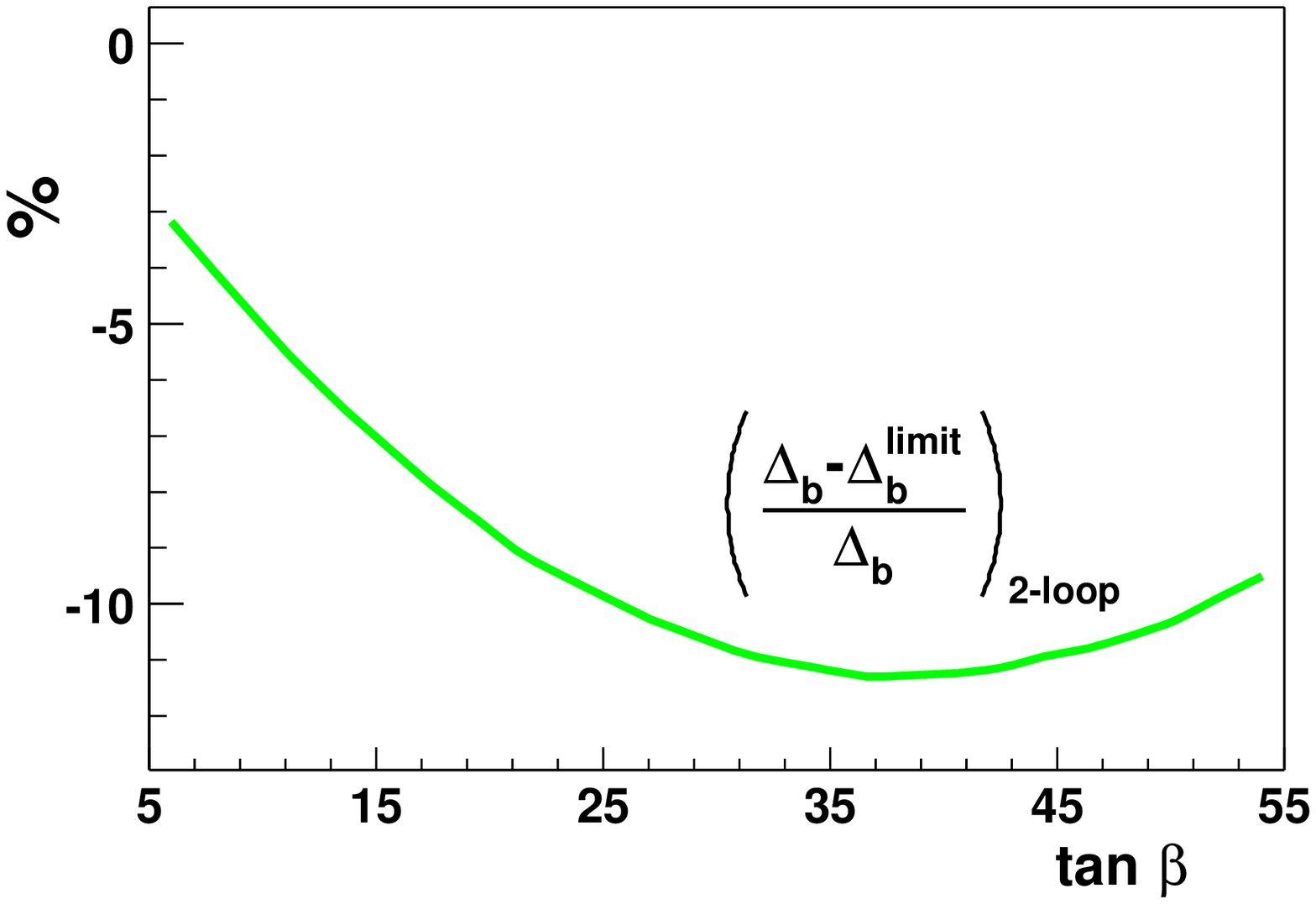,height=5.5cm,width=8.5cm}
\caption{Difference between 
$\Delta_p$ 
with the full two-loop
results and the limit from eq.(\ref{limit}). Upper plots corresponds to the
$t$-quark and bottom ones are for the $b$-quark. Left column 
shows the
dependence on $m_{1/2}$ of the difference between $\Delta_p$ and
$\Delta_p^{\mbox{limit}}$. Right column 
shows the relative
dependence on $\tan\beta$ of the difference between $\Delta_p$ and
$\Delta_p^{\mbox{limit}}$ for the same parameters as in text.
{\bf Line A} and {\bf Line C} correspond to Model Line A 
(see Fig.~\ref{fig:linea}) 
and Model Line C  (see Fig.~\ref{fig:linec}) }
\label{fig:limit}
\end{figure*}
\end{center}

Now let us discuss our limiting formula eq.(\ref{limit}). We have
studied the difference between our full result and this
approximate expression in different regions of MSSM parameter
space using the same strategy as above. As a result we find (see
Fig.~\ref{fig:limit}) that for the $b$-quark this difference does not
exceed 12\% for the parameters region considered above,
both for
the dependence on $\tan\beta$ and $m_{1/2}$. For the $t$-quark this is
even better.

\vspace*{-8.5mm}

\section{Conclusion}

We presented the results for two-loop MSSM corrections to the relation
between pole and running masses of heavy quarks up to the ${\cal
O}(\alpha_s^2)$ order. We provided a detail analysis of the value
of these corrections in different regions of parameter space and
discussed their impact on SUSY particle spectra. We would like to
note that our results presented here may also be included in the
codes of programs like Isajet \cite{isajet}, SuSpect \cite{suspect}
and used for predictive phenomenological analysis like
\cite{Ellis:2001ms}, \cite{deBoer:1995fr}, \cite{Blazek:1996yv},
\cite{Baer:2002gm}. In one of our next papers we suppose to
calculate missing corrections from stop-chargino loops to the
$b$-quark pole mass 
and to provide  more terms in the expansion
in the  relation between the top pole and \DR masses.

The authors would like to thank K.Chetyrkin, M.Kalmykov, D.Kazakov
and V.Smirnov for fruitful discussions and multiple comments. This
work of A.O. and O.V. was supported by DFG-Forschergruppe
"Quantenfeldtheorie, Computeralgebra and Monte-Carlo-Simulation"
(contract FOR 264/2-1) and by BMBF under grant No 05HT9VKB0.
Financial support for A.B. and V.V. from RFBR grants \#
02-02-16889 and \# 00-15-96691 is kindly acknowledged.



\begin{thebibliography}{**}

\bibitem{Haber:1984rc}
H.~E.~Haber and G.~L.~Kane,
Phys.\ Rept.\  {\bf 117} (1985) 75.

\bibitem{Nilles:1983ge}
H.~P.~Nilles,
Phys.\ Rept.\  {\bf 110} (1984) 1.

\bibitem{Barbieri:xf}
R.~Barbieri,
Riv.\ Nuovo Cim.\  {\bf 11N4} (1988) 1.

\bibitem{Siegel}
W. Siegel, Phys. Lett. B {\bf 84} (1979) 193.

\bibitem{Avdeev:1982xy}
L.~V.~Avdeev and A.~A.~Vladimirov,
Nucl.\ Phys.\ B {\bf 219} (1983) 262.

\bibitem{SoftSUSY}
B.~C.~Allanach,
Comput.\ Phys.\ Commun.\  {\bf 143} (2002) 305
[arXiv:hep-ph/0104145].

\bibitem{Avdeev:1997sz}
L.~V.~Avdeev and M.~Y.~Kalmykov,
Nucl.\ Phys.\ B {\bf 502} (1997) 419 [arXiv:hep-ph/9701308].

\bibitem{Gray:1990yh}
N.~Gray, D.~J.~Broadhurst, W.~Grafe and K.~Schilcher,
Z.\ Phys.\ C {\bf 48} (1990) 673.

\bibitem{Fleischer:1998dw}
J.~Fleischer, F.~Jegerlehner, O.~V.~Tarasov and O.~L.~Veretin,
Nucl.\ Phys.\ B {\bf 539} (1999) 671 [Erratum-ibid.\ B {\bf 571}
(1999) 511] [arXiv:hep-ph/9803493].

\bibitem{Chetyrkin:1999ys}
K.~G.~Chetyrkin and M.~Steinhauser,
Nucl.\ Phys.\ B {\bf 573} (2000) 617.

\bibitem{Melnikov:2000qh}
K.~Melnikov and T.~v.~Ritbergen,
Phys.\ Lett.\ B {\bf 482} (2000) 99.

\bibitem{Hempfling:1993kv}
R.~Hempfling,
Phys.\ Rev.\ D {\bf 49} (1994) 6168.

\bibitem{Hall:1993gn}
L.~J.~Hall, R.~Rattazzi and U.~Sarid,
Phys.\ Rev.\ D {\bf 50} (1994) 7048 [arXiv:hep-ph/9306309].

\bibitem{SUSY94&Wright} D. Pierce, [arXiv:hep-ph/9407202], in {\sl Proceedings
of the International Workshop on Supersymmetry and Unification of
Fundamental Interactions: SUSY 94}, \rm eds. C. Kolda and J.D.
Wells (1994);

\bibitem{Donini95}
A.~Donini,
Nucl.\ Phys.\ B {\bf 467} (1996) 3 [arXiv:hep-ph/9511289].

\bibitem{Pierce:1996zz}
D.~M.~Pierce, J.~A.~Bagger, K.~T.~Matchev and R.~j.~Zhang,
Nucl.\ Phys.\ B {\bf 491} (1997) 3 [arXiv:hep-ph/9606211].

\bibitem{else}
R. Tarrach, Nucl. Phys. B{\bf 183} (1981) 384.

\bibitem{onshell2}
J.~Fleischer and M.~Y.~Kalmykov,
Comput.\ Phys.\ Commun.\  {\bf 128}, 531 (2000)
[arXiv:hep-ph/9907431].


\bibitem{Davydychev}
A.~I.~Davydychev and J.~B.~Tausk,
Nucl.\ Phys.\ B {\bf 397} (1993) 123.

\bibitem{jones}
I. Jack and D. R. T. Jones,
Phys. Lett. B {\bf 333} (1994) 372; \\
I. Jack, D. R. T. Jones, S. P. Martin, M. T. Vaughn and Y. Yamada,
Phys. Rev. D {\bf 50} (1994) 5481.

\bibitem{spurion} L.~Girardello and M.~T.~Grisaru, {\em
Nucl. Phys.,}  {\bf B194} (1982) 65.

\bibitem{Kuroda}
M.~Kuroda, KEK-CP-080,
[arXiv:hep-ph/9902340].

\bibitem{Ellis:1983ed}
J.~R.~Ellis and S.~Rudaz,
Phys.\ Lett.\ B {\bf 128} (1983) 248.

\bibitem{FeynArts}
J.~Kublbeck, M.~Bohm and A.~Denner,
Comput.\ Phys.\ Commun.\  {\bf 60} (1990) 165; \\
T.~Hahn,
Comput.\ Phys.\ Commun.\  {\bf 140} (2001) 418
[arXiv:hep-ph/0012260]; \\
T.~Hahn and M.~Perez-Victoria   
Comput.\ Phys.\ Commun.\  {\bf 118} (1999) 153
[arXiv:hep-ph/9807565].

\bibitem{FAMSSM}
T.~Hahn and C.~Schappacher,
Comput.\ Phys.\ Commun.\  {\bf 143} (2002) 54
[arXiv:hep-ph/0105349].


\bibitem{BDEGMOPW}
M.~Battaglia, A.~De~Roeck, J.~Ellis, F.~Gianotti, K.T.~Matchev,
K.A.~Olive, L.~Pape and G.W.~Wilson,
Eur.\ Phys.\ J.\ C {\bf 22}, 535 (2001), [arXiv:hep-ph/0106204];
Snowmass P3-47, [arXiv:hep-ph/0112013].

\bibitem{aix}
A.~Djouadi et al., GDR SUSY Workshop Aix-la-Chapelle 2001,
see \\{\tt http://www.desy.de/$\sim$heinemey/LesPointsdAix.html}.

\bibitem{modelline}
S.P.~Martin, {\tt
http://zippy.physics.niu.edu/modellines.html};\\
S.P.~Martin, S.~Moretti, J.~Qian and G.W.~Wilson, Snowmass P3-46.

\bibitem{SPS}
B.~C.~Allanach {\it et al.},
in {\it Proc. of the APS/DPF/DPB Summer Study on the Future of Particle Physics (Snowmass 2001) } ed. R.~Davidson and C.~Quigg,
[arXiv:hep-ph/0202233].

\bibitem{Ellis:2001ms}
J.~R.~Ellis, T.~Falk, G.~Ganis, K.~A.~Olive and M.~Srednicki,
Phys.\ Lett.\ B {\bf 510} (2001) 236
[arXiv:hep-ph/0102098].

\bibitem{isajet}
H.~Baer, F.~E.~Paige, S.~D.~Protopopescu and X.~Tata,
[arXiv:hep-ph/0001086].

\bibitem{suspect}
A. Djouadi, J.L. Kneur and G. Moultaka, "{\bf
SuSpect}: a program for the supersymmetric spectrum", to appear.
A preliminary version of the users manual 
[arXiv:hep-ph/9901246]; 
A.~Djouadi, J.~L.~Kneur and G.~Moultaka,
[arXiv:hep-ph/0211331].
{\tt http://www.lpm.univ.montp2.fr:6714/\~{}kneur/suspect.html}

\bibitem{deBoer:1995fr}
W.~de Boer {\it et al.},
Z.\ Phys.\ C {\bf 71} (1996) 415
[arXiv:hep-ph/9603350]; \\
W.~de Boer, R.~Ehret and D.~I.~Kazakov,
Phys.\ Lett.\ B {\bf 334} (1994) 220
[arXiv:hep-ph/9405419].

\bibitem{Blazek:1996yv}
T.~Blazek, M.~Carena, S.~Raby and C.~E.~Wagner,
Phys.\ Rev.\ D {\bf 56} (1997) 6919
[arXiv:hep-ph/9611217].

\bibitem{Baer:2002gm}
H.~Baer, C.~Balazs, A.~Belyaev, J.~K.~Mizukoshi, X.~Tata and Y.~Wang,
JHEP {\bf 0207} (2002) 050
[arXiv:hep-ph/0205325].

\bibitem{Jegerlehner:2001fb}
F.~Jegerlehner, M.~Y.~Kalmykov and O.~Veretin,
Nucl.\ Phys.\ B {\bf 641} (2002) 285.

\end{thebibliography}
\end{document}